%% file: main.tex
\documentclass{article}
\usepackage[utf8]{inputenc}
\usepackage[english]{babel}

\usepackage{amsmath}
\usepackage{amsfonts}
\usepackage{graphicx}
\usepackage{subcaption}
\usepackage{wrapfig}
\usepackage{tikz}
\usepackage{rotating}
\usepackage[export]{adjustbox}
\usepackage{float}
\usepackage{mathrsfs}
\usepackage{amssymb}
\usepackage{slashed}
\usepackage{physics}
\usepackage{simpler-wick}
\usepackage[utf8]{inputenc}
\usepackage{amsthm}
\usepackage{microtype}
\usepackage{bm}
\usepackage{bbm}
\usepackage{cite}
\usepackage{soul}
\usepackage[LGR,T1]{fontenc}
\usepackage[utf8]{inputenc}  

\usepackage{pifont}

\usepackage{cancel}
\usepackage{pdfpages}
\allowdisplaybreaks
\makeatletter

\renewcommand*\env@matrix[1][*\c@MaxMatrixCols c]{%
	\hskip -\arraycolsep
	\let\@ifnextchar\new@ifnextchar
	\array{#1}}
\makeatother
\usepackage{axodraw2}

\newcommand{\nn}{\nonumber}

\newcommand{\M}{\mathcal{M}}

\newcommand{\s}{\hat{s}}
\newcommand{\eps}{\epsilon}

\newcommand{\aem}{\alpha}
\newcommand{\ord}{{\cal O}}

\newcommand\Eqn[1]     {Eq.\,(\ref{#1})}

\newcommand\eqn[1]     {eq.\,(\ref{#1})}
\newcommand\eqns[2]    {eqs.\,(\ref{#1}) and~(\ref{#2})}
\newcommand\eqnss[2]   {eqs.\,(\ref{#1})--(\ref{#2})}

\newcommand\refr[1]      {ref.\,\cite{#1}}
\newcommand\refrs[1]    {refs.\,\cite{#1}}

\def\beq{\begin{equation}}
\def\eeq{\end{equation}}
\def\beqa{\begin{eqnarray}}
\def\eeqa{\end{eqnarray}}

\newcommand{\be}{\begin{equation}}
\newcommand{\ee}{\end{equation}}
\newcommand{\bea}{\begin{eqnarray}}
\newcommand{\eea}{\end{eqnarray}}

\newcommand{\al}{\alpha}

\newcommand{\ga}{\gamma}

\newcommand{\la}{\lambda}
\newcommand{\si}{\sigma}

\newcommand{\cA}{\mathcal{A}}

\newcommand{\cM}{\mathcal{M}}

\newcommand{\cO}{\mathcal{O}}

\usepackage{jheppub}
\input{NLP_axodraw}

\title{Next-to-leading power jet functions in the small-mass limit in QED}
\author[a]{Robin van Bijleveld,}
\author[a,b,c]{Eric Laenen,} 
\author[a,b]{Coenraad Marinissen,}
\author[d]{Leonardo Vernazza}
\author[e]{and Guoxing Wang}

\affiliation[a]{Nikhef, Theory Group, Science Park 105, 1098 XG, Amsterdam, The Netherlands}
\affiliation[b]{IoP/ITFA, University of Amsterdam, Science Park 904, 1098 XH Amsterdam, The Netherlands}
\affiliation[c]{ITF, Utrecht University, Leuvenlaan 4, 3584 CE Utrecht, The Netherlands}
\affiliation[d]{INFN, Sezione di Torino, Via P. Giuria 1, I-10125 Torino, Italy}
\affiliation[e]{Laboratoire de Physique Th\'eorique et Hautes Energies (LPTHE), UMR 7589, Sorbonne Universit\'e et CNRS, 4 place Jussieu, 75252 Paris Cedex 05, France}

\abstract{
We investigate the factorization properties of the massive fermion form factor in QED, to next-to-leading power in the fermion mass, and up to two-loop order. For this purpose we define new jet functions that have multiple connections to the hard part as operator matrix elements, and compute them to second order in the coupling. We test our factorization formula using these new jet functions in a region-based analysis and find that 
factorization indeed holds. We address a number of subtle aspects such as rapidity regulators and external line corrections, and we find an interesting sequence of relations among the jet functions.
}
\begin{document}

\maketitle

\section{Introduction}\label{sec:intro}
\input{Sections/Introduction}

\section{Factorization in the small-mass limit}\label{sec:setup}
\input{Sections/Set_up}

\section{One-loop results up to NLP}\label{sec:oneloop}
\input{Sections/One_Loop}

\section{Two-loop results up to NLP}\label{sec:twoloop}
\input{Sections/Two_Loop}

\section{Conclusions}\label{sec:conclusions}
\input{Sections/Conclusion}

\appendix

\section{Sequence of jet functions}\label{app:seqjets}
\input{Appendices/seqjets}

\section{Proof of factorization for the collinear-anticollinear region}\label{app:regtofact}
\input{Appendices/regtofact}

\section{Two-loop integral expressions for jet functions}\label{app:2loopres}
\input{Appendices/2loopres}

\section{UV counterterms}\label{app:renorm}
\input{Appendices/renormalization}

\bibliography{spire.bib}
\bibliographystyle{JHEP.bst}

\end{document}

%% file: NLP_axodraw.tex
\def\Jf1{
	\begin{axopicture}(102,108)(0,-54)
		\Line[arrow](0,0)(60,30)
		\Line[arrow](60,30)(100,50)
		\Line(98,54)(102,46)
		
		\Photon(60,-30)(60,30){4}{4}
		\Line[double](0,0)(80,-40)
		
		\Line[arrow,arrowpos=1,arrowscale=0.8](84,32)(94,37)
		\Text(90,25){$p_1$}
	\end{axopicture}
}

\def\Jfg1{
	\begin{axopicture}(102,108)(0,-54)
		\Line[arrow](0,0)(60,30)
		\Line[arrow](60,30)(100,50)
		\Line(98,54)(102,46)
		
		\Photon(60,-30)(60,30){4}{4}
		\GCirc(56,-28){4}{0}
		
		\Line[arrow,arrowpos=1,arrowscale=0.8](84,32)(94,37)
		\Text(90,25){$p_1$}
	\end{axopicture}
}

\def\Hfgfb0{
	\begin{axopicture}(132,108)(-30,-54)
	\Photon(-30,0)(0,0){4}{3}
	\Line[arrow](0,0)(50,50)
	\Line[arrow,flip](0,0)(25,-25)
	\Line[arrow,flip](25,-25)(50,-50)
	\Line[dash,arrow,arrowpos=1](25,-25)(55,5)
	\Text(60,40){$p_1-\ell$}
	\Text(60,0){$\ell$}
	\Line[arrow,arrowpos=1,arrowscale=0.8](45,-32)(53,-40)
	\Text(55,-30){$p_2$}
	\end{axopicture}
}

\def\QEDtree{
	\begin{axopicture}(102,100)(-30,-50)
		\Photon(-30,0)(0,0){4}{3}
		\Line[arrow](0,0)(50,50)
		\Line[arrow,flip](0,0)(50,-50)
		\Line[arrow,arrowpos=1,arrowscale=0.8](45,32)(53,40)
		\Text(55,30){$p_1$}
		\Line[arrow,arrowpos=1,arrowscale=0.8](45,-32)(53,-40)
		\Text(55,-30){$p_2$}
	\end{axopicture}
}

\def\Hfgfbgf{
	\begin{axopicture}(90,80)(-30,-40)
		\Photon(-30,0)(0,0){4}{3}
		\Line[arrow](0,0)(50,50)
		\Line[arrow,flip](0,0)(20,-20)
		\Line[arrow,flip](20,-20)(40,-40)
		\Line[arrow,flip](40,-40)(50,-50)
		
		\Line[dash,arrow,arrowpos=1](40,-40)(60,-20)
		\Bezier[dash,arrow,arrowpos=1](20,-20)(23,-45)(30,-48)(40,-50)

		\Text(65,40){$p_1\!-\!\ell_1$}
		\Text(70,-15){$\ell_1$}
		\Text(65,-40){$\ell_2\!-\!p_2\!$}
		\Text(15,-40){$\ell_2$}
	\end{axopicture}
}
\def\Hfgfbgg{
	\begin{axopicture}(90,80)(-30,-40)
		\Photon(-30,0)(0,0){4}{3}
		\Line[arrow,flip](0,0)(50,-50)
		\Line[arrow](0,0)(20,20)
		\Line[arrow](20,20)(40,40)
		\Line[arrow](40,40)(50,50)
		
		\Line[dash,arrow,arrowpos=1](40,40)(60,20)
		\Bezier[dash,arrow,arrowpos=1](20,20)(23,45)(30,48)(40,50)

		\Text(65,-40){$\ell_2\!-\!p_2$}
		\Text(70,15){$\ell_2$}
		\Text(65,40){$p_1\!-\!\ell_1$}
		\Text(15,40){$\ell_1$}
	\end{axopicture}
}
\def\Hfgfbgh{
	\begin{axopicture}(90,80)(-30,-40)
		\Photon(-30,0)(0,0){4}{3}
		\Line[arrow](0,0)(30,30)
		\Line[arrow](30,30)(50,50)
		\Line[arrow,flip](0,0)(30,-30)
		\Line[arrow,flip](30,-30)(50,-50)
		
		\Line[dash,arrow,arrowpos=1](30,30)(50,10)
		\Line[dash,arrow,arrowpos=1](30,-30)(50,-10)

		\Text(65,40){$p_1\!-\!\ell_1$}
		\Text(60,15){$\ell_2$}
		\Text(60,-15){$\ell_1$}
		\Text(65,-40){$\ell_2\!-\!p_2\!$}
	\end{axopicture}
}

\def\Jfgrule{
    \begin{axopicture}(70,50)(0,-4)
	\Photon(35,35)(35,0){4}{4}
	\Text(35,40){$\si$}
	\GCirc(35,0){4}{0}
	\Text(35,-10){$\rho$}
        \Line[arrow,flip,arrowpos=0,arrowscale=0.8](45,25)(45,15)
	\Text(50,20){$k$}
\end{axopicture}
}

\def\Jfda{
	\begin{axopicture}(140,130)(0,-60)
		\Line[arrow](100,50)(140,70)
		\Line(138,74)(142,66)
		\Line[double](0,0)(120,-60)
		\Line[arrow](0,0)(100,50)
		\Photon(100,-50)(100,-17){4}{3}
		\Arc[dash,arrow](100,0)(17,0,360)
		\Photon(100,17)(100,50){4}{3}
		\Line[arrow,arrowpos=1,arrowscale=0.8](124,52)(134,57)
		\Text(130,45){$p_1$}
	\end{axopicture}
}
\def\Jfdc{
	\begin{axopicture}(140,130)(0,-60)
		\Line[arrow](100,50)(140,70)
		\Line(138,74)(142,66)
		\Line[double](0,0)(120,-60)
		\Line[arrow](0,0)(30,15)
		\Line[arrow](30,15)(70,35)
		\Line[arrow](70,35)(100,50)
		\PhotonArc(50,25)(23,-153,27){4}{6}
		\Photon(100,-50)(100,50){4}{7}
		\Line[arrow,arrowpos=1,arrowscale=0.8](124,52)(134,57)
		\Text(130,45){$p_1$}
	\end{axopicture}
}
\def\Jfdd{
	\begin{axopicture}(140,130)(0,-60)
		\Line[arrow](100,50)(140,70)
		\Line(138,74)(142,66)
		\Line[double](0,0)(120,-60)
		\Line[arrow](0,0)(50,25)
		\Line[arrow](50,25)(100,50)
		\Photon(50,-25)(50,25){4}{4}
		\Photon(100,-50)(100,50){4}{7}
		\Line[arrow,arrowpos=1,arrowscale=0.8](124,52)(134,57)
		\Text(130,45){$p_1$}
	\end{axopicture}
}
\def\Jfde{
	\begin{axopicture}(140,130)(0,-60)
		\Line[arrow](100,50)(140,70)
		\Line(138,74)(142,66)
		\Line[double](0,0)(120,-60)
		\Line[arrow](0,0)(100,50)
		\Photon(100,-50)(100,-17){4}{3}
		\Arc[arrow](100,0)(17,0,360)
		\Photon(100,17)(100,50){4}{3}
		\Line[arrow,arrowpos=1,arrowscale=0.8](124,52)(134,57)
		\Text(130,45){$p_1$}
	\end{axopicture}
}
\def\Jfdg{
	\begin{axopicture}(140,130)(0,-60)
		\Line[arrow](100,50)(140,70)
		\Line(138,74)(142,66)
		\Line[double](0,0)(100,-50)
		\Line[arrow](0,0)(50,25)
		\Photon(50,25)(100,50){4}{4}
		\Arc(75,37.5)(28,-153,27)
		\Photon(80,-40)(80,10){4}{5}
		\Line[arrow,arrowpos=1,arrowscale=0.8](124,52)(134,57)
		\Text(130,45){$p_1$}
	\end{axopicture}
}
\def\Jfdh{
	\begin{axopicture}(140,130)(0,-60)
		\Line[arrow](100,50)(140,70)
		\Line(138,74)(142,66)
		\Line[double](0,0)(120,-60)
		\Line[arrow](0,0)(50,25)
		\Line[arrow](50,25)(100,50)
		\Photon(50,-25)(100,50){4}{6}
		\Photon(50,25)(100,-50){4}{6}
		\Line[arrow,arrowpos=1,arrowscale=0.8](124,52)(134,57)
		\Text(130,45){$p_1$}
	\end{axopicture}
}

\def\Jfgda{
	\begin{axopicture}(140,130)(0,-60)
		\Line[arrow](100,50)(140,70)
		\Line(138,74)(142,66)
		\GCirc(96,-48){4}{0}
		\Line[arrow](0,0)(100,50)
		\Photon(100,-50)(100,-17){4}{3}
		\Arc[dash,arrow](100,0)(17,0,360)
		\Photon(100,17)(100,50){4}{3}
		\Line[arrow,arrowpos=1,arrowscale=0.8](124,52)(134,57)
		\Text(130,45){$p_1$}
	\end{axopicture}
}
\def\Jfgdc{
	\begin{axopicture}(140,130)(0,-60)
		\Line[arrow](100,50)(140,70)
		\Line(138,74)(142,66)
		\GCirc(96,-48){4}{0}
		\Line[arrow](0,0)(30,15)
		\Line[arrow](30,15)(70,35)
		\Line[arrow](70,35)(100,50)
		\PhotonArc(50,25)(23,-153,27){4}{6}
		\Photon(100,-50)(100,50){4}{7}
		\Line[arrow,arrowpos=1,arrowscale=0.8](124,52)(134,57)
		\Text(130,45){$p_1$}
	\end{axopicture}
}
\def\Jfgdd{
	\begin{axopicture}(140,130)(0,-60)
		\Line[arrow](100,50)(140,70)
		\Line(138,74)(142,66)
		\Line[double](0,0)(66,-33)
		\GCirc(96,-48){4}{0}
		\Line[arrow](0,0)(50,25)
		\Line[arrow](50,25)(100,50)
		\Photon(50,-25)(50,25){4}{4}
		\Photon(100,-50)(100,50){4}{7}
		\Line[arrow,arrowpos=1,arrowscale=0.8](124,52)(134,57)
		\Text(130,45){$p_1$}
	\end{axopicture}
}
\def\Jfgde{
	\begin{axopicture}(140,130)(0,-60)
		\Line[arrow](100,50)(140,70)
		\Line(138,74)(142,66)
		\GCirc(96,-48){4}{0}
		\Line[arrow](0,0)(100,50)
		\Photon(100,-50)(100,-17){4}{3}
		\Arc[arrow](100,0)(17,0,360)
		\Photon(100,17)(100,50){4}{3}
		\Line[arrow,arrowpos=1,arrowscale=0.8](124,52)(134,57)
		\Text(130,45){$p_1$}
	\end{axopicture}
}
\def\Jfgdg{
	\begin{axopicture}(140,130)(0,-60)
		\Line[arrow](100,50)(140,70)
		\Line(138,74)(142,66)
		\GCirc(76,-38){4}{0}
		\Line[arrow](0,0)(50,25)
		\Photon(50,25)(100,50){4}{4}
		\Arc(75,37.5)(28,-153,27)
		\Photon(80,-40)(80,10){4}{5}
		\Line[arrow,arrowpos=1,arrowscale=0.8](124,52)(134,57)
		\Text(130,45){$p_1$}
	\end{axopicture}
}
\def\Jfgdh{
	\begin{axopicture}(140,130)(0,-60)
		\Line[arrow](100,50)(140,70)
		\Line(138,74)(142,66)
		\Line[double](0,0)(66,-33)
		\GCirc(96,-48){4}{0}
		\Line[arrow](0,0)(50,25)
		\Line[arrow](50,25)(100,50)
		\Photon(50,-25)(100,50){4}{6}
		\Photon(50,25)(100,-50){4}{6}
		\Line[arrow,arrowpos=1,arrowscale=0.8](124,52)(134,57)
		\Text(130,45){$p_1$}
	\end{axopicture}
}

\def\Jfgga{
	\begin{axopicture}(140,130)(0,-60)
		\Line[arrow](100,50)(140,70)
		\Line(138,74)(142,66)
		\GCirc(46,-24){4}{0}
		\GCirc(96,-48){4}{0}
		\Line[arrow](0,0)(50,25)
		\Line[arrow](50,25)(100,50)
		\Photon(50,-25)(50,25){4}{4}
		\Photon(100,-50)(100,50){4}{7}
		\Line[arrow,arrowpos=1,arrowscale=0.8](124,52)(134,57)
		\Text(130,45){$p_1$}
	\end{axopicture}
}
\def\Jfggb{
	\begin{axopicture}(140,130)(0,-60)
		\Line[arrow](100,50)(140,70)
		\Line(138,74)(142,66)
		\GCirc(46,-24){4}{0}
		\GCirc(96,-48){4}{0}
		\Line[arrow](0,0)(50,25)
		\Line[arrow](50,25)(100,50)
		\Photon(50,-25)(100,50){4}{6}
		\Photon(50,25)(100,-50){4}{6}
		\Line[arrow,arrowpos=1,arrowscale=0.8](124,52)(134,57)
		\Text(130,45){$p_1$}
	\end{axopicture}
}

\def\Hfggfb{
	\begin{axopicture}(90,80)(-30,-40)
		\Photon(-30,0)(0,0){4}{3}
		\Line[arrow](0,0)(50,50)
		\Line[arrow,flip](0,0)(20,-20)
		\Line[arrow,flip](20,-20)(40,-40)
		\Line[arrow,flip](40,-40)(50,-50)
		
		\Line[dash,arrow,arrowpos=1](20,-20)(50,10)
		\Line[dash,arrow,arrowpos=1](40,-40)(70,-10)
		
		\Text(60,15){$\ell_2$}
		\Text(80,-5){$\ell_1$}
		\Text(75,40){$p_1-\ell_1-\ell_2$}
	\end{axopicture}
}

\def\Jfffdf{
	\begin{axopicture}(140,130)(0,-60)
		\Line[arrow](80,40)(140,70)
		\Line(138,74)(142,66)
		\Line[arrow](0,0)(80,40)
		\Text(0,-10){$a$}
		\Photon(80,40)(80,-10){4}{5}
		\Line[arrow,flip](50,-25)(80,-10)
		\Text(50,-35){$b$}
		\Line[arrow,flip](80,-10)(100,-50)
		\Text(100,-60){$c$}
		\Line[arrow,arrowpos=1,arrowscale=0.8](124,52)(134,57)
		\Text(130,45){$p_1$}
	\end{axopicture}
}
\def\Jfffdh{
	\begin{axopicture}(140,130)(0,-60)
		\Line[arrow](100,50)(140,70)
		\Line(138,74)(142,66)
		\Line[arrow,flip](0,0)(50,25)
		\Text(0,-10){$a$}
		\Photon(50,25)(100,50){4}{4}
		\Line[arrow](50,-25)(100,50)
		\Text(50,-35){$b$}
		\Line[arrow](100,-50)(50,25)
		\Text(100,-60){$c$}
		\Line[arrow,arrowpos=1,arrowscale=0.8](124,52)(134,57)
		\Text(130,45){$p_1$}
	\end{axopicture}
}
\def\JfffdFurrya{
	\begin{axopicture}(140,130)(0,-60)
		\Line[arrow](100,50)(140,70)
		\Line(138,74)(142,66)
		\Line[arrow](0,0)(50,25)
		\Text(0,-10){$a$}
		\Photon(50,25)(100,50){4}{4}
		\Line[arrow,flip](50,-25)(50,25)
		\Text(50,-35){$b$}
		\Line[arrow](100,-50)(100,50)
		\Text(100,-60){$c$}
		\Line[arrow,arrowpos=1,arrowscale=0.8](124,52)(134,57)
		\Text(130,45){$p_1$}
	\end{axopicture}
}
\def\JfffdFurryb{
	\begin{axopicture}(140,130)(0,-60)
		\Line[arrow](100,50)(140,70)
		\Line(138,74)(142,66)
		\Line[arrow,flip](0,0)(50,25)
		\Text(0,-10){$a$}
		\Photon(50,25)(100,50){4}{4}
		\Line[arrow](50,-25)(50,25)
		\Text(50,-35){$b$}
		\Line[arrow](100,-50)(100,50)
		\Text(100,-60){$c$}
		\Line[arrow,arrowpos=1,arrowscale=0.8](124,52)(134,57)
		\Text(130,45){$p_1$}
	\end{axopicture}
}

\def\HffffbI{
	\begin{axopicture}(90,80)(-30,-40)
		\Photon(-30,0)(0,0){4}{3}
		\Line[arrow](0,0)(50,50)
		\Line[arrow,flip](0,0)(20,-20)
		\Photon(20,-20)(40,-40){3}{2}
		\Line[arrow,flip](40,-40)(50,-50)
		
		\Line[dash,arrow,flip](20,-20)(50,10)
		\Line[dash,arrow](40,-40)(70,-10)
            \Text(55,45){$a$}
            \Text(55,5){$b$}
            \Text(75,-15){$c$}
            \Text(55,-45){$d$}
    \end{axopicture}

}
\def\HffffbII{
	\begin{axopicture}(90,80)(-30,-40)
		\Photon(-30,0)(0,0){4}{3}
		\Line[arrow,flip](0,0)(50,50)
		\Line[arrow](0,0)(20,-20)
		\Photon(20,-20)(40,-40){3}{2}
		\Line[arrow,flip](40,-40)(50,-50)
		
		\Line[dash,arrow](20,-20)(50,10)
		\Line[dash,arrow](40,-40)(70,-10)
            \Text(55,45){$a$}
            \Text(55,5){$b$}
            \Text(75,-15){$c$}
            \Text(55,-45){$d$}
	\end{axopicture}
}

\def\QEDda{
	\begin{axopicture}(170,145)(-30,-70)
		\Photon(-30,0)(0,0){4}{3}
		\Line[arrow](100,50)(140,70)
		\Line[arrow,flip](100,-50)(140,-70)
		\Line[arrow](0,0)(100,50)
		\Line[arrow,flip](0,0)(100,-50)
		\Photon(100,-50)(100,-17){4}{3}
		\Arc[dash,arrow](100,0)(17,0,360)
		\Photon(100,17)(100,50){4}{3}
		\Line[arrow,arrowpos=1,arrowscale=0.8](124,52)(134,57)
		\Text(130,45){$p_1$}
		\Line[arrow,arrowpos=1,arrowscale=0.8](124,-52)(134,-57)
		\Text(130,-45){$p_2$}
	\end{axopicture}
}
\def\QEDdb{
	\begin{axopicture}(170,145)(-30,-70)
		\Photon(-30,0)(0,0){4}{3}
		\Line[arrow](100,50)(140,70)
		\Line[arrow,flip](100,-50)(140,-70)
		\Line[arrow](0,0)(100,50)
		\Line[arrow,flip](0,0)(30,-15)
		\Line[arrow,flip](30,-15)(70,-35)
		\Line[arrow,flip](70,-35)(100,-50)
		\PhotonArc(50,-25)(23,-27,153){4}{6}
		\Photon(100,-50)(100,50){4}{7}
		\Line[arrow,arrowpos=1,arrowscale=0.8](124,52)(134,57)
		\Text(130,45){$p_1$}
		\Line[arrow,arrowpos=1,arrowscale=0.8](124,-52)(134,-57)
		\Text(130,-45){$p_2$}
	\end{axopicture}
}
\def\QEDdc{
	\begin{axopicture}(170,145)(-30,-70)
		\Photon(-30,0)(0,0){4}{3}
		\Line[arrow](100,50)(140,70)
		\Line[arrow,flip](100,-50)(140,-70)
		\Line[arrow](0,0)(30,15)
		\Line[arrow](30,15)(70,35)
		\Line[arrow](70,35)(100,50)
		\Line[arrow,flip](0,0)(100,-50)
		\PhotonArc(50,25)(23,-153,27){4}{6}
		\Photon(100,-50)(100,50){4}{7}
		\Line[arrow,arrowpos=1,arrowscale=0.8](124,52)(134,57)
		\Text(130,45){$p_1$}
		\Line[arrow,arrowpos=1,arrowscale=0.8](124,-52)(134,-57)
		\Text(130,-45){$p_2$}
	\end{axopicture}
}
\def\QEDdd{
	\begin{axopicture}(170,145)(-30,-70)
		\Photon(-30,0)(0,0){4}{3}
		\Line[arrow](100,50)(140,70)
		\Line[arrow,flip](100,-50)(140,-70)
		\Line[arrow](0,0)(50,25)
		\Line[arrow](50,25)(100,50)
		\Line[arrow,flip](0,0)(50,-25)
		\Line[arrow,flip](50,-25)(100,-50)
		\Photon(50,-25)(50,25){4}{4}
		\Photon(100,-50)(100,50){4}{7}
		\Line[arrow,arrowpos=1,arrowscale=0.8](124,52)(134,57)
		\Text(130,45){$p_1$}
		\Line[arrow,arrowpos=1,arrowscale=0.8](124,-52)(134,-57)
		\Text(130,-45){$p_2$}
	\end{axopicture}
}
\def\QEDde{
	\begin{axopicture}(170,145)(-30,-70)
		\Photon(-30,0)(0,0){4}{3}
		\Line[arrow](100,50)(140,70)
		\Line[arrow,flip](100,-50)(140,-70)
		\Line[arrow](0,0)(100,50)
		\Line[arrow,flip](0,0)(100,-50)
		\Photon(100,-50)(100,-17){4}{3}
		\Arc[arrow](100,0)(17,0,360)
		\Photon(100,17)(100,50){4}{3}
		\Line[arrow,arrowpos=1,arrowscale=0.8](124,52)(134,57)
		\Text(130,45){$p_1$}
		\Line[arrow,arrowpos=1,arrowscale=0.8](124,-52)(134,-57)
		\Text(130,-45){$p_2$}
	\end{axopicture}
}
\def\QEDdf{
	\begin{axopicture}(170,145)(-30,-70)
		\Photon(-30,0)(0,0){4}{3}
		\Line[arrow](80,40)(140,70)
		\Line[arrow,flip](100,-50)(140,-70)
		\Line[arrow](0,0)(80,40)
		\Line[arrow,flip](0,0)(50,-25)
		\Photon(50,-25)(100,-50){4}{4}
		\Arc(75,-37.5)(28,-27,153)
		\Photon(80,40)(80,-10){4}{5}
		\Line[arrow,arrowpos=1,arrowscale=0.8](124,52)(134,57)
		\Text(130,45){$p_1$}
		\Line[arrow,arrowpos=1,arrowscale=0.8](124,-52)(134,-57)
		\Text(130,-45){$p_2$}
	\end{axopicture}
}
\def\QEDdg{
	\begin{axopicture}(170,145)(-30,-70)
		\Photon(-30,0)(0,0){4}{3}
		\Line[arrow](100,50)(140,70)
		\Line[arrow,flip](80,-40)(140,-70)
		\Line[arrow](0,0)(50,25)
		\Photon(50,25)(100,50){4}{4}
		\Line[arrow,flip](0,0)(80,-40)
		\Arc(75,37.5)(28,-153,27)
		\Photon(80,-40)(80,10){4}{5}
		\Line[arrow,arrowpos=1,arrowscale=0.8](124,52)(134,57)
		\Text(130,45){$p_1$}
		\Line[arrow,arrowpos=1,arrowscale=0.8](124,-52)(134,-57)
		\Text(130,-45){$p_2$}
		\end{axopicture}
}
\def\QEDdh{
	\begin{axopicture}(170,145)(-30,-70)
		\Photon(-30,0)(0,0){4}{3}
		\Line[arrow](100,50)(140,70)
		\Line[arrow,flip](100,-50)(140,-70)
		\Line[arrow,flip](0,0)(50,25)
		\Photon(50,25)(100,50){4}{4}
		\Line[arrow](0,0)(50,-25)
		\Photon(50,-25)(100,-50){4}{4}
		\Line[arrow,arrowpos=0.6](50,-25)(100,50)
		\Line[arrow,arrowpos=0.6,flip](50,25)(100,-50)
		\Line[arrow,arrowpos=1,arrowscale=0.8](124,52)(134,57)
		\Text(130,45){$p_1$}
		\Line[arrow,arrowpos=1,arrowscale=0.8](124,-52)(134,-57)
		\Text(130,-45){$p_2$}
	\end{axopicture}
}
\def\QEDdFurrya{
	\begin{axopicture}(170,145)(-30,-70)
		\Photon(-30,0)(0,0){4}{3}
		\Line[arrow](100,50)(140,70)
		\Line[arrow,flip](100,-50)(140,-70)
		\Line[arrow](0,0)(50,25)
		\Photon(50,25)(100,50){4}{4}
		\Line[arrow,flip](0,0)(50,-25)
		\Photon(50,-25)(100,-50){4}{4}
		\Line[arrow,flip](50,-25)(50,25)
		\Line[arrow](100,-50)(100,50)
		\Line[arrow,arrowpos=1,arrowscale=0.8](124,52)(134,57)
		\Text(130,45){$p_1$}
		\Line[arrow,arrowpos=1,arrowscale=0.8](124,-52)(134,-57)
		\Text(130,-45){$p_2$}
	\end{axopicture}
}
\def\QEDdFurryb{
	\begin{axopicture}(170,145)(-30,-70)
		\Photon(-30,0)(0,0){4}{3}
		\Line[arrow](100,50)(140,70)
		\Line[arrow,flip](100,-50)(140,-70)
		\Line[arrow,flip](0,0)(50,25)
		\Photon(50,25)(100,50){4}{4}
		\Line[arrow](0,0)(50,-25)
		\Photon(50,-25)(100,-50){4}{4}
		\Line[arrow](50,-25)(50,25)
		\Line[arrow](100,-50)(100,50)
		\Line[arrow,arrowpos=1,arrowscale=0.8](124,52)(134,57)
		\Text(130,45){$p_1$}
		\Line[arrow,arrowpos=1,arrowscale=0.8](124,-52)(134,-57)
		\Text(130,-45){$p_2$}
	\end{axopicture}
}

\def\QEDOneLoop{
	\begin{axopicture}(130,100)(-30,-50)
		\Photon(-30,0)(0,0){4}{3}
		\Line[arrow](30,30)(45,45)
		\Line[arrow,flip](30,-30)(45,-45)
		\Line[arrow](0,0)(30,30)
		\Line[arrow,flip](0,0)(30,-30)
		\Photon(30,-30)(30,30){4}{6}
	\end{axopicture}
}

\def\QEDdFurrya{
	\begin{axopicture}(170,145)(-30,-70)
		\Photon(-30,0)(0,0){4}{3}
		\Line[arrow](100,50)(140,70)
		\Line[arrow,flip](100,-50)(140,-70)
		\Line[arrow](0,0)(50,25)
		\Photon(50,25)(100,50){4}{4}
		\Line[arrow,flip](0,0)(50,-25)
		\Photon(50,-25)(100,-50){4}{4}
		\Line[arrow,flip](50,-25)(50,25)
		\Line[arrow](100,-50)(100,50)
		\Line[arrow,arrowpos=1,arrowscale=0.8](124,52)(134,57)
		\Text(130,45){$p_1$}
		\Line[arrow,arrowpos=1,arrowscale=0.8](124,-52)(134,-57)
		\Text(130,-45){$p_2$}
	\end{axopicture}
}
\def\QEDdFurryb{
	\begin{axopicture}(170,145)(-30,-70)
		\Photon(-30,0)(0,0){4}{3}
		\Line[arrow](100,50)(140,70)
		\Line[arrow,flip](100,-50)(140,-70)
		\Line[arrow,flip](0,0)(50,25)
		\Photon(50,25)(100,50){4}{4}
		\Line[arrow](0,0)(50,-25)
		\Photon(50,-25)(100,-50){4}{4}
		\Line[arrow](50,-25)(50,25)
		\Line[arrow](100,-50)(100,50)
		\Line[arrow,arrowpos=1,arrowscale=0.8](124,52)(134,57)
		\Text(130,45){$p_1$}
		\Line[arrow,arrowpos=1,arrowscale=0.8](124,-52)(134,-57)
		\Text(130,-45){$p_2$}
	\end{axopicture}
}

%% file: Sections/Introduction.tex
Power corrections in scattering amplitudes 
have gained considerable attention in recent years
(see e.g.\ \refrs{Low:1958sn,Burnett:1967km,DelDuca:1990gz,Laenen:2008ux,Laenen:2010uz,Larkoski:2014bxa,Bonocore:2014wua,Ma:2014svb,Kang:2014pya,Bonocore:2015esa,Bonocore:2016awd,Moult:2016fqy,Feige:2017zci,Moult:2017rpl,Moult:2017jsg,Gervais:2017yxv,DelDuca:2017twk,Beneke:2017ztn,Moult:2018jjd,Beneke:2018rbh,Beneke:2018gvs,Ebert:2018lzn,Bahjat-Abbas:2019fqa,Moult:2019mog,Beneke:2019kgv,Beneke:2019mua,Liu:2019oav,Beneke:2019oqx,vanBeekveld:2019prq,vanBeekveld:2019cks,Moult:2019uhz,Liu:2020ydl,Liu:2020eqe,AH:2020iki,AH:2020xll,AH:2020qoa,AH:2021kvg,Beneke:2020ibj,Laenen:2020nrt,Liu:2020tzd,vanBeekveld:2021hhv,Broggio:2021fnr,vanBeekveld:2021mxn,Bonocore:2021qxh,Engel:2021ccn,Beneke:2022obx,Bell:2022ott,Liu:2022ajh,Engel:2023ifn,Engel:2023rxp,Broggio:2023pbu,vanBeekveld:2023gio,vanBeekveld:2023liw,Balsach:2023ema,Beneke:2024cpq,Beneke:2025ufd}). 
Besides their phenomenological relevance, the study 
of power corrections proves to be useful to deepen 
our understanding of gauge theory amplitudes. 
Scattering processes at hadron colliders 
often involve two or more widely separated scales, 
hence the corresponding amplitude can be expressed 
as a power expansion in the small ratio of such scales. 
In this work we focus in particular on the 
wide-angle scattering of $n$ energetic particles 
in quantum electrodynamics (QED), and consider them to have a small 
mass $m$ compared to the centre of mass energy $\sqrt{s}$, 
which represents a configuration to which most of 
the scattering processes occurring at high energy 
colliders can be traced back. The corresponding 
amplitude is expressed as a power expansion in 
$\lambda = m/\sqrt{s} \ll 1$, taking the 
form\footnote{At cross section level, 
NLP terms are $\ord(\lambda^2)$, thus 
we use the notation $\sqrt{\rm N}{\rm LP}$
for $\ord(\lambda)$ terms.}
\begin{equation}\label{powerseries}
\M = \M_{\textrm{LP}} +\M_{\rm\sqrt{N}LP} 
+ \M_{\textrm{NLP}} + \mathcal{O}(\lambda^3),
\end{equation}
where $\M_{\textrm{LP}}$ represents the leading 
power (LP) behaviour of this amplitude, while 
$\M_{\rm{\sqrt{N}LP}}$ and $\M_{\rm NLP}$
provide the first two power corrections in 
the expansion, of $\mathcal{O}(\lambda)$
and $\mathcal{O}(\lambda^2)$ respectively.

Central to this paper is the expectation that, 
as shown by a power counting analysis 
in \refr{Laenen:2020nrt}, each term on 
the r.h.s.\ of \eqn{powerseries} factorizes into a product of hard, soft 
and jet functions. These represent 
respectively virtual hard modes, soft 
modes, and modes collinear to each of 
the $n$ external particles. Focusing in 
particular on the terms describing hard 
and collinear modes up to NLP, and soft 
modes at leading power, one has the 
following schematic factorization formula \cite{Laenen:2020nrt}:
\begin{align} \label{NLPfactorization} \nn
\mathcal{M}_{\mathrm{coll.}} 
& = \left(\prod_{i=1}^n J_{f}^i\right)\otimes  H_{f} \, S  
+ \sum_{i=1}^n \bigg(\prod_{j\neq i}J_{f}^j\bigg)
\Big[J^i_{f\gamma} \otimes H^i_{f\gamma}
+J^i_{f\partial \gamma} \otimes H^i_{f\partial \gamma}\Big]\,S \\ \nn
&\quad + \sum_{i=1}^n \bigg(\prod_{j\neq i}J_{f}^j\bigg)
J^i_{f\gamma\gamma} \otimes H^i_{f\gamma\gamma}\, S
+ \sum_{i=1}^n \bigg(\prod_{j\neq i}J_{f}^j\bigg)
J^i_{f\!f\!f} \otimes H^i_{f\!f\!f}\, S  \\ 
&\quad +\sum_{1\leq i \leq j \leq n} 
\bigg(\prod_{k\neq i,j}J_{f}^k \bigg)
J^i_{f\gamma}J^j_{f\gamma} 
\otimes H^{ij}_{f\gamma,f\gamma}\, S 
+ \ord(\lambda^3)\,,
\end{align}
where $H_I$ are process-dependent hard functions, 
$J_I$ represent universal jet functions, the soft 
function $S$ describes leading power virtual soft 
radiation. The symbol $\otimes$ denotes 
convolution. 

The factorization at LP, given by the first term 
in \eqn{NLPfactorization}, has been known for a long 
time, see e.g.\ \refrs{Mueller:1979ih,Collins:1980ih,Collins:1981uk,Sen:1981sd,Collins:1989bt,Dixon:2008gr}.
For massless particles it can be shown that the 
LP term in \eqn{NLPfactorization} is in direct 
correspondence with the factorization structure
of soft and collinear divergences \cite{Aybat:2006mz,Gardi:2009qi,Becher:2009qa}, 
hence it provides the starting point for the 
resummation of large logarithms associated to 
the emission of soft and collinear radiation \cite{Magnea:1990zb,Contopanagos:1996nh,Kidonakis:1997gm,Kidonakis:1998bk,Magnea:2000ss}.
For massive particles, \eqn{NLPfactorization} 
contains relevant information for the resummation 
of large logarithms of $m^2/s$ \cite{Catani:2000ef,Penin:2005eh,Penin:2005kf,Mitov:2006xs,Becher:2007cu,Czakon:2007ej,Czakon:2007wk,Mitov:2009sv,Becher:2009kw,Banerjee:2020rww,Wang:2023qbf,Wang:2024pmv}. 
The factorization of scattering amplitudes 
(or cross sections) into single-scale objects 
is at the basis of resummation, as it allows 
one to resum the large logarithms by means 
of renormalization group equations.
In general, factorization theorems such as 
the one in \eqn{NLPfactorization} have been 
derived directly in the original theory
(QCD or, in the case considered here, QED)
\cite{Sterman:1987aj,Catani:1989ne,Catani:1990rp,Korchemsky:1993xv,Korchemsky:1993uz}, 
or within an approach based on the soft-collinear
effective field theory (SCET), where soft and 
collinear modes are separated at the Lagrangian level
\cite{Bauer:2000yr,Bauer:2001yt,Beneke:2002ph}. 

Much less is known concerning the factorization
properties of the terms appearing beyond LP. 
Within SCET, factorization theorems for
(massless) $n$-particle scattering 
amplitudes have been considered in 
the label formalism
\cite{Larkoski:2014bxa,Moult:2015aoa,Kolodrubetz:2016uim,Feige:2017zci} 
and in the position-space formulation 
of SCET \cite{Beneke:2017ztn,Beneke:2018rbh}.
Within the direct approach, factorization 
theorems have been discussed for $n$-particle 
amplitudes in the Yukawa theory 
\cite{Gervais:2017yxv} and QED 
\cite{Laenen:2020nrt}. An early discussion of factorization in Drell-Yan was presented in \refrs{Qiu:1990xxa,Qiu:1990xy}.

Although the power counting analysis 
in \refr{Laenen:2020nrt} has led to the factorization formula 
in \eqn{NLPfactorization}, the functions
appearing in this equation have been 
defined so far only diagrammatically, i.e.\ 
by listing the Feynman diagrams expected 
to contribute to each function in QED.
For the factorization theorem to be 
complete and serve as a basis for resummation, the jet functions in 
\eqn{NLPfactorization} need to be 
properly defined in terms of matrix 
elements of time-ordered operators 
in QED, while the corresponding hard 
functions are matching 
coefficients. The aim of this paper 
is to fill this gap. 

In this regard, let us recall 
that the factorization formula in 
\eqn{NLPfactorization} is valid
both for the case of massless and
massive particles (with $m^2 \ll s$). 
Indeed, while \eqn{NLPfactorization} describes 
the factorization of an $n$-particle
scattering amplitude, its relevance 
extends also to the case in which 
additional soft radiation is emitted. 
For this, as discussed in 
\refrs{Gervais:2017yxv,Laenen:2020nrt}, 
the jet functions are upgraded 
to radiative jet functions, and the 
emission of soft radiation from the hard 
functions is related to soft emissions from 
the other functions by Ward 
identities. 
Thus, the study we aim to conduct in this 
work is useful not only to complete and 
validate the factorization theorem in 
\eqn{NLPfactorization}, but is also 
preparatory to study the case of real 
radiation, which we leave for future 
analysis.

In order to validate the factorization
theorem we should consider an amplitude 
sufficiently non-trivial that all functions 
appearing in \eqn{NLPfactorization} contribute, 
yet simple enough that it can be calculated 
with standard methods known in literature. Moreover, functions such as $J_{f\gamma\gamma}$ and 
$J_{f\!f\!f}$ only contribute from two loops onwards, 
so that the validation of \eqn{NLPfactorization} 
requires a complete two-loop calculation. The decay of an off-shell photon into 
a massive fermion-antifermion pair, usually 
described in literature in terms of massive 
fermion form factors,  satisfies our 
requirements. 

Much work has been dedicated to the 
perturbative calculation of the massive 
quark form factors. The two-loop result 
has been obtained in \refrs{Bernreuther:2004ih,Gluza:2009yy} 
(see also \refr{Blumlein:2020jrf}
for a calculation of the corresponding 
contribution to the $e^+e^- \to \gamma^*/Z^{0*}$ 
cross section). In recent years a lot 
of effort has been devoted to the calculation 
of the three-loop correction \cite{Henn:2016kjz,Henn:2016tyf,Ablinger:2017hst,Lee:2018nxa,Lee:2018rgs,Ablinger:2018yae,Blumlein:2018tmz,Blumlein:2019oas,Fael:2022rgm,Fael:2022miw,Fael:2023zqr,Blumlein:2023uuq},
although no complete analytic result 
as yet exists. For our analysis we will 
need the two-loop region calculation 
obtained in \refr{terHoeve:2023ehm}, 
because a straightforward expansion of the 
results in
\refrs{Bernreuther:2004ih,Gluza:2009yy}
for $m^2 \ll s$ is not enough to 
disentangle hard and collinear
modes, which is needed to 
check the calculation of the 
jet functions in \eqn{NLPfactorization}. 

Given these premises, our first task 
is to define and calculate the jet 
functions appearing in \eqn{NLPfactorization}.
Then we will proceed to construct the 
corresponding photon-fermion-antifermion 
amplitude, and compare with the 
method of regions calculation. 
It may be worth mentioning here that the factorization formula 
in \eqn{NLPfactorization} is given in 
terms of functions with open Lorentz 
and Dirac indices; no attempt is done
to further project the jet functions
onto a basis of scalar functions.
This is not uncommon in case of 
factorization theorems beyond leading 
power, see e.g.\ \refr{Beneke:2019oqx},
as projecting to scalar functions would 
involve non-trivial projection operators,
which we preferably avoid in order not to 
hide the relatively simple structure of 
\eqn{NLPfactorization}. In any case, 
the jet functions 
appearing in \eqn{NLPfactorization} 
are universal objects, thus the results 
obtained in this paper are valid for 
any $n$-particle scattering amplitude.

The paper is structured as follows. In 
sec.\ \ref{sec:setup} we describe the 
process of interest, set up our notation, 
recall factorization at leading power and 
then describe in detail the factorization 
structure of the terms contributing beyond 
leading power. In sec.\ \ref{sec:oneloop} 
we give matrix element definitions for the 
new jet functions, and compute them at one 
loop, up to NLP. Sec.\ \ref{sec:twoloop} 
then deals with factorization at two loops, 
which also includes the calculation of the 
relevant jet functions up to two loops. We 
compare our results to \refr{terHoeve:2023ehm}, 
region by region. We conclude and discuss 
our findings in sec.\ \ref{sec:conclusions}. 
App.\ \ref{app:seqjets} then
discusses the interesting interplay between 
the $J_f$, $J_{f\gamma}$ and $J_{f\ga\ga}$ 
jet functions, where we demonstrate factorization 
for the double-collinear region at the 
integrand level. The derivation also shows 
that there is no double counting between 
the different (subleading) jet functions. 
In app.\ \ref{app:regtofact} we discuss 
the collinear-anticollinear factorization, 
and some of its subtle aspects. 
App.\ \ref{app:2loopres} reports the 
integral expressions for all the jet 
functions at two-loop order, while 
app.\ \ref{app:renorm} discusses UV 
counterterms for the (factorized) 
amplitude, paying particular attention to mass 
renormalization.

%% file: Sections/Set_up.tex

\subsection{Fermion-antifermion amplitude in QED}
\label{sec:QEDAmpDef}

Let us consider the QED process
\begin{equation}\label{process}
\gamma^{*}(q) \to f(p_1) + \bar f(p_2), 
\end{equation}
in which an off-shell photon with 
momentum $q$ produces 
a fermion-antifermion pair. Keeping cross 
sections in mind we also denote $q^2 = s$. 
The fermions have mass $m$ 
so that $p_1^2= p_2^2 = m^2$. In what 
follows we consider the corresponding 
unrenormalized, or bare\footnote{We discuss UV 
counterterms in app.~\ref{app:renorm}.},
amplitude
\begin{equation}\label{VtoGamma}
V^\mu(p_1,p_2) = \bar{u}(p_1)\, 
\Gamma^\mu(p_1,p_2)\, v(p_2),
\end{equation}
where we make the important note that the amplitude also contains external line contributions, i.e.\ non-1PI diagrams. External line propagators are amputated according to the LSZ formalism. In the equation above we defined $\Gamma^\mu(p_1,p_2)$, which 
represents the amplitude with stripped-off spinors. At lowest order we have
\be\label{GammaTree}
\Gamma^{\mu(0)}(p_1,p_2) = i \, e\, e_f \, 
\gamma^{\mu},
\ee
where $e_f$ is the fermion electric 
charge in units of the positron charge 
$e>0$. In the small-mass limit $m \ll \sqrt{s}$, 
the amplitude $V^{\mu}(p_1, p_2)$ can 
be evaluated as a power expansion in the 
small parameter $\lambda \sim m/\sqrt{s} \ll 1$:
\be \label{AmpPowerExpansion}
V^{\mu}\big(p_1,p_2\big) = 
V_{\rm LP}^{\mu}\big(p_1,p_2\big)
+ V_{\!\! \rm \sqrt{N}LP}^{\mu}\big(p_1,p_2\big)
+ V_{\rm NLP}^{\mu}\big(p_1,p_2\big) 
+\ord\big(\lambda^3\big),
\ee
where $V_{\rm LP}^{\mu}$ is $\ord(\lambda^0)$,
$V_{\!\! \rm \sqrt{N}LP}^{\mu}$ is $\ord(\lambda)$, 
$V_{\rm NLP}^{\mu}$ is $\ord(\lambda^2)$, and 
so on. Central to our discussion is the expectation 
that each term on the r.h.s.\ of \eqn{AmpPowerExpansion} 
should factorize into hard, jet-like and soft 
structures, where the virtual momenta are respectively 
hard, i.e.\ of order $\sqrt{s}$, collinear to one of the 
two external fermions, or soft, i.e.\ of order
$m^2$ \cite{Laenen:2020nrt,terHoeve:2023ehm}. 
Our goal is to verify that 
$V^{\mu}\big(p_1,p_2)$ factorizes 
according to \eqn{NLPfactorization}. 
To this end we will construct the factorized 
amplitude, and compare it with the recent 
calculation obtained by means of the method 
of regions \cite{terHoeve:2023ehm}.

Let us start by setting 
the kinematic notation. We consider 
the centre-of-mass reference frame, 
where the outgoing momenta $p_1$ and $p_2$ 
read
\begin{equation}\label{eq:p1p2def}
p_1^\mu = \left(\sqrt{p^2+m^2},0,0,p\right),
\qquad \quad
p_2^\mu = \left(\sqrt{p^2+m^2},0,0,-p\right).
\end{equation}
It proves useful to introduce 
two light-like vectors\footnote{
Ref. \cite{terHoeve:2023ehm} uses the
light-like vectors $n_- = \sqrt{2}\,n$
and $n_+ = \sqrt{2}\,\bar n$ with 
$n_+ \cdot n_- = 2$.
Here we follow the conventions of \refr{Laenen:2020nrt},
and decompose momenta along $n$ and 
$\bar n$.}
$n$ and $\bar n$, with $n^2 = \bar n^2 = 0$ 
and $n \cdot \bar n = 1$, with $p_1$ mostly 
along $n$ and $p_2$ mostly along $\bar n$, 
i.e.
\beq\label{ExternalMomentumDef1}
p^{\mu}_1 = p_1^+ \, n^{\mu} + p_1^- \, \bar n^{\mu}, 
\qquad \quad
p^{\mu}_2 = p_2^+ \, n^{\mu} + p_2^-\, \bar n^{\mu} \,,
\eeq
with 
\begin{align} \nn
p_1^+ & = \bar n \cdot p_1 
= p_2^- = n \cdot p_2 = \frac{1}{\sqrt{2}}
\Big(\sqrt{p^2+m^2}+p \Big) \sim \sqrt{s}, \\
p_1^- &=  n \cdot p_1 
= p_2^+ = \bar n \cdot p_2 = \frac{1}{\sqrt{2}}
\Big( \sqrt{p^2+m^2}-p\Big) \sim \la^2\sqrt{s}. 
\label{eq:scaling_rel}
\end{align}
The mass-shell condition implies 
\begin{equation}\label{smallcomponents}
p_i^2=2p_i^+p_i^-=m^2 \implies 
p_1^-=\frac{m^2}{2p_1^+},\quad p_2^+=\frac{m^2}{2p_2^-}.
\end{equation}
We then have
\begin{equation}\label{shatImplicit}
s=(p_1+p_2)^2= 2m^2 + 2p_1^+ p_2^- 
+2 p_1^- p_2^+ = 
\s\left(1+\frac{2m^2}{\s}+\frac{m^4}{\s^2}\right),
\end{equation}
where we defined the $\mathcal{O}(\la^0)$ variable 
$\s \equiv 2 p_1^+p_2^-$.
In what follows we will refer to the direction 
identified by the vector $n$ as \emph{collinear} 
($c$), while the direction identified by $\bar n$ 
will be labelled as \emph{anticollinear} ($\bar c$). 
Where needed we will indicate the large components 
of $p_1$ and $p_2$ by 
\begin{align}
\hat p_1^{\mu} &= p_1^+ \, n^{\mu}, 
\qquad \qquad \qquad 
\hat p_2^{\mu} = p_2^- \, \bar n^{\mu},
\label{LargeComponentsExternal}
\intertext{
while the corresponding small components 
will be indicated by}
\tilde p_1^{\mu} &= p_1^- \, \bar n^{\mu}, 
\qquad \qquad \qquad 
\tilde p_2^{\mu} = p_2^+ \, n^{\mu}.
\label{SmallComponentsExternal}
\end{align}
A generic momentum is decomposed 
along $n$, $\bar n$ as follows:
\be\label{eq:SudakovDecomposition}
k^{\mu} = k^+ \, n^{\mu} + k^-\, \bar n^{\mu} + k_{\perp}^{\mu}, 
\qquad \qquad \quad 
k^{\mu} = (k^+, k^-, k_{\perp}),
\ee
where $k^+ = \bar n \cdot k$, $k^- = n \cdot k$. 
This notation can be used to express scaling 
relations. For instance,
$p^{\mu}_1 = (p_1^+,p_1^{-},p_{1\perp}) \sim 
\sqrt{s}(\la^0,\lambda^2,0)$,
$p^{\mu}_2 = (p_2^+,p_2^{-},p_{2\perp})
\sim \sqrt{s}(\lambda^2,\la^0,0)$. In general,
virtual momenta leading to non-scaleless 
contributions have the scaling properties\footnote{Note that 
we did not include the soft scaling 
$\sqrt{s}(\lambda^2,\lambda^2,\lambda^2)$. 
In our case soft integrals turn out to be scaleless, which
was already observed in the region analysis 
\cite{terHoeve:2023ehm}.}
\begin{figure}[t]
\centering
\includegraphics[width=.8\linewidth]{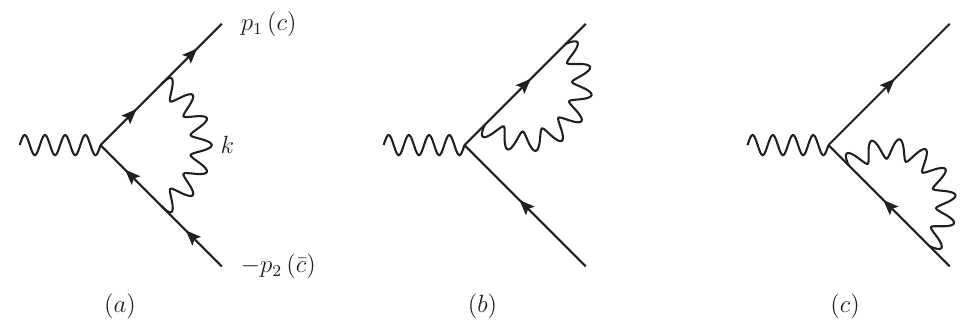}
\caption{One-loop diagrams contributing to the $\gamma\to f \bar f$
amplitude, according to the conventions discussed in the main text.
The momenta $p_1$ and $p_2$ are outgoing.}
\label{fig:Amp1loop}
\end{figure}
\begin{align}
\text{hard ($h$):} \qquad 
&k\sim\sqrt{s}\,
(\lambda^0,\lambda^0,\lambda^0)\,, \nn \\
\text{collinear ($c$):} \qquad  
&k\sim\sqrt{s}\,
(\lambda^0,\lambda^2,\lambda^1)\,, \label{eq:momentumModes1L} \\ 
\text{anticollinear ($\bar{c}$):} \qquad 
&k\sim\sqrt{s}\,
(\lambda^2,\lambda^0,\lambda^1)\,. \nn 
\end{align}

With these conventions we are almost ready 
to discuss the factorization of the amplitude
for the case $m \ll \sqrt{s}$. Before proceeding 
we highlight the fact that, as stated below 
\eqn{VtoGamma}, 
we are considering here the full amplitude 
and therefore include self-energy contributions 
on the external legs, order by order in 
perturbation theory. 
As will become clear in sec.\ \ref{sec:oneloop}, 
this is because the jet functions defined
within the factorization theorem in 
\eqn{NLPfactorization} naturally include 
such contributions. Therefore, we need to 
include the self-energy corrections in 
the region calculation in QED as well, 
for a proper comparison. Thus
the one-loop amplitude 
$V^{\mu (1)}\big(p_1,p_2\big)$ 
will receive contributions from all 
diagrams listed in fig.\ \ref{fig:Amp1loop}. 
The inclusion of self-energy corrections
modifies the equation of motion and the mass-shell condition, and corrections can be included order by order in perturbation theory. 
The self-energy contribution is 
usually taken into account by means of 
a Dyson sum:
\be
\frac{i}{\slashed{p}-m}\bigg\{ 1 + \Big[ i \, \Sigma(p) \Big] 
\frac{i}{\slashed{p}-m} + \mathcal{O}(\aem^2) \bigg\} 
= \frac{i}{\slashed{p}-m + \Sigma(p)}.
\ee
Here $\Sigma(p)$ has the general form 
\be \label{selfEn1}
\Sigma(p) = A(p^2) \, \slashed{p}+ B(p^2) \, m,
\ee
such that the propagator can be rewritten as
\begin{equation}\label{newprop}
\frac{i R}{\slashed{p}-m_p}, \qquad 
R^{-1} = 1+A(m_p^2)+\mathcal{O}(p^2-m_p^2), \qquad 
m_p =m\left(\frac{1-B(m_p^2)}{1+A(m_p^2)}\right),
\end{equation}
where $m_p$ represents the unrenormalized pole mass.
At one loop one has \cite{Honemann:2018mrb}
\bea \label{selfEn2} \nn 
A(m^2) &=& \frac{\aem \,e_f^2}{4\pi}
\left(\frac{\bar\mu^2}{m^2}\right)^\eps 
\left[\frac{1}{\eps}+2+\bigg(4+\frac{\zeta_2}{2}\bigg)\eps
+\bigg(8+\zeta_2-\frac{\zeta_3}{3}\bigg)\eps^2+\mathcal{O}(\eps^3)\right] + \ord(\alpha^2), \\ 
B(m^2) &=& \frac{\aem \,e_f^2}{4\pi}
\left(\frac{\bar\mu^2}{m^2}\right)^\eps 
\left[-\frac{4}{\eps}-6-(12+2\zeta_2)\eps
-\bigg(24+3\zeta_2-\frac{4\zeta_3}{3}\bigg)\eps^2+\mathcal{O}(\eps^3)\right]+\ord(\alpha^2),
\eea
where $\alpha\equiv \alpha_{\rm EM}=e^2/4\pi$. Here we evaluated $A$ and $B$ in $m^2$, and not in $m_p^2$, since differences are of higher order in $\aem$.
As a consequence the equation of motion are modified 
as 
\bea\label{eom} \nn
\bar{u}(p_1) \slashed{p}_1 &=& 
\bar{u}(p_1) \, m \Big[1 - A(m^2) - B(m^2) 
+ \mathcal{O}(\aem^2) \Big], \\[0.1cm]
\slashed{p}_2 v(p_2) &=& 
- m \Big[1 - A(m^2) - B(m^2) 
+ \mathcal{O}(\aem^2) \Big] \, v(p_2),
\eea
or upon  expanding in powers 
of $\lambda$:
\bea\label{eomexpanded} \nn 
\bar u(p_1) \, \slashed{n} &=& 
\bar u(p_1) \frac{1}{p_1^+} \bigg\{ m \Big[1 - A(m^2) - B(m^2) 
+ \mathcal{O}(\aem^2) \Big] 
- p_1^-\, \slashed{\bar n} \bigg\}, \\
\slashed{\bar n} \, v(p_2) &=& 
- \bigg\{m \Big[1 - A(m^2) - B(m^2) 
+ \mathcal{O}(\aem^2) \Big] 
+ p_2^+\, \slashed{n} \bigg\} 
\frac{1}{p_2^-} v(p_2). 
\eea
Similarly, beyond tree level the mass-shell condition in \eqn{smallcomponents} 
is modified accordingly. Up to one loop 
one has 
\be \label{smallcomponents2}
p_i^2 = m^2\Big[1- 2 A(m^2) -2 B(m^2) 
+ \mathcal{O}(\aem^2)  \Big].
\ee
We are now ready to discuss the factorization
of the power expansion of $V^\mu(p_1,p_2)$ for $m \ll \sqrt s$.


\subsection{Factorization at LP}
\label{sec:LPFact}

In order to set up the factorization formula 
up to NLP, we start with the leading power term. The case of massless 
fermions has been discussed at length in
\refr{Dixon:2008gr}. The factorization 
formula for the massive case is formally 
equivalent. For an outgoing fermion-antifermion 
pair it reads
\be\label{FF-LP-ren}
V_{\rm LP}^{\mu}(p_1,p_2,\mu) = 
H_{f,\bar f}^{\mu}(\hat p_1,\hat p_2,\mu) \, 
\frac{J_{f}(p_1,\bar n,\mu)}{
{\cal J}_{f}(n,\bar n,\mu)}\, 
\frac{J_{\bar f}(p_2,n,\mu)}{
{\cal J}_{\bar f}(\bar n, n,\mu)}
\, S_{f,\bar f}(n \cdot \bar n,\mu),
\ee
where $\mu$ represents the factorization scale.
In what follows we will keep the dependence on 
$\mu$ implicit, unless needed. 
The convention for now and what follows is that $f$ is for the fermion line 
(in the $c$-collinear direction) with outgoing momentum $p_1$, and the $\bar f$ is for the antifermion line (in the $\bar c$-collinear, i.e.\ anticollinear, direction) with outgoing momentum $p_2$.
In \eqn{FF-LP-ren} the hard function $H_{f,\bar f}^{\mu}(\hat p_1,\hat p_2)$
represents the virtual hard modes, thus 
it depends on the large momentum components 
$\hat p_1^{\mu}$, $\hat p_2^{\mu}$ defined 
in \eqn{LargeComponentsExternal}.
Next, the jet functions $J_{f}(p_1,\bar n)$ 
and $J_{\bar f}(p_2,n)$ reproduce virtual 
collinear and anticollinear modes respectively, 
and are defined in terms of matrix 
elements of (implicitly) time-ordered operators\footnote{Here we follow the 
convention adopted in \refr{Laenen:2020nrt}, 
which incorporate the spinor into the jet 
function. In this way, the leading order 
jet functions coincide with the 
corresponding spinor: 
$J^{(0)}_{f}(p_1,n_1) = \bar u(p_1)$, 
$J^{(0)}_{\bar f}(p_2,n_2) = v(p_2)$,
respectively for outgoing fermions and 
antifermions.}
\be \label{JetDef}
J_{f}(p_1,n_1,\mu) 
= \big\langle p_1 \big| \bar \psi(0) \, 
\Phi_{n_1} (0,\infty) \big| 0 \big\rangle, 
\ee
and 
\be \label{JetBarDef}
J_{\bar f}(p_2,n_2,\mu)
= \big\langle p_2 \big| \Phi_{n_2} (\infty,0)\, 
\psi(0) \big| 0 \big\rangle,
\ee
where $n_1$ and $n_2$ represent gauge-link 
vectors associated to the definition of the 
Wilson line 
\be\label{WilsonLineDef}
\Phi_{n} (y,x) = {\cal P} 
\exp \bigg[ i \,e \, e_f \int_{x}^{y} ds \, 
n \cdot A(s n)\bigg].
\ee
The direction of the Wilson line is largely arbitrary, but given that their function is 
to mimic the coupling of photons collinear 
to the outgoing parton (say, the fermion) to 
the opposite moving hard parton (i.e., the 
antifermion), we choose $n_1 = \bar n$, 
$n_2 = n$, as indicated in \eqn{FF-LP-ren}.
Lastly, \eqn{FF-LP-ren} contains the
soft function $S_{f,\bar f}(n \cdot \bar n)$, 
as well as the eikonal approximation 
of the two jet functions, all defined
in terms of the vacuum expectation 
value of Wilson lines. One has
\be \label{SoftfDef}
S_{f,\bar f}(\hat \beta_1 \cdot \hat \beta_2,\mu) 
= \big\langle 0 \big| 
\Phi_{\hat \beta_1} (\infty,0) \, 
\Phi_{\hat \beta_2} (0,\infty) 
\big| 0 \big\rangle
\ee
for the soft function, and
\bea \label{EikJetDef} \nn
{\cal J}_{f}(\hat \beta_1,n_1,\mu) 
&=& \big\langle 0 \big| 
\Phi_{\hat \beta_1} (\infty,0) \, 
\Phi_{n_1} (0,\infty) 
\big| 0 \big\rangle, \\[0.2cm] 
{\cal J}_{\bar f}(\hat \beta_2,n_2,\mu) 
&=& \big\langle 0 \big| 
\Phi_{n_2} (\infty,0) \, 
\Phi_{\hat \beta_2} (0,\infty) 
\big| 0 \big\rangle
\eea
for the two eikonal jet functions. Note that these eikonal jet functions prevent overcounting between the jet function and the soft function.
Observe also that, for simplicity, in 
\eqns{SoftfDef}{EikJetDef} we 
approximate the velocities of 
the fermion and antifermion pair, 
$\beta_1$ and $\beta_2$, with the 
directions of large momentum flow, 
i.e. $\beta_1 \sim \hat \beta_1 = n$, 
$\beta_2 \sim \hat \beta_2 = \bar n$,
as already indicated implicitly in 
\eqn{FF-LP-ren}. 

The most important aspect to underline 
concerning \eqn{FF-LP-ren} is that it 
describes the factorization of the 
amplitude \emph{after} soft and collinear 
singularities have been moved from the 
hard function to the jet and soft functions. 
Let us clarify this relevant point. Loop 
corrections to the amplitude $V^{\mu}\big(p_1,p_2)$
in general contain soft and collinear divergences
in case of massless fermions, and just soft 
divergences in case of massive fermions,
as considered here. (There 
are ultraviolet divergences as well;
these can be removed by standard UV 
renormalization. We discuss these in app.~\ref{app:renorm}.) Calculating 
$V^{\mu}\big(p_1,p_2)$ 
by means of the method of regions
generates spurious singularities 
within each region, such that 
the original singularities are 
reproduced only after the sum of 
all regions has been considered.
In the case at hand, we know 
\cite{terHoeve:2023ehm} that 
the massive form factor receives 
contributions from the hard, 
collinear and anticollinear 
regions.\footnote{As shown in 
\refr{terHoeve:2023ehm}, 
additional ultra-collinear 
and ultra-anticollinear regions,
respectively with scaling 
$k \sim \sqrt{s}(\lambda^2, \lambda^4,\lambda^3)$
$k \sim \sqrt{s}(\lambda^4, \lambda^2,\lambda^3)$,
give non-scaleless contributions 
to single diagrams, but cancel at 
the level of the amplitude 
$V^{\mu}\big(p_1,p_2,\mu)$.}
The hard region develops soft 
and (spurious) collinear singularities, 
while the collinear and anticollinear 
regions develop soft and (spurious) 
ultraviolet singularities, such that in the 
sum collinear and ultraviolet singularities 
cancel, leaving the soft singularities. 
In general, it is preferred to define 
the functions appearing in the factorization 
\eqn{FF-LP-ren} such that only the ``true''
singularities appear explicitly: in other 
words, one wants to define a hard function
free of collinear and soft singularities, 
and associate collinear divergences 
to the jet functions (in case of 
massless particles) and soft
divergences to the soft function. 
For the massless case, this is 
discussed at length in \refr{Dixon:2008gr},
to which we refer for further details.
Focusing on the problem at hand, both 
the bare soft function and eikonal 
jets are scaleless. The standard 
interpretation of scaleless regions 
in dimensional regularization is that 
one has infrared and ultraviolet 
divergences cancelling each other, 
such that $1/\eps_{\rm IR} 
- 1/\eps_{\rm UV} = 0$. Thus, in order 
to achieve the factorization structure
in \eqn{FF-LP-ren}, one splits this 
singularity structure: the ultraviolet pole
moves into the hard function. The soft 
function then contains the leftover infrared 
divergence, and it becomes a pure counterterm.
In the problem at hand, given that only the 
hard, collinear and anticollinear regions 
are not scaleless, both the soft function 
$S_{f,\bar f}(\beta_1 \cdot \beta_2)$ and 
the eikonal jets ${\cal J}_{f}(\beta_1)$,
${\cal J}_{\bar f}(\beta_2)$ in \eqn{FF-LP-ren}
contribute as pure counterterms. 

However, in our approach we do not redistribute the poles over the various functions, but rather compute them at face value. Thus we aim to check that the various jet functions that appear up to subleading power are able to reproduce the contribution given by the collinear and anticollinear regions.
Hence, we aim to reproduce 
the factorization theorem in 
\eqn{FF-LP-ren} (and its generalization
at subleading power, to be discussed)
in its ``bare'' form, i.e., before 
soft (and collinear) singularities 
are reshuffled into pure-counterterm 
soft and eikonal jet functions. For the 
leading power contribution this 
implies that we aim to prove 
\be\label{FF-LP-bare}
V_{\rm LP}^{\mu}(p_1,p_2,\mu) = 
H_{f,\bar f}^{\mu,b}(\hat p_1,\hat p_2,\mu) \, 
J_{f}^{b}(\hat p_1,\bar n,\mu) \, 
J_{\bar f}^{b}(\hat p_2,n,\mu),
\ee
where the index $b$, for ``bare'',
means that no reshuffling of 
soft and collinear singularities 
has been considered. In what follows 
we will always intend a factorization 
theorem as written in \eqn{FF-LP-bare}, 
and drop the index $b$ for simplicity. 


\subsection{Factorization at NLP}
\label{sec:NLPFact}

\begin{figure}[t]
\centering
\includegraphics[width=.55\linewidth]{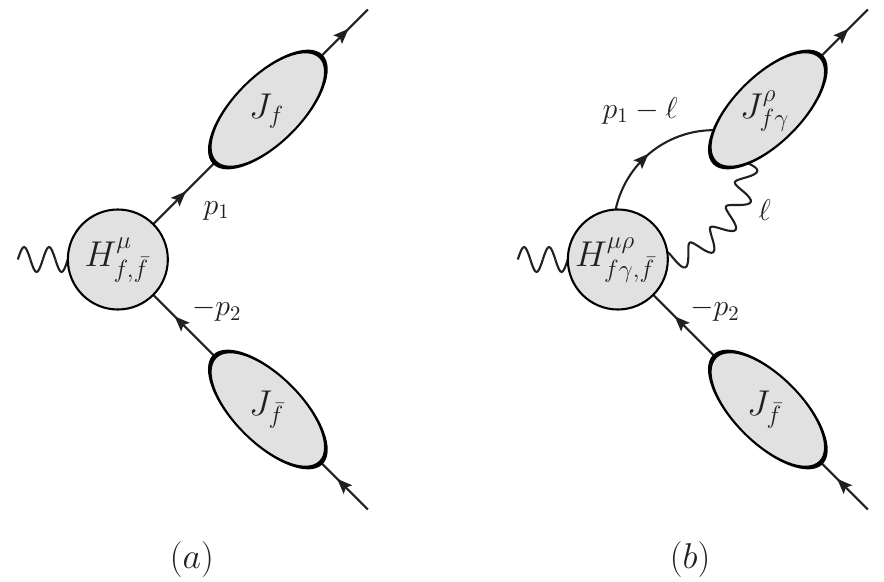}
\caption{Diagrammatic representation of 
$V_{f,\bar f}^{\mu}$ (a) and $V_{f\gamma,\bar f}^{\mu}$ (b),
defined respectively in \eqns{NLPfactorizationf}{NLPfactorizationfgamma}.}
\label{fig:FactNLPf-fgamma}
\end{figure}

We are now ready to specialize the factorization
formula in \eqn{NLPfactorization} to the case of 
a fermion-antifermion pair. In order to keep equations 
short, we assign a name to the various terms in 
\eqn{NLPfactorization}. The factorization formula
then becomes a sum over factorized terms: 
\bea\label{NLPfactorizationQQbar} \nn
V^{\mu}\big(p_1,p_2) &=& 
V_{f,\bar f}^{\mu}(p_1,p_2)
+V_{f\gamma,\bar f}^{\mu}(p_1,p_2)
+V_{f\partial\gamma,\bar f}^{\mu}(p_1,p_2)
+V_{f,\bar f\gamma}^{\mu}(p_1,p_2)
+V_{f,\bar f\partial\gamma}^{\mu}(p_1,p_2) \\[0.1cm] \nn
&&+\, V_{f\gamma\gamma,\bar f}^{\mu}(p_1,p_2)  
+ V_{f,\bar f\gamma \gamma}^{\mu}(p_1,p_2)
+V_{f\! f\!\bar f,\bar f}^{\mu}(p_1,p_2)
+V_{f,\bar f\! f\! \bar f}^{\mu}(p_1,p_2) \\[0.1cm]
&&+\, V_{f\gamma,\bar f\gamma}^{\mu}(p_1,p_2)
+ \ord\big(\lambda^{3}\big), 
\eea
which we discuss one by one in what follows. 
First of all we have 
\be\label{NLPfactorizationf}
V_{f,\bar f}^{\mu}(p_1,p_2) 
= J_{f}(p_1,\bar n)\, 
H_{f,\bar f}^{\mu}(p_1; p_2) \, 
J_{\bar f}(p_2,n),
\ee
which represents a configuration involving 
a fermion along the collinear direction and an antifermion along the anticollinear
direction, as depicted in fig.\ 
\ref{fig:FactNLPf-fgamma} (a). 
All additional virtual radiation is 
either hard, factorized into the hard 
function, or (anti)collinear along the 
two directions ($\bar n$) $n$, and 
factorized into the two jet functions. 
This term starts at $\ord(\lambda^0)$ 
and it is obviously a generalization 
of the leading power term in 
\eqn{FF-LP-bare}. In this regard, let us 
highlight that the hard and jet functions 
in the LP factorization formula depend on 
the leading momentum components $\hat p_i$. 
In contrast, in order to reach 
$\ord(\lambda^2)$ (i.e., NLP) accuracy 
in \eqn{NLPfactorizationf}, we start by 
retaining the dependence on the full momenta 
$p_i$, and expand the jet and hard functions 
in \eqn{NLPfactorizationf} into the mass 
$m \sim \ord(\lambda)$ and the small 
momentum components $\tilde p_i \sim 
\ord(\lambda^2)$. Indeed, exploiting 
the mass-shell condition in 
\eqns{smallcomponents}{smallcomponents2}, 
we can express the small momentum component 
as a function of the mass. Each 
function in \eqn{NLPfactorizationf} becomes
a power series in $m$. 
Combining the large components $p_1^+$, 
$p_2^-$ into $\hat s$ according to 
\eqn{shatImplicit} we have   
\bea\label{powerexpansionHf} \nn
H_{f,\bar f}^{\mu}(p_1; p_2)
= H_{f,\bar f}^{\mu}(\hat s, m) 
&=& \bigg( 1 + m \frac{\partial}{\partial m}
+ \frac{m^2}{2} \frac{\partial^2}{\partial m^2} \bigg)
H_{f,\bar f}^{\mu}(\hat s,m)\Big|_{m = 0}
+ \ord(\lambda^3) \\
&=& 
H_{f,\bar f}^{\mu}(\hat s)\big|_{\rm LP}
+ m \, H_{f,\bar f}^{\mu}(\hat s)\big|_{\rm \sqrt{N}LP}
+ \frac{m^2}{2} \, H_{f,\bar f}^{\mu}(\hat s)\big|_{\rm NLP}
+ \ord(\lambda^3),
\eea
and 
\be\label{powerexpansionJf} 
J_{f}(p_1,\bar n) = 
J_{f}(m,\bar n)\big|_{\rm LP} 
+ \frac{m}{p_1^+} J_{f}(m,\bar n)\big|_{\rm \sqrt{N}LP}
+ \frac{m^2}{(p_1^+)^2} J_{f}(m,\bar n)\big|_{\rm NLP}
+ \ord(\lambda^3),
\ee
with an equivalent expansion for 
$J_{\bar f}(p_2,n)$. The series
in \eqns{powerexpansionHf}{powerexpansionJf}
can then be inserted in 
\eqn{NLPfactorizationf} to 
get a systematic expansion of 
$V_{f,\bar f}^{\mu}$. 

The next terms in \eqn{NLPfactorizationQQbar} 
are given respectively by 
$V_{f\gamma,\bar f}^{\mu}$ and
$V_{f\partial\gamma,\bar f}^{\mu}$. They 
describe a configuration involving a fermion 
and a \emph{transverse}\footnote{
A transverse photon $A_{\perp}$ is suppressed 
by a power of $\lambda$ compared to longitudinal 
photons. As discussed in section II.A of 
\refr{Laenen:2020nrt}, this is best seen
in axial gauges, where longitudinal photons 
do not propagate. In turn, the scaling 
$A_{\perp} \sim \lambda$ determines 
the suppression of $J_{f\gamma}$ by a power 
of $\lambda$ compared to the $J_f$ jet. 
Unsuppressed longitudinal collinear and 
anticollinear photons, $A_c^+ = \bar n \cdot A_c
\sim \lambda^0$ and $A_{\bar c}^- = n\cdot A_{\bar c} 
\sim \lambda^0$ have been absorbed respectively into 
$J_{f}$ and $J_{\bar f}$ by means of Wilson lines, 
as discussed in sec. \ref{JfCalcOneLoop}.} 
photon along the collinear direction, and an 
antifermion along the anticollinear direction, 
as represented in fig.\ \ref{fig:FactNLPf-fgamma} 
(b). Similarly, the terms 
$V_{f,\bar f\gamma}^{\mu}$
and $V_{f,\bar f\partial\gamma}^{\mu}$
involve a fermion along the collinear direction, 
and an antifermion plus a transverse photon along 
the anticollinear direction. They can be obtained 
by symmetry from $V_{f\gamma,\bar f}^{\mu}$ 
and $V_{f\partial\gamma,\bar f}^{\mu}$,
by exchanging $p_1 \leftrightarrow -p_2$,\footnote{The 
minus sign can be understood because one switches from 
an outgoing fermion to an outgoing antifermion. This 
effectively changes $p_1$ into $-p_2$ and vice versa in all the propagators.} thus we will not 
discuss them further. 

In order to write down the explicit form 
of $V_{f\gamma,\bar f}^{\mu}$ and
$V_{f\partial\gamma,\bar f}^{\mu}$,
let us label the photon momentum by $\ell$, as 
indicated in fig.\ \ref{fig:FactNLPf-fgamma} (b). Consequently, the hard function for this process, 
$H_{f\gamma,\bar f}^{\mu\rho}$, involves a 
collinear fermion and a photon, respectively 
with momenta $p_1 - \ell$ and $\ell$, with 
$\ell  = (\ell^+,\ell^{-},\ell_{\perp}) \sim 
\sqrt{s}\,(\lambda^0,\lambda^2,\lambda^1)$.
As argued in \refr{Laenen:2020nrt}, a consistent 
factorization order by order in the power expansion 
requires the hard function to depend only on the 
large momentum component $\hat \ell = \ell^+ n$ 
(as well as $\hat p_1$). This is achieved by 
expanding $H_{f\gamma,\bar f}^{\mu}$ in the 
small components of $\ell$. Anticipating that 
the jet function $J^{\rho}_{f\gamma}$ starts 
at $\ord(\lambda)$, we need to expand the 
hard function up to $\ord(\lambda)$ as 
well. We thus have
\bea\label{powerexpansionHfgammaB} \nn
H_{f\gamma,\bar f}^{\mu\rho}(p_1-\ell,\ell;p_2)
&=& \bigg( 1 + \ell^{\sigma}_{\perp} 
\frac{\partial}{\partial \ell^{\sigma}_{\perp}} \bigg)
H_{f\gamma,\bar f}^{\mu\rho}(p_1-\ell,\ell;p_2)
\Big|_{\ell_\perp = 0} + \ord(\lambda^2) \\[0.2cm]
&=& 
H_{f\gamma,\bar f}^{\mu\rho}(p_1-\hat \ell,\hat \ell;p_2)
+\ell_{\perp \sigma} \, H_{f\partial\gamma,\bar f}^{\mu\rho\sigma}
(\hat p_1-\hat \ell,\hat \ell;\hat p_2)
+ \ord(\lambda^2),
\eea
where in the second line we have implicitly defined  
$H_{f\partial\gamma,\bar f}^{\mu\rho\sigma}$. Given 
\eqn{powerexpansionHfgammaB}, $V_{f\gamma,\bar f}^{\mu}$ 
and $V_{f\partial\gamma,\bar f}^{\mu}$ still 
involve a convolution over the large momentum
component $l^+$. We have respectively\footnote{We will 
see in sec.~\ref{sec:oneloop} that the integral over 
$\ell^+$ is indeed bounded, while in principle the 
integration is unbounded from above and below.}
\be\label{NLPfactorizationfgamma} 
V_{f\gamma,\bar f}^{\mu}(p_1,p_2)
=\int_0^{p_1^+} d \ell^+ \,
J_{f\gamma\, \rho}(p_1-\hat\ell,\hat\ell,\bar n)\,
H_{f\gamma,\bar f}^{\mu\rho}(p_1-\hat\ell,\hat\ell;p_2) 
\, J_{\bar f}(p_2,n), 
\ee
and
\be\label{NLPfactorizationfpartialgamma} 
V_{f\partial\gamma,\bar f}^{\mu}(p_1,p_2)
=\int_0^{p_1^+} d \ell^+ \,
J_{f\partial\gamma\, \rho\sigma}(\hat p_1-\hat\ell,\hat\ell,\bar n)\,
H_{f\partial\gamma,\bar f}^{\mu\rho\sigma}(\hat p_1-\hat\ell,\hat\ell;\hat p_2) 
\, J_{\bar f}(\hat p_2,n). 
\ee
A few comments are in order. First, 
$V_{f\gamma,\bar f}^{\mu}$ starts 
at $\ord(\lambda)$. Thus, the jet and 
hard functions still need to be expanded
in powers of $m$, for a consistent 
expansion up to $\ord(\lambda^2)$. 
Similarly to the series definitions
in \eqns{powerexpansionHf}{powerexpansionJf}, 
we have 
\bea\label{powerexpansionHfgamma} \nn
H_{f\gamma,\bar f}^{\mu\rho}(p_1-\hat\ell,\hat\ell;p_2)
&=& \bigg( 1 + m \frac{\partial}{\partial m} \bigg)
H_{f\gamma,\bar f}^{\mu\rho}(p_1-\hat\ell,\hat\ell;p_2)
\bigg|_{m = 0} + \ord(\lambda^2) \\[0.2cm] 
&=& 
H_{f\gamma,\bar f}^{\mu\rho}(\hat p_1-\hat \ell,\hat \ell;\hat p_2)\big|_{\rm LP}
+ m \, H_{f\gamma,\bar f}^{\mu\rho} 
(\hat p_1-\hat \ell,\hat \ell;\hat p_2)\big|_{\rm \sqrt{N}LP}
+ \ord(\lambda^2),
\eea
and 
\be\label{powerexpansionJfgamma} 
J^{\rho}_{f\gamma}(p_1-\hat\ell,\hat\ell,\bar n) = 
J^{\rho}_{f\gamma}(\hat p_1-\hat\ell,\hat\ell,\bar n)\big|_{\rm \sqrt{N}LP}
+ \frac{m}{p_1^+} J^{\rho}_{f\gamma}
(\hat p_1-\hat\ell,\hat\ell,\bar n)\big|_{\rm NLP}
+ \ord(\lambda^3).
\ee
The series in 
\eqns{powerexpansionHfgamma}{powerexpansionJfgamma}
can be inserted in 
\eqn{NLPfactorizationfgamma} to 
get a systematic expansion of 
$V_{f\gamma,\bar f}^{\mu}$. 
Concerning \eqn{NLPfactorizationfpartialgamma},
we point out that we have absorbed the factor 
$\ell^{\sigma}_{\perp}$ in \eqn{powerexpansionHfgammaB} 
into the definition of 
$J^{\rho \sigma}_{f\partial\gamma}$. 
Schematically we may write 
\be \label{JfpartialgammaDef1}
J^{\rho\sigma}_{f\partial\gamma}(\hat p_1-\hat\ell,\hat\ell,\bar n)
= J^{\rho}_{f\gamma}(\hat p_1-\hat \ell,\hat\ell,\bar n) 
\, \ell^{\sigma}_{\perp},
\ee
where we stress that $J^{\rho}_{f\gamma}$ involves
an integration over $\ell_\perp$, thus the explicit 
factor of $\ell^{\sigma}_{\perp}$ in \eqn{JfpartialgammaDef1}
is integrated over as well. In sec.\ 
\ref{sec:Jfgamma-fdgamma-1loop} we will see that 
$J^{\rho\sigma}_{f\partial\gamma}$
can be rigorously defined in terms of 
an operator matrix element.
$V_{f\partial\gamma,\bar f}^{\mu}$
starts at $\ord(\lambda^2)$, so there is no 
need to further expand in powers of $m$, 
as in \eqns{powerexpansionHfgamma}{powerexpansionJfgamma}.
Thus, the functions $J_{f\partial\gamma}^{\rho\sigma}$
and $H_{f\partial\gamma,\bar f}^{\mu\rho\sigma}$ 
are taken at leading order in the power expansion,
and depend only on the leading momentum components,
as indicated by the argument of the functions 
in \eqn{NLPfactorizationfpartialgamma}.

\begin{figure}[t]
\centering
\includegraphics[width=.80\linewidth]{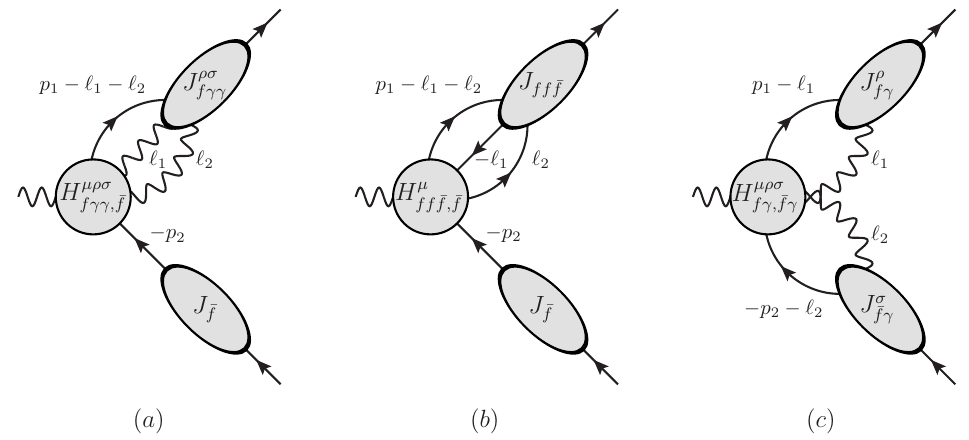}
\caption{Diagrammatic representation of 
$V_{f\gamma\gamma,\bar f}^{\mu}$ (a), 
$V_{f\!f\!\bar{f},\bar f}^{\mu}$ (b)
and $V_{f\gamma,\bar f\gamma}^{\mu}$ (c),
defined respectively in eqs.~(\ref{NLPfactorizationfgammagamma}),
(\ref{NLPfactorizationfff})
and~(\ref{NLPfactorizationfgammafgamma}).}
\label{fig:FactNLPlambda2}
\end{figure}
The remaining terms in \eqn{NLPfactorizationQQbar}
are $V_{f\gamma\gamma,\bar f}^{\mu}$, 
$V_{f\! f\!\bar f,\bar f}^{\mu}$ and 
$V_{f\gamma,\bar f\gamma}^{\mu}$, as well 
as the corresponding $p_1 \leftrightarrow -p_2$ 
symmetric terms $V_{f,\bar f\gamma \gamma}^{\mu}$ 
and $V_{f,\bar f\! \bar f\! f}^{\mu}$. 
All these terms start at $\ord(\lambda^2)$,
therefore we can consistently drop all 
dependence on $\ell_{1\perp}$, $\ell_{2\perp}$ and
all other small momentum components. The 
terms $V_{f\gamma\gamma,\bar f}^{\mu}$ 
and $V_{f\! f\!\bar f,\bar f}^{\mu}$
involve a configuration respectively with 
a fermion and two (transverse) photons or a fermion and a 
fermion-antifermion pair along the collinear 
direction, and an antifermion along the 
anticollinear direction, as represented 
in fig.\ \ref{fig:FactNLPlambda2} (a) 
and (b). They involve a convolution 
over the large momentum components 
$\hat \ell_1$, $\hat \ell_2$ as follows: 
\be\label{NLPfactorizationfgammagamma} 
V_{f\gamma\gamma,\bar f}^{\mu}\big(p_1,p_2) 
= \int_0^{p_1^+} d \ell_{1}^+ \, d \ell_{2}^+ 
\, J_{f\gamma\gamma\, \rho\sigma}
(\hat p_1-\hat\ell_{1}-\hat\ell_{2},
\hat\ell_{1},\hat\ell_{2},\bar n) \,
H_{f\gamma\gamma,\bar f}^{\mu\rho\sigma}
(\hat p_1-\hat\ell_{1}-\hat\ell_{2},
\hat\ell_{1},\hat\ell_{2}; \hat p_2) 
\, J_{\bar f}(\hat p_2,n),
\ee
and
\be\label{NLPfactorizationfff} 
V_{f\!f\!\bar f,\bar f}^{\mu}\big(p_1,p_2) 
=\int_0^{p_1^+} d \ell_{1}^+ \, d \ell_{2}^+ 
\, J_{f\!f\!\bar{f}}(\hat p_1-\hat\ell_{1}-\hat\ell_{2},
\hat\ell_{1},\hat\ell_{2},\bar n) \,
H^{\mu}_{f\!f\!\bar{f},\bar f}
(\hat p_1-\hat\ell_{1}-\hat\ell_{2},
\hat\ell_{1},\hat\ell_{2};\hat p_2) \, J_{\bar f}(\hat p_2,n).
\ee
Here we use simplified notation. In sec.\ \ref{sec:Jfff} we distinguish two contributions that differ by charge flow.
The last term, i.e. 
$V_{f\gamma,\bar f\gamma}^{\mu}$,
involves a configuration with a 
collinear fermion and photon, as well 
as an anticollinear antifermion and photon,
as depicted in fig.\ \ref{fig:FactNLPlambda2}
(c). As such it also involves two 
convolutions, over $\ell_{1}^+$ and $\ell_{2}^-$.
It reads
\bea\label{NLPfactorizationfgammafgamma}  \nn
V_{f\gamma,\bar f\gamma}^{\mu}\big(p_1,p_2)
&=& \int_0^{p_1^+} d \ell_{1}^+ 
\int_0^{p_2^-} d \ell_{2}^- \, 
J_{f\gamma\, \rho}
(\hat p_1-\hat\ell_{1},\hat\ell_{1},\bar n) \\ 
&&\hspace{1.0cm}\times\, 
H_{f\gamma,\bar f\gamma}^{\mu\rho\sigma}
(\hat p_1-\hat \ell_{1},\hat \ell_{1},
-\hat p_2-\hat\ell_{2},\hat\ell_{2}) \,
J_{\bar f\gamma\, \sigma}
(-\hat p_2-\hat\ell_{2},\hat\ell_{2},n).
\eea

With these definitions at hand, 
each term in \eqn{NLPfactorizationQQbar}
has been properly defined in terms of 
convolutions between hard and jet 
functions. In this regard, let us 
notice that, although we defined 
the convolutions in \eqns{NLPfactorizationfgamma}{NLPfactorizationfpartialgamma} in terms 
of the large photon momentum component 
$\ell^+$, and the convolutions in 
eqs.~(\ref{NLPfactorizationfgammagamma}),
(\ref{NLPfactorizationfff}) and 
(\ref{NLPfactorizationfgammafgamma})
in terms of the large components $\ell_1^+$
and $\ell_2^+$ (or $\ell_2^-$ for the case in eq.~\eqref{NLPfactorizationfgammafgamma}), in practice it will be more 
convenient to express the convolutions 
in terms of dimensionless momentum
fractions, for instance 
\be\label{NLPfactorizationfgammaB} 
V_{f\gamma,\bar f}^{\mu}(p_1,p_2)
=p_1^+ \int_0^{1} d x \,
J_{f\gamma\, \rho}(p_1-\hat\ell,\hat\ell,\bar n)\,
H_{f\gamma,\bar f}^{\mu\rho}(p_1-\hat\ell,\hat\ell;p_2) 
\, J_{\bar f}(p_2,n), 
\ee
with $x \equiv \ell^+/p_1^+$.

Our next goal is to show 
that each jet function can be defined
in terms of an operator matrix 
element in QED. In this context, 
each hard function can be considered 
as the corresponding short-distance
(Wilson) coefficient, which can be
obtained by matching with the full 
amplitude in QED. We devote secs.\
\ref{sec:oneloop} and \ref{sec:twoloop}
to the jet function definitions in terms
of operator matrix elements, and their
calculation in perturbation theory up 
to the perturbative order required to
reproduce the full amplitude up to 
$\ord(\alpha^2)$.


\subsection{\texorpdfstring{Form factors in the limit $m^2 \ll s$}{}}
\label{sec:FFCalcDef}

In order to check that the factorized amplitude 
in \eqn{NLPfactorizationQQbar} correctly reproduces 
the QED amplitude in the limit $m^2\ll s$, we 
compare our calculation in terms of hard and 
jet functions with the massive form factors 
calculation obtained in \refr{terHoeve:2023ehm} 
by means of the method of regions. To this end, 
let us briefly recall that the $\gamma\to f\bar{f}$ 
amplitude can be expressed in terms of two 
form factors $F_1$ and $F_2$ such that
\begin{equation}\label{FFdef1}
\Gamma^{\mu}(p_1,p_2) = i \, e \, e_f 
\bigg[F_{1}(s,m)\gamma^{\mu} 
+ \frac{1}{2m} F_{2}(s,m) 
\, i \, \sigma^{\mu\nu} (p_1+p_2)_\nu \bigg],
\end{equation}
where $\sigma^{\mu\nu} = \frac{i}{2}
[\gamma^{\mu},\gamma^{\nu}]$, and the electric 
charge is consistent with \eqn{GammaTree}.
In the rest of the paper we will take $e_f
= -1$, except those equations where the 
factor $e_f$ is left explicit. The form 
factors computed in \refr{terHoeve:2023ehm}
up to $\ord(\alpha^2)$ correspond to the 
unrenormalized, 
1PI amplitude, 
and can thus be extracted by means of the projection operators 
\begin{equation}\label{eq:form_factor_i}
F_{i}\big(s,m^2\big) 
= {\rm Tr}\big[P_{i}^{\mu}(m,p_1,p_2) 
\, \Gamma_{\mu}(p_1,p_2) \big],
\end{equation}
where\footnote{The projection operators
in \eqn{eq:form_factor_i} are slightly 
different from those defined in 
eq.~(2.5) of \refr{terHoeve:2023ehm}, 
because there the process $f(p_1) + 
\bar f(p_2) \to \gamma^{*}(q)$ was 
studied, while here we consider the 
process $\gamma^{*}(q) \to f(p_1) 
+ \bar f(p_2)$. Although the projections 
are different, the form factors are of 
course the same.} \cite{Bernreuther:2004ih}
\begin{equation} \label{FFprojectorsDef}
P_{i}^{\mu}(m,p_1,p_2)
= \frac{\slashed{p}_2 - m}{m}
\bigg[i\, g_1^{(i)} \gamma^{\mu} 
+ \frac{i}{2m} g_{2}^{(i)} 
\big( p_2^{\mu} - p_1^{\mu}\big) \bigg]
\frac{\slashed{p}_1 + m}{m},
\end{equation}
and in turn
\begin{align}\label{projectors}
g_1^{(1)} &= \frac{1}{e \,e_f} 
\frac{1}{4(1-\eps)} \frac{1}{(s/m^2-4)}, \nn \\ \nn
g_2^{(1)} &= -\frac{1}{e \,e_f} 
\frac{3-2\eps}{(1-\eps)} \frac{1}{(s/m^2-4)^2}, \\ \nn
g_1^{(2)} &= -\frac{1}{e \,e_f} 
\frac{1}{(1-\eps)} \frac{1}{s/m^2(s/m^2-4)}, \\ 
g_2^{(2)} &= \frac{1}{e \,e_f} 
\frac{1}{(1-\eps)} \frac{1}{(s/m^2-4)^2}
\bigg[\frac{4 m^2}{s} + 2 - 2\eps \bigg].
\end{align}
Given that the factorized amplitude in 
\eqn{NLPfactorizationQQbar} corresponds 
to the unrenormalized amplitude including self-energy corrections, comparison with \refr{terHoeve:2023ehm}
would require to modify the projection
operators in \eqn{FFprojectorsDef} in 
order to take into account the 
modified equations of motion and mass-shell condition order by order in perturbation 
theory. This results in $\mathcal{O}(\alpha)$ corrections to the projection constants in \eqn{projectors}. Moreover, a residue factor of $\sqrt{R}$ on each external leg needs to be incorporated,\footnote{\Eqn{newprop} contains a factor of $R$, but this is reduced to $\sqrt{R}$ by the LSZ formula.} which would yield
\begin{equation}\label{gammaamp}
    {\Gamma}^\mu_{}(p_1,p_2) = \left(\sqrt{R}\right)^2\,\Gamma^\mu_{\rm 1PI}(p_1,p_2),
\end{equation}
where $\Gamma^\mu$ on the l.h.s.\ includes also the diagrams that are not 1PI. Note that the r.h.s.\ of \eqn{gammaamp} now needs to be computed with the equations of motion and mass-shell condition as in \eqns{eom}{smallcomponents2} respectively.
In practice, we will see in secs.\ 
\ref{sec:oneloop} and \ref{sec:twoloop}
that the contribution due to the correction
on the external legs can be easily factorized 
and dropped both in the factorized approach 
and in the region computation, except for the 
collinear-anticollinear region contribution 
at two loops. Thus, we will be mostly able
to compare directly with the computation in 
\refr{terHoeve:2023ehm}, dropping the correction 
on the external legs from $V^{\mu}(p_1,p_2)$
calculated by means of \eqn{NLPfactorizationQQbar}
and obtaining the corresponding form factors by 
means of \eqn{eq:form_factor_i}. 

To conclude, 
we recall that we expand the form factors 
in powers of $\alpha/4\pi$:
\begin{equation}
F_{i}\left(\s,m^2,\mu^2\right) = 
F^{(0)}_{i}\left(\s,m^2\right) 
+ \sum_{k=1}^\infty\left(\frac{\aem}{4\pi}\right)^k \, 
F^{(k)}_{i}\left(\s,m^2,\mu^2\right) \, ,
\end{equation}
where the tree-level form factors read
\begin{equation}
F^{(0)}_{1}\left(\s,m^2\right) = 1, \qquad \qquad \qquad 
F^{(0)}_{2}\left(\s,m^2\right) = 0.
\end{equation}
Furthermore, each form factor is 
expanded in powers of $m^2/\s$ as follows:
\bea \label{FFPowerExpansion}\nn
F_1\big(\s,m^2\big) &=& 
F_{\rm 1,LP}^{\mu}\big(\s,m^2\big)
+ F_{\rm 1,NLP}^{\mu}\big(\s,m^2\big) 
+\ord\big(\lambda^4\big), \\[0.2cm]
F_2\big(\s,m^2\big) &=& 
F_{\rm 2,NLP}^{\mu}\big(\s,m^2\big) 
+\ord\big(\lambda^4\big),
\eea
where $F_{2,\rm NLP}^\mu$ gives the $\mathcal{O}(\la)$ part of the amplitude, since $F_2$ is divided by a factor of $m$ in eq.~\eqref{FFdef1}, and $F_{1,\rm LP}$ and $F_{1,\rm NLP}$ give the $\mathcal{O}(\la^0)$ and $\mathcal{O}(\la^2)$ parts respectively.


%% file: Sections/One_Loop.tex

With the factorization formula 
\eqn{NLPfactorizationQQbar} in 
place, we now need to properly 
define the jet functions as 
matrix elements of time-ordered 
operators in QED. In this section 
we start by considering the jet 
functions whose contribution starts
at LO and NLO in perturbation theory, 
namely $J_{f}$, $J_{f\gamma}$ and 
$J_{f\partial\gamma}$. In the 
following we define and calculate 
these functions at one loop and up to 
NLP, i.e.\ up to $\mathcal{O}(\lambda^2)$ 
in the small-mass limit. 
The jet functions describe jets 
of one or more collimated particles.
As a consequence, we expect jet functions 
defined along the collinear direction 
to reproduce the contribution given by 
the collinear region in an expansion 
by region calculation, and jet functions
along the anticollinear direction to 
reproduce the anticollinear region. 
To this end, we will compare our result 
with the form factor calculation in
\refr{terHoeve:2023ehm}, which uses 
the method of regions. In this regard
we note that at NLO in the coupling, 
non-1PI diagrams only 
enter via the multiplicative residue 
factor $\sqrt{R}$ for each external 
leg. This is the same on the factorization 
and the region side, so that it can be
ignored when comparing the results 
to one another. 


\subsection{\texorpdfstring{$J_{f}$}{} jet function}
\label{JfCalcOneLoop}

\begin{figure}[t]
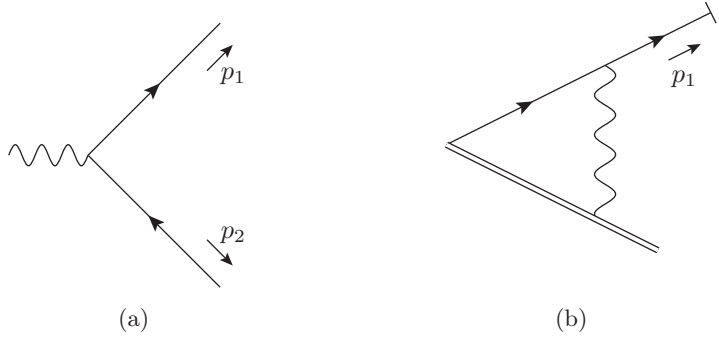

\centering
\subfloat[]{\QEDtree\label{Jf_diagb}}
\hspace{2cm}
\subfloat[]{\Jf1
\label{Jf_diaga}}
\caption{In (a) we show the diagrammatic representation of the leading order hard function $H_{f,\bar{f}}$. In (b) we show the one-loop $J_f$ jet contribution.}
\label{fig:Jf_diagrams}
\end{figure}

As anticipated in sec.~\ref{sec:NLPFact}, 
the jet function $J_f$ consists of a single
(outgoing) $c$-collinear fermion, surrounded 
by a cloud of photons (and virtual fermion-antifermion pairs). The 
corresponding matrix element (as anticipated 
in \eqn{JetDef}) is thus given by a single 
fermion field, dressed by a collinear Wilson 
line, which leads to a matrix element of a gauge-invariant operator
\cite{Collins:1989bt,Dixon:2008gr}:
\begin{equation}\label{eq: defJf}
J_{f}(p_1,\bar n) = \bra{p_1}\bar\psi(0)\, 
\Phi_{\bar{n}}(0,\infty)\ket{0},
\end{equation}
where $\bra{p_1}$ denotes the state of the 
outgoing fermion with momentum $p_1$. 
The Wilson line $\Phi_{\bar{n}}$ has 
been defined for the general case in 
\eqn{WilsonLineDef}. For the specific 
case at hand we have 
\begin{equation}
\Phi_{\bar{n}}(x,\infty) 
= \exp\left[-ie\int_\infty^0 d\lambda\, 
\bar{n}\cdot A(x+\lambda \bar{n})\right].
\end{equation}
An analogous definition can be given 
for a $\bar c$-collinear jet with an 
outgoing antifermion, and will be 
denoted by $J_{\bar f}$. It reads
\begin{equation}\label{eq: defJfcbar}
J_{\bar f}(p_2,n) = \bra{p_2}
\Phi_{n}(\infty,0)\,\psi(0)\ket{0},
\end{equation}
where $\bra{p_2}$ denotes the state 
of the outgoing antifermion with 
momentum $p_2$. At leading order, 
we have 
\begin{equation}\label{eq:JfLO}
J_{f}^{(0)}(p_1) = \bar{u}(p_1),
\qquad \qquad
J_{\bar f}^{(0)}(p_2) = v(p_2).
\end{equation}
The spinors $\bar{u}(p_1)$ and $v(p_2)$ 
here respectively depend on the full 
momentum $p_1$ and $p_2$, and hence 
obey the
equations of motion from \eqn{eomexpanded}, 
where we can ignore the terms $A$ and $B$ 
since they only start to modify results 
at two loops. Indeed, the tree-level $J_f$ jet and $H_{f,\bar{f}}^\mu$ hard function (which multiply the one-loop expressions of $A$ and $B$) do not contain factors of $\slashed{p}_1$ and $\slashed{p}_2$ that act on the spinors.
Comparison with \eqns{VtoGamma}{GammaTree}
immediately gives the leading order hard 
function, which reads (with $e_f = -1$, 
see comment after \eqn{FFdef1})
\begin{equation}\label{eq:hardff0}
H_{f,\bar f}^{(0)\mu} =-i\, e\, \gamma^\mu,
\end{equation}
with a diagrammatic representation given in fig.\ \ref{Jf_diagb}. Here and in what 
follows we note that although formally a 
hard function, being a matching coefficient, is not described by Feynman 
diagrams, a diagrammatic representation does prove useful 
and intuitive. For the 
form factors at leading order we have 
$F_1^{(0)} = 1,$ and $F_2^{(0)}=0$.

At one loop, we consider the vertex 
correction, which is the only 1PI 
contribution that occurs after 
performing the Wick contractions. 
A diagrammatic representation is 
given in fig.\ \ref{Jf_diaga}. The 
double line represents here an eikonal 
Feynman rule. We note that this jet 
function also produces the (non-1PI) self-energy 
diagram. 
As mentioned before, we do not include 
this at the one-loop level when comparing with the region calculation. The one-loop 
expression for this jet reads
\begin{align} \label{eq:jet_jf}
J_{f}^{(1)}(p_1,\bar{n}) &= 
i 16\pi^2\bar{u}(p_1)\int[dk]\, 
\frac{\slashed{\bar{n}}
(\slashed{p}_1-\slashed{k}+m)}{k^2
[(p_1-k)^2-m^2][\bar{n}\cdot k]} \\
\label{eq:Jf1H0Jf0}
&= \left(\frac{\bar\mu^2}{m^2}\right)^\epsilon 
\bar{u}(p_1)\, \frac{\Gamma(\epsilon)
e^{\epsilon\gamma_E}}{\epsilon(1-2\epsilon)}
\bigg(1-\frac{m\eps}{p_1^+} \slashed{\bar{n}}\bigg),
\end{align}
which indeed admits an expansion like in 
\eqn{powerexpansionJf}. All propagators come 
with a standard $+i\eta$-prescription, which 
will be left implicit throughout the paper. 
We have chosen the Feynman gauge for the photon 
propagator. We also defined the 
measure 
\begin{equation}
\int [dk] \equiv 
\left(\frac{\bar\mu^2e^{\gamma_E}}{4\pi}\right)^\eps
\int\frac{d^dk}{(2\pi)^d},
\end{equation}
where $d=4-2\eps$ and 
$\bar\mu^2=4\pi e^{-\gamma_E}\mu^2$ 
is the $\overline{\text{MS}}$-renormalization 
scale. The result for the $\bar c$-collinear 
jet function at one loop is similar; it reads
\begin{equation}
J_{\bar f}^{(1)}(p_2,n) 
=\left(\frac{\bar\mu^2}{m^2}\right)^\epsilon 
\frac{\Gamma(\epsilon)e^{\epsilon\gamma_E}}{\epsilon(1-2\epsilon)}
\bigg(1+\frac{m\epsilon}{p_2^-} \slashed{n}\bigg)v(p_2).
\end{equation}

We are now ready to check the factorization
formula \eqn{NLPfactorizationf} at one loop. 
Suppressing for simplicity the argument of the 
functions, at this perturbative order we have 
\bea\label{NLPfactorizationfNLP} \nn
V^{\mu (1)}_{f,\bar f}
&=& V^{\mu (1)}_{f,\bar f}\Big|_h 
+ V^{\mu (1)}_{f,\bar f}\Big|_c 
+ V^{\mu (1)}_{f,\bar f}\Big|_{\bar c} \\[0.2cm]
&=& J_{f}^{(0)}\, 
H_{f,\bar f}^{\mu(1)} \, 
J_{\bar f}^{(0)}
+ J_{f}^{(1)}\, 
H_{f,\bar f}^{\mu(0)} \, 
J_{\bar f}^{(0)}
+J_{f}^{(0)}\, 
H_{f,\bar f}^{\mu(0)} \, 
J_{\bar f}^{(1)},
\eea
where in the first line we have implicitly 
defined the hard, collinear and anticollinear
contribution to $V^{\mu (1)}_{f,\bar f}$ with 
the corresponding terms in the second line. 
Let us start by comparing 
$V^{\mu(1)}_{f,\bar f}|_{c}$ 
with the collinear region calculated 
in \refr{terHoeve:2023ehm}. Upon 
projecting onto $F_1$, we find
\begin{equation}
F_{1\, f,\bar f}^{(1)}\Big|_{c} 
= \left(\frac{\bar\mu^2}{m^2}\right)^\epsilon
\bigg[\frac{1}{\epsilon^2}+\frac{2}{\epsilon}+4+\frac{\zeta_2}{2}
+\frac{m^2}{\s}\left(\frac{2}{\eps}+4\right)
+\mathcal{O}(\lambda^4,\epsilon)\bigg],
\end{equation}
and for $F_2$ one finds
\begin{equation}
F_{2\, f,\bar f}^{(2)}\Big|_{c} 
= \left(\frac{\bar\mu^2}{m^2}\right)^\epsilon
\bigg[\frac{m^2}{\s}\left(-\frac{4}{\eps}-8\right)
+\mathcal{O}(\lambda^4,\epsilon)\bigg].
\end{equation}
Indeed the leading power term in 
$F_{1\, f,\bar f}^{(1)}|_{c}$ reproduces 
the full leading power collinear region in 
$F_1^{(1)}|_{c}$, given that 
$V^{\mu(1)}_{f,\bar f}|_{c}$ provides 
the only LP contribution in the factorization 
formula \eqn{NLPfactorizationQQbar}. 
Instead, the $\mathcal{O}(\lambda)$ 
and $\mathcal{O}(\lambda^2)$ part of 
$V^{\mu(1)}_{f,\bar f}|_{c}$, which 
contribute respectively to $F_2$ and $F_1$,
do not yet reproduce the full collinear region
contribution to the form factor. This is also 
expected, because at $\mathcal{O}(\lambda)$ 
and $\mathcal{O}(\lambda^2)$ we are still 
missing the contributions due to the $J_{f\ga}$ 
jet and the $J_{f\partial\ga}$ jet. The result 
for the anticollinear region is equivalent, i.e.\
$F_{i\, f,\bar f}^{(1)}|_{\bar c} 
= F_{i\, f,\bar f}^{(1)}|_{c}$. 
The hard function at one loop, namely 
$H_{f,\bar f}^{\mu(1)}$, is obtained by 
matching, thus this is by construction
equal to the hard region contribution. In 
what follows, we will not discuss further
the pure hard region contribution to the 
form factor.


\subsection{\texorpdfstring{$J_{f\gamma}$}{} 
and \texorpdfstring{$J_{f\partial\gamma}$}{} jet functions}
\label{sec:Jfgamma-fdgamma-1loop}

As discussed in sec.~\ref{sec:NLPFact}, the 
jet function $J_{f\gamma}$ consists of a fermion 
and a transverse photon along the collinear direction. 
As for the $J_f$ jet function, we need to construct 
a gauge-invariant time-ordered operator in QED. 
Concerning the fermion field, it is natural to 
take the gauge-invariant building block 
$\bar\psi(0)\, \Phi_{\bar{n}}(0,\infty)$. 
For the photon field more definitions are 
in principle possible. One can choose to 
describe the photon either through the gauge field 
$A^{\mu}$, or through the gauge field strength 
$F^{\mu\nu}$. We choose a gauge-invariant definition of the gauge field\footnote{We take this occasion to clarify the
following point: gauge-invariant building blocks
such as $\bar\xi(0)\, \Phi_{\bar{n}}(0,\infty)$
and $\Phi_{\bar{n}}(\infty,x)\big(i D_{c\,\perp}^\rho(x) \,
\Phi_{\bar{n}}(x,\infty)\big)$ appear 
in SCET as well. The main difference is that 
the SCET fields are systematically expanded 
in powers of $\lambda$, thus fields 
such as 
$\xi_c= (\slashed{n}\slashed{\bar{n}}/2) \psi$
and $A_{c\perp}^{\mu}$ have homogeneous scaling, 
and the same is true for time-ordered operators
built out of such fields. Here we are defining our 
operators in terms of the full fields in QED, thus 
each matrix element lacks homogenous scaling. 
Instead, they involve a tower of power corrections, 
as seen for instance for $J_f$ in 
\eqns{powerexpansionJf}{eq:Jf1H0Jf0}. The advantage 
of this approach is that the matrix elements
can be evaluated more easily in terms of standard
QED Feynman rules (in addition to eikonal 
interactions originating from Wilson lines). 
A possible drawback is that the various 
matrix elements corresponding to $J_f$, 
$J_{f\gamma}$, $J_{f\gamma\gamma}$, etc.\ 
could have overlapping contributions, which 
one would need to subtract. As it turns out, 
we do not find overlap among the various jet 
functions contributing to the factorization
of the massive form factor. We devote 
app.~\ref{app:seqjets} to discuss this 
point in some detail.}\textsuperscript{,}\footnote{At lowest order in $e$, one finds $\cA^\mu(x) = e\, e_f A^\mu(x)$ supplemented by Wilson line terms that render this building block gauge invariant. This provides some intuition that this building block indeed describes a photon.}
\cite{Bauer:2001yt,Hill:2002vw,Magnea:2018ebr}
\be\label{Acovariant}
{\cal A}^{\mu}(x) = 
\Phi_{\bar{n}}(\infty,x)\big(i D^\mu(x) \,
\Phi_{\bar{n}}(x,\infty)\big),
\ee
where $i D^\mu(x) = i \partial^{\mu} + e \,e_f \,A^{\mu}$
is the covariant derivative. Upon 
describing the photon by means of a gauge 
field strength, the corresponding gauge-invariant building block is 
\begin{equation}\label{Fcovariant}
\mathcal{F}^{\mu+}(x) = 
\Phi_{\bar{n}}(\infty, x)\, F^{\mu+}(x) 
\, \Phi_{\bar{n}}(x,\infty),
\end{equation}
which in QED 
reduces back to $F^{\mu+}$. One can show that the two choices are related, since
\begin{equation}
\mathcal{F}_{\mu\nu} = 
\partial_\mu \mathcal{A}_\nu
-\partial_\nu \mathcal{A}_\mu
-i\left[\mathcal{A}_\mu,\mathcal{A}_\nu\right].
\end{equation}
It is not hard to show that
using either of the two gauge-invariant building
blocks in \eqns{Acovariant}{Fcovariant} gives 
the same $J_{f\gamma}$ function at $\ord(\alpha)$,
up to an overall momentum factor $\ell^+$, which
can be absorbed by a redefinition of the hard 
function. In what follows, we will use the 
gauge-invariant building block given by 
${\cal A}^{\mu}(x)$. Thus we propose the definition 
\begin{equation}\label{eq:Jfgammadef}
J_{f\gamma}^\rho(p_1,\bar{n},\bar{n}\cdot\ell) 
=\int_{-\infty}^\infty \frac{d\xi}{2\pi}\,
e^{-i\ell(\xi \bar{n})}\bra{p_1}
\Big[\bar\psi(0) \Phi_{\bar{n}}(0,\infty)\Big]
\Big[\Phi_{\bar{n}}(\infty,\xi \bar{n})
\Big(iD^\rho \, \Phi_{\bar{n}}(\xi \bar{n},\infty)\Big)
\Big]\ket{0}\,.
\end{equation}
Here $D^\mu$ is understood to be evaluated 
at the point $\xi \bar{n}$, after the 
derivative has been performed. Note that
$\bar{n}\cdot J_{f\gamma} = 0$, since 
$\bar{n}\cdot D \Phi_{\bar{n}}=0$, which 
is a standard Wilson-line identity. Hence, 
the dominant term in \eqn{eq:Jfgammadef} 
is the transverse component $D_{\perp}^\rho$, 
just as for the building block typically 
introduced in SCET, see e.g.\ 
\refrs{Becher:2009th,Becher:2010pd}.
Given that $D_{\perp}^\rho \sim 
\ord(\lambda)$, the function $J_{f\gamma}$ is
suppressed by a power of $\lambda$ 
compared to $J_f$.
Before moving 
to the calculation of $J_{f\gamma}^\rho$
at one loop, let us mention that the 
corresponding jet $J_{\bar f\gamma}^\rho$
for an antifermion and photon along the 
anticollinear direction can be readily 
obtained by replacing the building
blocks in \eqn{eq:Jfgammadef} as 
follows: 
\begin{align}
\begin{cases} \bar\psi(x)\Phi_{\bar{n}}(x,\infty)
\leftrightarrow \Phi_{n}(\infty,x)\psi(x)
, \\[0.1cm]
\Phi_{\bar n}(\infty,x)\left[iD^\rho \Phi_{\bar n}(x,\infty)\right]
\leftrightarrow 
\Phi_n(\infty,x)\left[iD^\rho \Phi_n(x,\infty)\right] . 
\end{cases}
\end{align}

The function $J_{f\gamma}^\rho$ starts at
$\ord(\aem)$. At this order in perturbation 
theory the non-vanishing contribution is
\bea\label{Jfgamma1LcalcA} \nn
J_{f\gamma}^\rho(p_1,\bar{n},\bar{n}\cdot\ell) &=& 
ie^2\int_{-\infty}^\infty \frac{d\xi}{2\pi}\, 
e^{-i\ell(\xi \bar{n})}\bra{p_1}\bar\psi(0)
\left(A^\rho(\xi \bar{n})-\partial_{\xi \bar{n}}^\rho
\int_\infty^0 d\lambda\,\bar{n}\cdot A(\xi \bar{n}+\lambda \bar{n})\right)\\
&&\hspace{5.0cm}\times\, 
\int d^dy\,\left(\bar\psi A^\sigma\gamma_\sigma\psi\right)(y)\ket{0}
+ \ord(\alpha^2).
\eea
\begin{figure}[t]
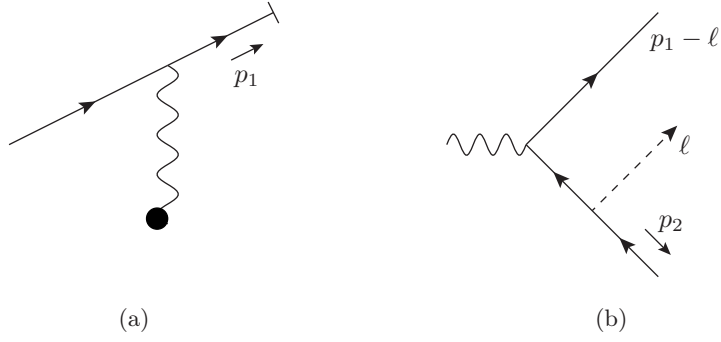

\centering
\subfloat[]{\Jfg1\label{Jfgamma_diaga}}
\hspace{2cm}
\subfloat[]{\Hfgfb0
\label{Jfgamma_diagb}}
\caption{In (a) we show the one-loop $J_{f\ga}$ jet contribution, 
and (b) shows the leading order hard function for this jet. The dashed line in (b) corresponds to the collinear photon 
that connects the hard function with the dot of the jet.}
\label{fig:Jfgamma_diagrams}
\end{figure}
From this we derive an effective Feynman rule for 
the photon propagator:
\begin{equation}\label{eq:Jfgrule}
\begin{gathered}\Jfgrule\end{gathered} = \frac{-i}{k^2}\left(\eta^{\rho\sigma}
-\frac{\bar{n}^\sigma k^\rho}{\bar{n}\cdot k}\right)
\delta(\bar{n}\cdot k - \ell^+).
\vspace{0.2cm}
\end{equation}
This is depicted in fig.\ \ref{Jfgamma_diaga}.
In particular, note that the effective rule
has an open index $\rho$, indicated by the 
dot at one end of the photon propagator, 
which is to be contracted with the open index 
from the photon in the hard function
(discussed below). 
Elaborating \eqn{Jfgamma1LcalcA} 
we get 
\begin{align} \label{eq: Jfgamma1}
J_{f\gamma}^{(1)\rho}(p_1,\bar{n},\ell^+) 
&=i 16\pi^2\, \bar{u}(p_1)\int[dk]
\left(\eta^{\rho\sigma}
-\frac{\bar{n}^\sigma k^\rho}{\bar{n}\cdot k}\right)
\frac{\gamma_\sigma\left(\slashed{p}_1-\slashed{k}
+m\right)}{k^2[(k-p_1)^2-m^2]}
\delta(\bar{n}\cdot k -\ell^+)\,.
\end{align}
As anticipated in sec.~\ref{sec:NLPFact}, 
$J_{f\ga}$ depends on the external momentum 
$p_1$ as well as the dominant loop momentum 
component $\ell^+ $, which flows into the 
jet function and out of the corresponding hard 
function. Given that both the large component 
of $p_1$ and $\ell$, i.e.\ $\hat p^{\mu}_1 = p_1^+ n^{\mu}$
and $\hat l^{\mu} = l^+ n^{\mu}$ respectively, are collinear 
along $n^{\mu}$, it proves useful to write the 
jet function in terms of the momentum fraction
$x=\ell^+/p_1^+$ (see \eqn{NLPfactorizationfgammaB}).
Evaluating the loop integration in \eqn{eq: Jfgamma1}
and expanding up to $\mathcal{O}(\lambda^2)$, we get 
\begin{align} \nn
J_{f\gamma}^{(1)\rho}(p_1,\bar{n},x) 
&= \left(\frac{\bar\mu^2}{m^2}\right)^{\epsilon}
\Gamma(\epsilon)e^{\eps\gamma_E}\, \bar{u}(p_1) 
\, m \, \Bigg\{x^{1-2\eps}
\left(-\gamma^\rho+\frac{\slashed{\bar{n}}\, 
\hat{p}_1^\rho}{p_1^+}\right) \\
&+\frac{m}{p_1^+} \bigg[\frac{1}{2(1-\eps)}
\left(\delta(1-x)-(1-2\eps)x^{1-2\eps}\right)\gamma^\rho \slashed{\bar{n}}-2x^{-2\eps}(1-x)\bar{n}^\rho\bigg]
\Bigg\}. \label{eq:Jfgamma1}
\end{align}
We see that the jet function $J_{f\ga}$
starts at $\ord(\lambda)$. Furthermore, the 
contribution at $\ord(\lambda^2)$ originates 
from expansion into the small components of $p_1$, 
as anticipated in \eqn{powerexpansionJfgamma}.
The contour integration over $k^-$ determines
the integration domain of $x$ to be in the range 
$0\leq x \leq 1$. Outside of this unit interval, 
the contour integral over $k^-$ yields zero, 
because the two propagator poles would be on the same side 
of the integration contour. The result in 
\eqn{eq:Jfgamma1} is equivalent to the one 
obtained in \refr{Laenen:2020nrt}, see 
in particular eq.~(48) there. However, 
we stress that the jet function
calculated in \refr{Laenen:2020nrt}
was based on a diagrammatic 
definition, while the jet function 
in \eqn{eq:Jfgamma1} follows from 
a well-defined matrix element.

As discussed in sec.~\ref{sec:NLPFact}, 
there is a second jet function involving 
a fermion and a transverse photon along 
the collinear direction, namely
$J^{\rho\sigma}_{f\partial\gamma}$, 
as defined in \eqn{JfpartialgammaDef1}. 
The function $J^{\rho\sigma}_{f\partial\gamma}$
emerges upon expansion of the corresponding hard
functions in powers of $\ell_{\perp}$, see
\eqns{powerexpansionHfgammaB}{NLPfactorizationfpartialgamma}.
It is possible to incorporate the 
factor of $\ell_{\perp}$ in \eqn{JfpartialgammaDef1}
directly into a matrix element definition for 
$J^{\rho\sigma}_{f\partial\gamma}$, which is 
given by 
\begin{equation}\label{eq:Jfdelgammadef} 
J_{f\partial\gamma}^{\rho\sigma}(p_1,\bar{n},\bar{n}\cdot\ell) 
=\int_{-\infty}^\infty \frac{d\xi}{2\pi}\,
e^{-i\ell(\xi \bar{n})} \bra{p_1}\bigg[\bar\psi(0) 
\Phi_{\bar{n}}(0,\infty)\bigg] \, i\partial^\sigma_{\perp}\, 
\bigg[\Phi_{\bar{n}}(\infty,\xi \bar{n})
\left(iD^\rho \Phi_{\bar{n}}(\xi \bar{n},\infty)
\right)\bigg]\ket{0}.
\end{equation}
Notice that the labelling of the jet function, 
``$f\partial\gamma$'', is chosen such as 
to indicate that the factor of $i\partial^\sigma_{\perp}$ 
in \eqn{eq:Jfdelgammadef} acts on the photon building block.
The transverse derivative generates the factor of
$\ell_\perp$ introduced in \eqn{JfpartialgammaDef1}, 
and as such makes the jet function 
$J_{f\partial\gamma}^{\rho\sigma} \sim \ord(\lambda^2)$.
One can readily show that
\begin{align}
J^{(1)\rho\sigma}_{f\partial\gamma}(p_1,\bar{n},x)&=\left(\frac{\bar\mu^2}{m^2}\right)^{\epsilon}\Gamma(\epsilon)e^{\eps\gamma_E}\bar{u}(p_1)\frac{m^2x^{2-2\epsilon}}{2(1-\epsilon)}\left[\frac{2}{x}\eta^{\rho\sigma}_\perp +\gamma_\perp^\sigma\left(\frac{\slashed{\bar{n}}\, \hat{p}_1^\rho}{p_1^+}-\gamma^\rho\right)\right],\label{eq:Jfdelg}
\end{align}
which coincides with the result found earlier 
in \refr{Laenen:2020nrt}, but, again, \eqn{eq:Jfdelg} now follows 
from a matrix element definition. 

In order to evaluate the collinear 
contribution due to the two factorized 
terms in \eqns{NLPfactorizationfgamma}{NLPfactorizationfpartialgamma}
we still need to determine the corresponding 
hard functions at lowest order. These are represented by the diagram in fig.~\ref{Jfgamma_diagb}, 
where the antifermion propagator has hard 
momentum $-p_2 - \ell$. Before expansion in 
$l_{\perp}$, the full hard function reads
\begin{equation}\label{Hfgammacalc1}
H_{f\gamma,\bar f\, \rho}^{\mu}(p_1-\ell,\ell;p_2) = 
\frac{1}{-e}(-ie\gamma^\mu)
\frac{i(-\slashed{p}_2-\slashed{\ell}+m)}{(p_2+\ell)^2-m^2}
(-ie\gamma_\rho),
\end{equation}
where we divided by $-e$ in order to match 
consistently with the jet functions defined 
in \eqns{eq:Jfgammadef}{eq:Jfdelgammadef}, 
which already contain such factor. 
Expanding \eqn{Hfgammacalc1} according to 
\eqn{powerexpansionHfgammaB}, we get the two 
coefficients 
\begin{align} \label{eq:hardfgamma}
H^{(0)\mu}_{f\gamma,\bar f \rho}(\hat{p}_1,\hat{p}_2,x) 
&=\frac{-ie}{x\s}\gamma^\mu \left(\hat{\slashed{p}}_2 
+ x\, \hat{\slashed{p}}_1-m\right) \gamma_\rho, \\
\label{eq:hardfdgamma}
H^{(0)\mu}_{f\partial\gamma,\bar f \rho\sigma}(\hat{p}_1,\hat{p}_2,x)
&=\frac{-ie}{x\s}\gamma^\mu\gamma_{\perp\sigma} \gamma_\rho.
\end{align}
Notice that the factor of $1/x$ appearing in both 
functions could potentially give rise to an endpoint 
singularity, for $x\to 0$, in the convolution with 
the jet function. As it turns out, at one loop 
this potential divergence is regularized in 
dimensional regularization by
$J_{f\gamma}^{\rho}$ and 
$J_{f\partial\gamma}^{\rho\sigma}$, 
see \eqns{eq:Jfgamma1}{eq:Jfdelg}. It is 
therefore not necessary at this order to 
additionally introduce an analytic regulator.
The emergence of endpoint singularities (regulated 
by dimensional or analytic regularization) has been 
observed in SCET as well, 
see \refrs{Moult:2019uhz,Liu:2019oav,Beneke:2020ibj,Beneke:2022obx}. 
The potential endpoint divergences we encounter here 
are entirely equivalent to those found in SCET.
In this regard, let us mention that the presence of 
endpoint divergences prevents the naive renormalization 
of the hard and jet function separately, and consequently 
the resummation of large logarithms by means of the 
renormalization group \cite{Liu:2019oav,Beneke:2020ibj,Beneke:2022obx}. Even though this reasoning applies both to SCET 
and to the approach considered here,
it has been shown that the resummation of large 
logarithms in presence of endpoint divergences
can be obtained also by diagrammatic exponentiation,
see \refrs{vanBeekveld:2021mxn,vanBeekveld:2023liw}.
Indeed, diagrammatic exponentiation can be formulated 
directly in dimensional regularization, and would 
thus be particularly suitable for the factorization
approach considered here.

It may be interesting at this point to notice 
that the jet function $J_{f\gamma}$ 
in \eqn{eq: Jfgamma1} contains two terms: the first 
one, proportional to $\eta^{\rho\si}$, when contracted 
with the hard function is equal to the full $c$-region. 
However, as is shown in app.\ \ref{app:seqjets}, the 
second term yields the contribution given by the $J_f$ 
jet function. Therefore, we see that in the $J_{f\ga}$ 
jet function the contribution from the LP $J_f$ jet 
function gets cancelled, preventing any double counting. Furthermore, it is clear 
that this subtraction in eq.~\eqref{eq: Jfgamma1} 
of the second term makes the $J_{f\ga}$ jet function 
of order $\cO(\la)$. Hence the effective photon propagator 
in \eqn{eq:Jfgrule} is subleading, i.e.\ of 
$\mathcal{O}(\la)$. As shown in app.\ \ref{app:seqjets},
by means of the Ward identity, this is a more general 
structure that is also present in the $J_{f\ga\ga}$ jet function. 

We can now verify the complete factorization formula 
at one loop, up to $\mathcal{O}(\lambda^2)$. In particular, 
we can check now that we are able to reproduce the full 
$c$-region. Adding $V^{\mu(1)}_{f,\bar f}\vert_{c}$, 
 $V^{\mu(1)}_{f\gamma,\bar f}\vert_{c}$ and $V^{\mu(1)}_{f\partial\gamma,\bar f}\vert_{c}$, and 
projecting onto the form factors, we obtain
\begin{align}
\left.F_1^{(1)}\right\vert_{\rm c}&=\left(\frac{\bar\mu^2}{m^2}\right)^\epsilon\bigg[\frac{1}{\epsilon^2}+\frac{2}{\epsilon}+4+\frac{\zeta_2}{2}+ \frac{m^2}{\s}\bigg(\frac{1}{\eps}+5\bigg)+\mathcal{O}(\lambda^4,\eps)\bigg],\\
\left.F_2^{(1)}\right\vert_{\rm c}&=\left(\frac{\bar\mu^2}{m^2}\right)^\epsilon\bigg[\frac{m^2}{\s}\left(-\frac{2}{\epsilon}-8\right)+\mathcal{O}(\lambda^4,\eps)\bigg],
\end{align}
which is in agreement with the region result in the $c$-collinear region. The $\bar c$-collinear region follows readily when calculating $J_{\bar f\gamma}$ and $J_{\bar f\partial \gamma}$, which have similar matrix element definitions as their $c$-collinear counterpart. This concludes the discussion of the factorization theorem at one loop, in which we have shown that it reproduces the QED form factor exactly up to $\mathcal{O}(\la^2)$.

%% file: Sections/Two_Loop.tex
\begin{figure}[H]
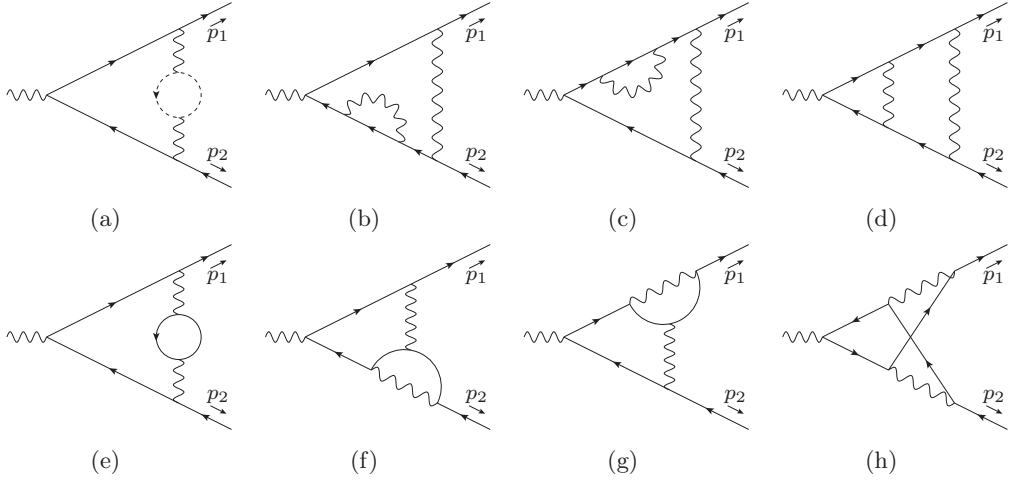

\centering
\subfloat[]{
    \SetScale{0.5}\text{\footnotesize\QEDda}
}
\label{fig:QED_a}
\subfloat[]{
   \SetScale{0.5} \text{\footnotesize\QEDdb}
}
\label{fig:QED_b}
\subfloat[]{
  \SetScale{0.5}  \text{\footnotesize\QEDdc}
}
\label{fig:QED_c}
\subfloat[]{
  \SetScale{0.5}  \text{\footnotesize\QEDdd}
}
\label{fig:QED_d}
\subfloat[]{
    \SetScale{0.5}\text{\footnotesize\QEDde}
}
\label{fig:QED_e}
\subfloat[]{
   \SetScale{0.5} \text{\footnotesize\QEDdf}
}
\label{fig:QED_f}
\subfloat[]{
    \SetScale{0.5}\text{\footnotesize\QEDdg}
}
\label{fig:QED_g}
\subfloat[]{
   \SetScale{0.5} \text{\footnotesize\QEDdh}
}
\label{fig:QED_h}
\caption{Diagrams that contribute to the massive form factor at two loops in QED. Dashed lines represent massless fermions.}
\label{fig:QED_diagrams}
\end{figure}
\begin{figure}[H]
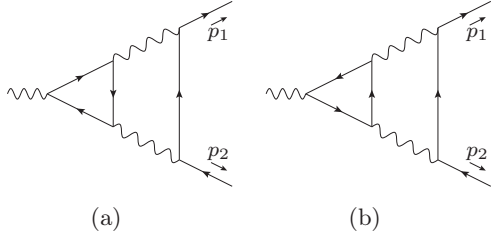

\centering
\subfloat[]{
    \SetScale{0.5}\text{\footnotesize\QEDdFurrya}
}
\label{fig:furry_a}
\subfloat[]{
   \SetScale{0.5} \text{\footnotesize\QEDdFurryb}
}
\label{fig:furry_b}
\caption{The two diagrams involving triangle fermion loops that cancel by Furry's theorem.}
\label{fig:Furry_diagrams}
\end{figure}

We move on to check the factorization theorem at two loop order. Now the $\mathcal{O}(\lambda^2)$ $J_{f\gamma\gamma}$ and $J_{f\!f\!\bar{f}}$ jet functions appear, since their leading-order contributions start at $\mathcal{O}(\aem^2)$. These jet functions, together with the $J_{f}$ jet, the $J_{f\ga}$ jet and the $J_{f\partial\ga}$ jet at two loops, are the necessary ingredients to check the double-collinear ($cc$ and $\bar c\bar c)$ region. Beside this check, one can already verify the two-loop  collinear-anticollinear ($c\bar c$) and the collinear-hard ($ch$ and $\bar c h$) factorization with the one-loop expressions for the $J_{f}$ jet and the $J_{f\gamma}$ and $J_{f\partial\ga}$ jet. We will indeed first check these two regions, and only then we will consider the full two-loop calculations of the (subleading) jet functions. The QED diagrams that appear at two loops are given in fig.\ \ref{fig:QED_diagrams}. There are also another two diagrams involving triangle fermion loops, but they cancel each other by Furry's theorem \cite{PhysRev.51.125}, see fig.\ \ref{fig:Furry_diagrams}.

\subsection{Verifying collinear-anticollinear factorization}\label{sec:ccbar}

At two loops, the terms in the factorization theorem 
\eqn{NLPfactorizationQQbar} contributing to the 
collinear-anticollinear region are given by 
\bea\label{eq:ccbarfact} \nn
\left.V^{\mu(2)}\right\vert_{c\bar c} &=& 
\left.V^{\mu(2)}_{f,\bar f}\right\vert_{c\bar c}
+\left.V^{\mu(2)}_{f\gamma,\bar f}\right\vert_{c\bar c}
+\left.V^{\mu(2)}_{f\partial\gamma,\bar f}\right\vert_{c\bar c}
+\left.V^{\mu(2)}_{f,\bar f\gamma}\right\vert_{c\bar c}
+\left.V^{\mu(2)}_{f,\bar f\partial\gamma}\right\vert_{c\bar c}
+\left.V^{\mu(2)}_{f\gamma,\bar f\gamma}\right\vert_{c\bar c} 
+ \ord(\lambda^3) \\[0.3cm] \nn
&=&  J_{f}^{(1)}\, H_{f,\bar f}^{\mu(0)} \, J_{\bar f}^{(1)}
+ J_{f\gamma\, \rho}^{(1)}\otimes H_{f\gamma,\bar f}^{\mu\rho(0)} \, J_{\bar f}^{(1)}
+ J_{f\partial\gamma\, \rho\sigma}^{(1)} \otimes 
H_{f\partial\gamma,\bar f}^{\mu\rho\sigma(0)} \, J_{\bar f}^{(1)} \\[0.2cm] 
&&+\, J_{f}^{(1)}\, H_{f,\bar f\gamma}^{\mu\rho(0)} \otimes J_{\bar f\gamma\, \rho}^{(1)}
+ J_{f}^{(1)}\, H_{f,\bar f\partial\gamma}^{\mu\rho\sigma(0)} 
\otimes J_{\bar f\partial\gamma\, \rho\sigma}^{(1)}
+ J_{f\gamma\, \rho}^{(1)}\otimes H_{f\gamma,\bar f\gamma}^{\mu\rho\sigma(0)} 
\otimes J_{\bar f\gamma\, \sigma}^{(1)} + \ord(\lambda^3) ,\,\,
\eea
where we indicated 
convolution with the symbol $\otimes$. 
All terms in this equation are known from 
sec.\ \ref{sec:oneloop}, except for 
$H_{f\gamma,\bar f\gamma}^{\mu\rho\sigma(0)}$. 
The corresponding term in the factorization 
theorem has been introduced in 
\eqn{NLPfactorizationfgammafgamma}, and 
contributes for the first time at 
two loops, since the corresponding jet functions 
start at $\mathcal{O}(\aem)$. 
$H_{f\gamma,\bar f\gamma}$ consists 
of three contributions, namely
\vspace{0.3cm}
\begin{equation}\label{Hfgfg}
H_{f\gamma,\bar f\gamma }^{\mu\rho\sigma}
(\hat{p}_1,\hat{p}_2,x,\bar x) = \; \begin{gathered}
    \SetScale{0.8}\Hfgfbgf \hspace{0.5cm}
    \end{gathered}+
    \begin{gathered}
        \hspace{0.2cm}\SetScale{0.8}\Hfgfbgg\hspace{0.5cm}
    \end{gathered}+
    \begin{gathered}
       \hspace{0.2cm} \SetScale{0.8}\Hfgfbgh
    \end{gathered}\quad,
\vspace{0.3cm}
\end{equation}
where $x=\ell_1^+/p_1^+$ and $\bar x = \ell_2^-/p_2^-$ 
are the relevant momentum fractions. $H_{f\gamma,\bar f\gamma}$
is easily expressed in terms of propagators and vertices by using the standard Feynman rules in QED, thus we do not report the 
exact expression here. 

With the calculation of $H_{f\gamma,\bar f\gamma}$
in place, we have all terms needed to evaluate 
\eqn{eq:ccbarfact}, which we then use to evaluate 
the collinear-anticollinear contribution to the 
form factors. A check, including only 
1PI diagrams, shows that the leading power 
contribution to $F_1$ equals
\begin{equation}\label{ccbarLP}
\left.F_1^{(2)}\right\vert_{\mathrm{ LP},c\bar c}=
\left(\frac{\bar\mu^2}{m^2}\right)^{2\eps}
\left[\frac{1}{\eps^4}+\frac{4}{\eps^3}
+\frac{12+\zeta_2}{\eps^2}
+\frac{32+4\zeta_2-\frac{2\zeta_3}{3}}{\eps}
+80+12\zeta_2-\frac{8\zeta_3}{3}
+\frac{7\zeta_2^2}{10}+\mathcal{O}(\eps)\right],
\end{equation}
which agrees with the region calculation.

One may be inclined to think that the generalization to subleading powers is now readily achieved by calculating all the terms in \eqn{eq:ccbarfact}, but that is not the case. Indeed, if one calculates the subleading power contributions as one would expect at first, there is a mismatch between the region approach and our factorization approach. The reason behind this is a subtle point, to which app.\ \ref{app:regtofact} is devoted. As already discussed in sec.\ \ref{sec:setup}, it is crucial to include self-energy contributions to have a valid factorization theorem. In a massless gauge theory this subtlety does not appear as the self-energy contribution consists of scaleless integrals, which vanish. However, since we are working with a nonzero mass $m$, the inclusion of self energies also modifies the mass-shell condition and the Dirac equations of motion, which alter the expressions. There is also a residue factor, although that will not change whether factorization holds or not, since it is multiplicative.

The additional contributions for the collinear-anticollinear result then come from using the one-loop collinear (anticollinear) result, both on the region side and on the factorization side, and using the new one-loop equations of motion and mass-shell condition on the anticollinear (collinear) leg, as given respectively in \eqns{eom}{smallcomponents2}: this is equivalent to including the self-energy contribution on the anticollinear (collinear) leg. It does not alter the ${\cal O}(\alpha)$ result, but gives a new ${\cal O}(\alpha^2)$ contribution on the region side as well as on the factorization side. These new contributions from both sides are not identical. At $\mathcal{O}(\lambda)$, these additional contributions precisely cure the existing mismatch in the naive factorization check of the 1PI amplitude. We report here the total contribution to $F_2$ including self-energy corrections, namely
\begin{equation}
    \left.F_2^{(2)}\right\vert_{\mathrm{NLP}, c\bar c}= \left(\frac{\bar\mu^2}{m^2}\right)^{2\eps}\frac{m^2}{\s}\left[\frac{8}{\eps^3}+\frac{18}{\eps^2}+\frac{44+8\zeta_2}{\eps}+104+18\zeta_2-\frac{16\zeta_3}{3}+\mathcal{O}(\eps)\right].
\end{equation}

At $\mathcal{O}(\lambda^2)$, the modified equations of motion and mass-shell condition are also the missing ingredients compared to the naive calculation. However another subtlety arises here, namely the choice of analytic regulator. If we mimic the use of analytic regulator as in the region approach, we do not yet achieve a match between the factorization approach and the region approach. This can be explained by app.\ \ref{app:regtofact}. The derivation done there would be invalid if one uses the analytic regulators from the region approach. Indeed, previously scaleless integrals would receive a scale and thus not vanish, nor would the important cancellations happen. The way out of this is choosing a different propagator for the analytic regulator. If one puts the regulator on photon lines only, and thus not on fermion lines, the Ward identity, frequently used in app.\ \ref{app:regtofact}, is operative. Moreover, scaleless integrals remain scaleless. As a result, the $J_{f\ga}$ jet develops a factor $(1-x)^\nu$, which can be readily derived, together with the known $x^{-2\eps}$, cf.\ \eqn{eq:Jfgamma1}. Hence both the endpoint divergences at $x=0$ and $x=1$ are now regularized. As a result, the $H_{f\gamma,\bar{f}}$ and $H_{f\gamma,\bar{f}\gamma}$ hard functions do not contain any analytic regulator. 

We now report the total contribution to $F_1$ including the self-energy correction, as was done for $F_2$. We put analytic regulators $\nu_1$ and $\nu_2$ on respectively the $k_1^2$ and $k_2^2$ photon propagator, and obtain\footnote{Compared to eq.~\eqref{ccbarLP}, also the LP contribution is modified, as a result of including the self-energy corrections.}
\begin{align}
     \left.F_1^{(2)}\right\vert_{c\bar c} &= \left(\frac{\bar\mu^2}{m^2}\right)^{2\eps}\left(\frac{\tilde \mu^2}{m^2}\right)^{2\nu}\bigg\{\frac{1}{\eps^4}+\frac{3}{\eps^3}+\frac{8+\zeta_2}{\eps^2}+\frac{20+3\zeta_2-\frac{2\zeta_3}{3}}{\eps}+48+8\zeta_2-2\zeta_3+\frac{7\zeta_2^2}{10}\nn\\
     &+\frac{m^2}{\s}\bigg[\frac{4}{\eps^4}+\frac{1}{\eps^3}\left(-\frac{4}{\nu_1}-\frac{4}{\nu_2}\right)+\frac{1}{\eps^2}\left(\frac{4}{\nu_1\nu_2}-13-20\zeta_2\right)\nn\\
     &\hspace{1cm}+\frac{1}{\eps}\left(-\frac{4}{\nu_1\nu_2}+\frac{8\zeta_2}{\nu_1}+\frac{8\zeta_2}{\nu_2}-35-\frac{80\zeta_3}{3}\right)+\frac{-4+4\zeta_2}{\nu_1\nu_2}+\frac{1}{\nu_1}\left(-8-12\zeta_2+\frac{44\zeta_3}{3}\right)\nn\\
     &\hspace{2cm}+\frac{1}{\nu_2}\left(-8-12\zeta_2+\frac{44\zeta_3}{3}\right)-91-13\zeta_2-14\zeta_2^2\bigg]+\mathcal{O}(\lambda^4,\eps,\nu_1,\nu_2)\bigg\},
\end{align}
which agrees with the region calculation if one accounts for the new equations of motion, mass-shell condition, and the different use of the analytic regulator on that side as well.

For the other checks of factorization at the two-loop level in the next subsections, it is not strictly necessary to include self-energy corrections. It suffices to calculate the form factor of the 1PI amplitude and compare that with the region calculation, where the self-energy diagrams were omitted.

\subsection{Verifying hard-collinear factorization}

Next, we select in \eqn{NLPfactorizationQQbar} 
the contribution where one loop is collinear 
and the other one is hard, i.e., up to 
$\mathcal{O}(\lambda^2)$ we consider 
\bea\label{Vch2Loops} \nn
\left.V^{\mu(2)}\right\vert_{ch} &=& 
\left.V^{\mu(2)}_{f,\bar f}\right\vert_{ch}
+\left.V^{\mu(2)}_{f\gamma,\bar f}\right\vert_{ch}
+\left.V^{\mu(2)}_{f\partial\gamma,\bar f}\right\vert_{ch} + \ord(\lambda^3) \\[0.3cm] 
&=& \Big( J_{f}^{(1)}\, H_{f,\bar f}^{\mu(1)} 
+ J_{f\gamma\, \rho}^{(1)}\otimes H_{f\gamma,\bar f}^{\mu\rho(1)}
+ J_{f\partial\gamma\, \rho\sigma}^{(1)} \otimes 
H_{f\partial\gamma,\bar f}^{\mu\rho\sigma(1)} \Big) J_{\bar f}^{(0)} 
+ \ord(\lambda^3).
\eea
In this case, the only missing factors are 
the one-loop hard functions. For the hard 
function corresponding to the $J_{f}$ jet 
one finds
\begin{align}\label{eq:Hf1}
H_{f,\bar f}^{(1)\mu}(p_1,p_2) 
&\,=\,\begin{gathered}
\SetScale{0.6}\QEDOneLoop\end{gathered}
\hspace{-1.0cm}
=\,e^3\int [dk]\,\gamma^\mu
\frac{i(\slashed{p}_1-\slashed{k}+m)}{k^2-2\hat{p}_1\cdot k}
\left(1+\frac{2m^2}{\s}\frac{\hat{p}_2\cdot k}{k^2-2\hat{p}_1\cdot k}\right) \nn \\
&\hspace{4.0cm}\times\, 
\gamma^\mu\frac{i(-\slashed{p}_2-\slashed{k}+m)}{k^2+2\hat{p}_2\cdot k}\left(1-\frac{2m^2}{\s}\frac{\hat{p}_1\cdot k}{k^2+2\hat{p}_2\cdot k}\right)\gamma_\mu\frac{1}{k^2},
\end{align}
where we have expanded the propagators up 
to $\mathcal{O}(\lambda^2)$. The leading 
power contribution arises from 
$V^{\mu(2)}_{f,\bar f}\vert_{ch}$ only.
Projecting onto the form factors we get at 
leading power
\begin{align}
\left.F_1^{(2)}\right\vert_{\mathrm{LP}, ch} 
&= \left(\frac{\bar\mu^2}{-\s}\right)^\eps
\left(\frac{\bar\mu^2}{m^2}\right)^\eps
\bigg[-\frac{2}{\eps^4}-\frac{7}{\eps^3}
-\frac{22}{\eps^2}+\frac{1}{\eps}\left(-60+\frac{16\zeta_3}{3}\right) \nn \\
&\hspace{4cm}-152+\frac{56\zeta_3}{3}+\frac{12\zeta_2^2}{5}
+\mathcal{O}(\eps)\bigg],
\end{align}
which matches the region result.

Considering now the one-loop hard function 
corresponding to the $J_{f\ga}$  and $J_{f\partial\gamma}$ 
jet functions, it becomes rather cumbersome to write down 
the full expression, hence we resort to a 
diagrammatic representation. There are five 
contributions, corresponding to the five 
diagrams that contribute to the $ch$-region, 
which add up to the hard function
\begin{align} \nn
H_{f\gamma,\bar f}^{\mu\rho}(p_1-\hat \ell,\hat \ell;p_2)
+\ell_{\perp \sigma} \, H_{f\partial\gamma,\bar f}^{\mu\rho\sigma}
(\hat p_1-\hat \ell,\hat \ell;\hat p_2) & \\
&\hspace{-6.0cm}=\,\begin{gathered}
\includegraphics[width=0.17\textwidth]{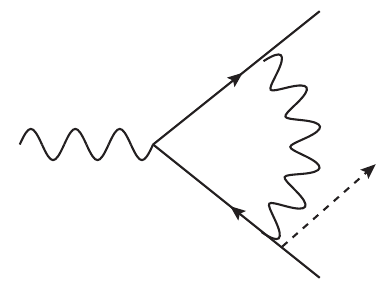}\end{gathered}
+\begin{gathered}
\includegraphics[width=0.17\textwidth]{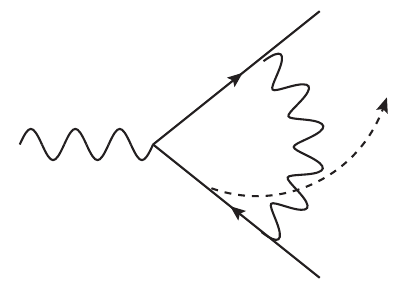}\end{gathered}
+\begin{gathered}
\includegraphics[width=0.17\textwidth]{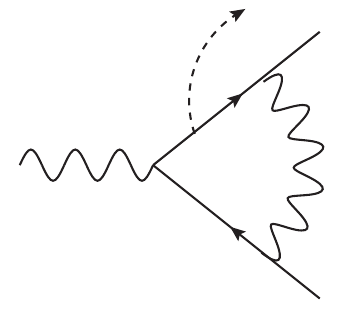}\end{gathered}  \nn \\
&\hspace{-4.0cm}+\begin{gathered}
\includegraphics[width=0.17\textwidth]{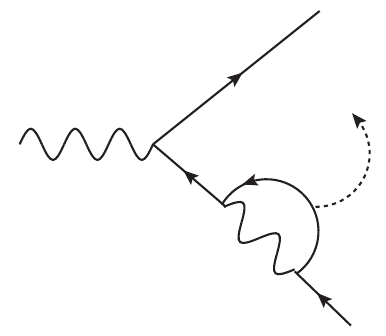}\end{gathered}
+\begin{gathered}
\includegraphics[width=0.17\textwidth]{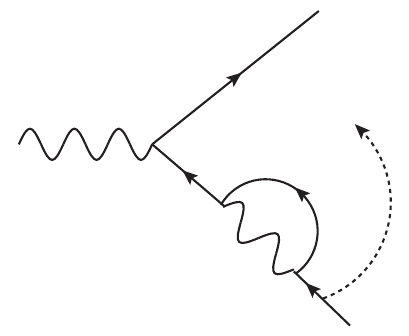}\end{gathered}\, ,
\label{eq:Hfgamma1}
\end{align}
where the dashed line represents a 
$c$-collinear photon that is to be 
connected to the $J_{f\gamma}$ 
($J_{f\partial\gamma}$) jet function. 
Hence momentum $\ell$ flows
through this line, while the remaining loop 
is hard. The relevant propagators need 
to be expanded up to $\mathcal{O}(\lambda)$ 
in order to be accurate up to 
$\mathcal{O}(\lambda^2)$ when combined 
with the corresponding jet function. 
For example, one needs to do the expansion 
\begin{equation}  
\frac{\slashed{k}+\slashed{\ell}+\slashed{p}_2+m}{(k+\ell+p_2)^2-m^2} 
= \frac{\slashed{k}+\hat{\slashed{\ell}}
+\slashed{\ell}_{\perp}+\hat{\slashed{p}}_2+m
+\mathcal{O}(\lambda^2)}{(k+\hat{\ell}+\hat{p}_2)^2}
\left(1-\frac{2k_{\perp}\cdot \ell_{\perp}}{(k+\hat{\ell}+\hat{p}_2)^2}
+\mathcal{O}(\lambda^2)\right),
\end{equation}
where $k$ is hard and $\ell$ is $c$-collinear. 
To obtain the $H_{f\ga,\bar{f}}$ hard function, one 
evaluates the relevant propagators at $\ell_{\perp}=0$. 
For the $H_{f\partial\gamma,\bar{f}}$ hard function, 
one takes these perpendicular components into account 
by calculating the derivative of the $H_{f\gamma,\bar{f}}$ 
hard function with respect to $\ell_{\perp}$, cf.\ the 
calculation done at one loop for the $J_{f\ga}$ jet.

With the calculation of the one-loop hard functions
in place, it is then possible to evaluate each term
in \eqn{Vch2Loops} and project onto the form factors,
obtaining
\begin{align}
\left.F_2^{(2)}\right\vert_{\mathrm{NLP}, ch} 
&=\left(\frac{\bar\mu^2}{-\s}\right)^\eps
\left(\frac{\bar\mu^2}{m^2}\right)^\eps\frac{m^2}{\s}
\bigg[-\frac{4}{\eps^3}-\frac{8}{\eps^2}
+\frac{-56+32\zeta_2}{\eps}-280+136\zeta_2
+\frac{320\zeta_3}{3}+\mathcal{O}(\eps)\bigg]
\end{align}
for $F_2$, which originates from the $\mathcal{O}(\lambda)$
contribution in \eqn{Vch2Loops}. At $\mathcal{O}(\lambda^2)$,
$\left.V^{\mu(2)}\right\vert_{ch}$ contributes to $F_1$, 
and we get 
\begin{align}
\left.F_1^{(2)}\right\vert_{\mathrm{NLP}, ch} 
&=\left(\frac{\bar\mu^2}{-\s}\right)^\eps
\left(\frac{\bar\mu^2}{m^2}\right)^\eps\frac{m^2}{\s}
\bigg[-\frac{4}{\eps^4}+\frac{2}{\eps^3}
+\frac{-8+20\zeta_2}{\eps^2}+\frac{1}{\eps}
\left(-22-30\zeta_2+\frac{284\zeta_3}{3}\right) \nn\\
&\hspace{4cm}+6-92\zeta_3-\frac{322\zeta_3}{3}
+\frac{568\zeta_2^2}{5}+\mathcal{O}(\eps)\bigg],
\end{align}
which is exactly what we should find according to 
\refr{terHoeve:2023ehm}. Equivalent results can be 
obtained for the $\bar c h$-region. This concludes 
the verification of the factorization approach for 
the hard-collinear region.

\subsection{Verifying double-collinear factorization}

In order to complete the verification of 
the factorization formula in \eqn{NLPfactorizationQQbar}
we need to check one last contribution, which is
the double-collinear region at two loops. 
Specializing \eqn{NLPfactorizationQQbar}
to this region we have
\bea\label{NLPfactorizationQQbarcc} \nn
V^{\mu(2)}\Big|_{cc} &=& 
V_{f,\bar f}^{\mu(2)}\Big|_{cc}
+V_{f\gamma,\bar f}^{\mu(2)}\Big|_{cc}
+V_{f\partial\gamma,\bar f}^{\mu(2)}\Big|_{cc}
+ V_{f\gamma\gamma,\bar f}^{\mu(2)}\Big|_{cc}  
+V_{f\! f\!\bar f,\bar f}^{\mu(2)}\Big|_{cc}
+ \ord\big(\lambda^{3}\big) \\[0.3cm] \nn 
&=& \Big( J_{f}^{(2)}\, H_{f,\bar f}^{\mu(0)} 
+ J_{f\gamma\, \rho}^{(2)}\otimes H_{f\gamma,\bar f}^{\mu\rho(0)} 
+ J_{f\partial\gamma\, \rho\sigma}^{(2)} \otimes 
H_{f\partial\gamma,\bar f}^{\mu\rho\sigma(0)} \\[0.2cm] 
&&\hspace{2.0cm}
+\,J_{f\gamma\gamma\, \rho\sigma}^{(2)}
\otimes H_{f\gamma\gamma,\bar f}^{\mu\rho\sigma(0)} 
+ J_{f\!f\!\bar f}^{(2)}\otimes H_{f\!f\!\bar f,\bar f}^{\mu(0)} \Big)
J_{\bar f}^{(0)} + \ord(\lambda^3).
\eea
At this order we need to evaluate both the $J_{f}$, 
the $J_{f\ga}$ and the $J_{f\partial \gamma}$ jet functions 
at second order in perturbation theory. Furthermore,
for the first time at this order, we get the contribution
due to the $J_{f\ga\ga}$ and $J_{f\!f\!\bar{f}}$ jet functions,
introduced respectively in 
\eqns{NLPfactorizationfgammagamma}{NLPfactorizationfff}.
In what follows we proceed first to the calculation of 
$J_{f}$, $J_{f\ga}$ and $J_{f\partial \gamma}$ at two 
loops. Then we will consider the remaining $J_{f\ga\ga}$ 
and $J_{f\!f\!\bar{f}}$ jet functions, provide their matrix 
element definition and proceed to calculate them 
at $\ord(\alpha^2)$.

\subsubsection{\texorpdfstring{$J_{f}$ jet function}{}}
\label{sec:f-jetfunction}

Starting from the definition in \eqn{eq: defJf} and  
upon performing all Wick contractions up to 
$\mathcal{O}(e^4)$, we find that the jet function 
$J_{f}$ at two loops is given by the following diagrams:
\begin{align}
    J_{f}^{(2)}(p_1,\bar{n}) &= \begin{gathered}
    \SetScale{0.6}\hspace{0.5cm}
   \text{\footnotesize\Jfda}\end{gathered}+\begin{gathered}
    \SetScale{0.6}\hspace{0.5cm}\text{\footnotesize\Jfdc}\end{gathered}+\begin{gathered}
   \SetScale{0.6}\hspace{0.5cm}\text{\footnotesize\Jfdd}\end{gathered}\nn\\
   &+\begin{gathered}
   \SetScale{0.6}\hspace{0.5cm}\text{\footnotesize\Jfde}\end{gathered}+\begin{gathered}
    \SetScale{0.6}\hspace{0.5cm}\text{\footnotesize\Jfdg}\end{gathered}+ \begin{gathered}
        \SetScale{0.6}\hspace{0.5cm}\text{\footnotesize\Jfdh}
    \end{gathered}.\label{eq:Jf2diag}
\end{align}
The corresponding expression in terms of two-loop integrals can be found in app.\ \ref{app:2loopres}. The dotted line in the first diagram represents a light fermion loop, cf.\ diagram (a) in fig.~\ref{fig:QED_diagrams}. This contribution should be multiplied by a factor of $n_f$ when there are $n_f$ light flavours, and by a factor of $e_f^2$, where $e_f$ is the (fractional) charge of the fermion in the bubble. The fourth contribution also has a fermion loop, but that fermion is massive with mass $m$, which is proportional to the small scale around which we expand.

Since this jet function does not have any open Lorentz indices, nor is there an unintegrated momentum fraction, it is quite straightforward to calculate the contribution from this jet. Just like in \refr{terHoeve:2023ehm}, the integrals for these diagrams can be written in terms of three different topologies, namely topology $A$, $B$ and $X$. Since the different contributions in eq.~\eqref{eq:Jf2diag} have resemblance with the various two-loop diagrams that exist, it is straightforward to see which part of the expression can be written in terms of which topology, see \refr{terHoeve:2023ehm} for this classification. We take the two loop momenta to be collinear, and one now defines topology $A_{cc}$ to be 
\begin{align}
    I_{A_{cc};\{n_i\}} &\equiv \int [dk_1][dk_2]\frac{1}{[k_1^{2}]^{n_1}}\frac{1}{[k_2^{2}]^{n_2}}\frac{1}{[(k_1-k_2)^{2}]^{n_3}}\frac{1}{[(k_1+p_1)^2-m^2]^{n_4}}\frac{1}{[(k_2+p_1)^2-m^2]^{n_5}}\nn\\
    &\quad \times \frac{1}{[-2k_1\cdot \hat{p}_2]^{n_6}}\frac{1}{[-2k_2\cdot \hat{p}_2]^{n_7}}\,,\label{eq:topA}
\intertext{
topology $B_{cc}$ to be}
    I_{B_{cc};\{n_i\}} &\equiv \int[dk_1]\int[dk_2]\frac{1}{[k_1^2-m^2]^{n_1}}\frac{1}{[k_2^2]^{n_2}}\frac{1}{[(k_1-k_2)^2-m^2]^{n_3}}\frac{1}{[(k_1+p_1)^2]^{n_4}}\frac{1}{[(k_2+p_1)^2-m^2]^{n_5}}\nn\\
    &\quad\times\frac{1}{[-2k_1\cdot \hat{p}_2]^{n_6}}\frac{1}{[-2k_2\cdot \hat{p}_2]^{n_7}}\label{eq:topB}\,,
\intertext{
and topology $X_{cc}$ is given by }
    I_{X_{cc};\{n_i\}} &\equiv \int [dk_1][dk_2]\frac{1}{[k_1^{2}]^{n_1}}\frac{1}{[k_2^{2}]^{n_2}}\frac{1}{[(k_1+k_2)^{2}]^{n_3}}\frac{1}{[(k_2+p_1)^2-m^2]^{n_4}}\frac{1}{[(k_1+k_2+p_1)^2-m^2]^{n_5}}\nn\\
    &\quad \times \frac{1}{[-2k_1\cdot \hat{p}_2]^{n_6}}\frac{1}{[-2(k_1+k_2)\cdot \hat{p}_2]^{n_7}}\label{eq:topX}\,.
\end{align}
We first project the integral expression for the jet to the two available structures, viz.\ $\bar{u}(p_1)$ and $\bar{u}(p_1)\slashed{\bar{n}}$. The integrals can be written in terms of the three topologies as defined in \eqnss{eq:topA}{eq:topX}. One then projects onto the relevant form factors, performs an IBP reduction with \texttt{Kira} \cite{Klappert:2020nbg,Maierhofer:2017gsa} and computes the master integrals. These manipulations are done in \texttt{FORM} \cite{Vermaseren:2000nd,Ruijl:2017dtg}, in combination with \texttt{SUMMER} \cite{Vermaseren:1998uu} and some \texttt{Mathematica} packages \cite{Czakon:2005rk,Ochman:2015fho,Huber:2005yg} to calculate the master integrals. For the massive bubble diagram it is not enough to regularize the integrals with dimensional regularization; one needs to put an analytic regulator $\nu$ on the last propagator for topology $B_{cc}$ in \eqn{eq:topB}, as was done in \refr{terHoeve:2023ehm}. The result yields
\begin{align}
    J_{f}^{(2)}(p_1,\bar{n}) &= \left(\frac{\bar\mu^2}{m^2}\right)^{2\eps}\left(\frac{\tilde{\mu}^2}{\s}\right)^{\nu}\bar{u}(p_1)\nn\\
    &\times\bigg\{\frac{1}{2\eps^4}+\frac{2}{3\eps^3}+\frac{1}{\eps^2}\left(\frac{4}{3\nu}+\frac{32}{3}+\frac{\zeta_2}{2}\right)+\frac{1}{\eps}\left(-\frac{20}{9\nu}+\frac{52}{9}-\frac{11\zeta_2}{3}+\frac{17\zeta_3}{3}\right)\nn\\
    &\qquad +\frac{1}{\nu}\left(\frac{112}{27}+\frac{4\zeta_2}{3}\right)+\frac{3040}{81}+\frac{218\zeta_2}{3}-72\log(2)\zeta_2-\frac{161\zeta_2^2}{20}+\frac{89\zeta_3}{9}\nn\\
    &\qquad+ n_f e_f^2 \left(-\frac{1}{3\eps^3}-\frac{17}{9\eps^2}-\frac{1}{\eps}\left(\frac{196}{27} +\frac{5\zeta_2}{3}\right)-\frac{2012}{81}-\frac{85\zeta_2}{9}-\frac{22\zeta_3}{9}\right)+ \nn\\
    &\quad +\frac{m\slashed{\bar{n}}}{p_1^+}\bigg[\frac{1}{2\eps^3}-\frac{7}{3\eps^2}+\frac{1}{\eps}\left(\frac{10}{9}-\frac{5\zeta_2}{2}\right)+\frac{398}{27}-27\zeta_2-\frac{46\zeta_3}{3}+24\zeta_2\log(2)\nn\\
    &\qquad + n_f e_f^2\left(\frac{2}{3\eps^2}+\frac{28}{9\eps}+\frac{308}{27}+\frac{10\zeta_2}{3}\right)\bigg]+\mathcal{O}(\eps)\bigg\}.\label{eq:Jf2result}
\end{align}
This is, to the best of our knowledge, the first time that the massive leading power jet function has been written down up to two loops. Let us comment on our use of the analytic regulator. In the way it has been used now, only the last propagator of topology $B_{cc}$ has been $\nu$-regularized, which is not symmetric in $p_1$ and $p_2$. We therefore do not expect a symmetry between the $c$-collinear and the $\bar{c}$-collinear jet at this point. In that sense, the use of the analytic regulator is quite ad hoc. Indeed, its sole purpose is to mimic the region calculation, to which we compare our results. In the spirit of a factorization approach, it might be better to employ this regulator in a symmetric way in order to keep the symmetry between $p_1$ and $p_2$. One possibility would be to put the analytic regulator on the photon propagators $k_1^2$ and $k_2^2$, which is symmetric in $p_1$ and $p_2$. We already explored this option for the collinear-anticollinear factorization. However, as we will see for the double-collinear factorization, this choice does not always work. A different choice could be to use an analytic regulator that modifies the integration measure, as considered in \refrs{Becher:2011dz,Gritschacher:2013tza}, instead of modifying the amplitude by raising the power of a propagator by $\nu$, as considered here. The measure could be modified in a way that retains the symmetry between $p_1$ and $p_2$, while still regularizing the problematic integrals. A downside to this approach is that loop integrals become more complicated, since there are more propagators to be integrated over as the modified measure is the same for all the regions. We leave further investigations about the use of analytic regulators (at subleading powers) for future work.

This result in \eqn{eq:Jf2result} already suffices to check double-collinear factorization at leading power. One can even perform the check on a diagram-by-diagram basis, since the region results of the six contributing diagrams exactly correspond to the six contributions of this $J_{f}$ jet. This check is indeed satisfied. The leading power result of the sum is already in the jet function expression of \eqn{eq:Jf2result}, namely in the $\mathcal{O}(\la^0)$ part (given by the first four lines there), hence we do not repeat it here.

\subsubsection{\texorpdfstring{$J_{f\gamma}$ and $J_{f\partial\gamma}$ jet functions}{}}
\label{sec:JfgandJfgg}

Let us now move on to the $J_{f\gamma}$ and 
$J_{f\partial\gamma}$ jet functions, defined respectively 
in \eqns{eq:Jfgammadef}{eq:Jfdelgammadef}. 
By means of Wick contractions we have
\begin{align}
    J_{f\gamma}^{(2)\rho}(p_1,\bar{n},\ell^+)&= \begin{gathered}
    \SetScale{0.6}\hspace{0.5cm}
   \text{\footnotesize\Jfgda}\end{gathered}+\begin{gathered}
    \SetScale{0.6}\hspace{0.5cm}\text{\footnotesize\Jfgdc}\end{gathered}+\begin{gathered}
   \SetScale{0.6}\hspace{0.5cm}\text{\footnotesize\Jfgdd}\end{gathered}\nn\\
   &+\begin{gathered}
   \SetScale{0.6}\hspace{0.5cm}\text{\footnotesize\Jfgde}\end{gathered}+\begin{gathered}
    \SetScale{0.6}\hspace{0.5cm}\text{\footnotesize\Jfgdg}\end{gathered}+ \begin{gathered}
        \SetScale{0.6}\hspace{0.5cm}\text{\footnotesize\Jfgdh}
    \end{gathered}\,.\label{eq:Jfg2diag}
\end{align}
The corresponding analytic expression 
can again be found in app.\ \ref{app:2loopres}. 
Starting from \eqn{eq:Jfg2diag}, we can also 
obtain an expression for the $J_{f\partial\ga}$ 
jet, see the discussion around \eqn{eq:Jfdelgammadef}. 

The most relevant observation concerning the 
calculation of the diagrams in \eqn{eq:Jfg2diag}, 
is related to the fact that the momentum $k_1$
is integrated only over the $d-2$ components of $k_{1\perp}$
and over $k_1^-$. The $k_1^+$ component is fixed by the delta function $\delta(\bar{n}\cdot k_1-\ell^+)$, 
cf.\ \eqn{eq:Jfg2int}. This complicates the calculation 
of the jet function. Here we are mostly interested to
check that the matrix element definition introduced in 
\eqn{eq:Jfgammadef} provides the correct contribution 
to reproduce the form factors $F_1$ and $F_2$ up to two 
loops and to $\mathcal{O}(\lambda^2)$. For this purpose we 
do not need the unintegrated jet function, whose 
calculation is left for future work. Instead, we 
observe that the convolution with the momentum 
fraction $x = \ell^+/p_1^+$, i.e.\ 
\begin{equation}
\int dx \, J_{f\gamma}^{(2)\rho}(x) H_{f\gamma,\bar f \rho}^{(0)\mu}(x),
\end{equation}
is essentially a two-loop integral when the 
unintegrated jet and hard function are combined. 
Moreover, let us recall that
\begin{equation}
x = \frac{1}{p_1^+}\bar{n}\cdot k_1 
= \frac{1}{p_1^+p_2^-}\hat{p}_2\cdot k_1 
= \frac{2}{\s}\hat{p}_2\cdot k_1,
\end{equation}
where $\hat{p}_2\cdot k_1$ is typically one of 
the denominators for our topologies that we 
defined before. We can therefore calculate 
the form factors in a standard way, without 
first calculating the specific $x$-dependence 
of the jet function, which is generally a 
harder task. This procedure could be dubbed a ``reversed factorization'' approach. Before moving on, let us also 
notice that one can proceed in a similar 
fashion for the remaining jet functions needed 
to evaluate the two-loop double-collinear 
region, namely the $J_{f\partial\ga}$, 
the $J_{f\ga\ga}$ and the $J_{f\!f\!\bar{f}}$ 
jet functions. In what follows we will therefore
always calculate these jet functions 
in a convolution with the corresponding 
hard function, and check that this reproduces
correctly the form factors. 

We conclude this section by noticing that 
the calculation of the $J_f$ and $J_{f\gamma}$ 
jet functions alone lets us check the factorization
formula up to $\ord(\lambda)$, given that 
\begin{equation}
\left.V^{\mu(2)}\right\vert_{cc} 
=  \left.V^{\mu(2)}_{f,\bar f}\right\vert_{cc} 
+ \left.V^{\mu(2)}_{f\gamma,\bar f}\right\vert_{cc} 
+ \ord(\lambda^2).
\end{equation}
Again, the check can be done on a diagram-by-diagram basis, 
with the note that the ladder and the crossed-ladder diagrams 
(i.e.\ diagram (d) and (h)) should be considered together as 
the diagrams mix due to the eikonal identity. This is explored in more detail in app.~\ref{app:seqjets}. 
The $\mathcal{O}(\lambda)$ result is fully captured by $F_2$, 
and the sum reads
\begin{align}\label{F2cc2Loops}
\left.F_2^{(2)}\right\vert_{\mathrm{NLP}, cc}&= \left(\frac{\bar\mu^2}{m^2}\right)^{2\eps}\frac{m^2}{\s}\Bigg[\frac{4}{3\eps^3}+\frac{13}{3\eps^2}+\frac{1}{\eps}\left(\frac{583}{18}-20\zeta_2\right)+\frac{17191}{108}-101\zeta_2+92\log(2)\zeta_2-\frac{296\zeta_3}{3}\nn\\
    &\hspace{3cm}+n_f e_f^2\left(\frac{4}{3\eps^2}+\frac{98}{9\eps}+\frac{1249}{27}+\frac{20\zeta_2}{3}\right)+\mathcal{O}(\eps)\Bigg],
\end{align}
which is in perfect agreement with the region result. 
We are now ready to go one step further and
check factorization at $\mathcal{O}(\lambda^2)$. For this we need 
two jet functions that we have not discussed before. 
They will be the topic of the next sections.

\subsubsection{\texorpdfstring{$J_{f\gamma\gamma}$}{} jet function}

\begin{figure}[H]
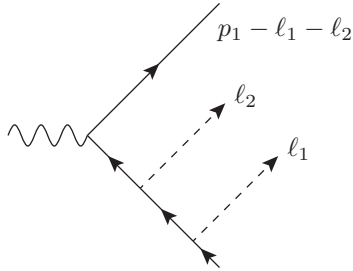

\centering
\vspace{0.5cm}
\Hfggfb
\caption{Diagrammatic representation of the leading 
order hard function for the $J_{f\ga\ga}$ jet. The 
dashed lines represent photons that have to be 
connected to the jet function.}
\label{fig:Hfgg}
\end{figure}

The $J_{f\ga\ga}$ jet is predicted by the factorization 
theorem to contribute starting at two loops, and to
start at $\mathcal{O}(\lambda^2)$. Compared to the 
$J_{f\ga}$ jet, there is an additional photon that connects to the hard part. We define the $J_{f\ga\ga}$ $c$-collinear 
jet as 
\begin{align}
J_{f\gamma\gamma}^{\rho\sigma}(p_1,\bar{n},\ell_1^+,\ell_2^+) 
&= \int\frac{d\xi_1}{2\pi}\int\frac{d\xi_2}{2\pi}e^{-i\ell_1(\xi_1 \bar{n})}
e^{-i\ell_2(\xi_2\bar{n})}\bra{p_1}\bigg[\bar\psi(0)\Phi_{\bar{n}}(0,\infty)\bigg] \nn \\
&\times \bigg[\Phi_{\bar{n}}(\infty,\xi_1 \bar{n})
\left(i D^\rho \Phi_{\bar{n}}(\xi_1 \bar{n},\infty)\right)\bigg]
\bigg[\Phi_{\bar{n}}(\infty,\xi_2 \bar{n})
\left(i D^\sigma \Phi_{\bar{n}}(\xi_2 \bar{n},\infty)\right)
\bigg]\ket{0},\label{eq:defJfgaga}
\end{align}
which is again built out of gauge-invariant building blocks. 
Note that we now have two large momentum components that 
are fixed, namely $\ell_1^+$ and $\ell_2^+$. 
We thus define two corresponding momentum fractions, 
\begin{equation}
    x_1 = \ell_1^+/p_1^+,\qquad x_2 = \ell_2^+/p_1^+.
\end{equation}
At leading order there are two contributions, which one can denote as the ladder and the crossed-ladder diagrams These are the only 
diagrams where two photon propagators can connect to the 
hard vertex. Diagrammatically one has
\begin{equation}\label{eq:Jfggdiag}
J_{f\gamma\gamma}^{(2)\rho\sigma}(p_1,\bar{n},\ell_1^+,\ell_2^+) = 
\begin{gathered}\hspace{0.5cm}\SetScale{0.5}\text{\footnotesize\Jfgga}\end{gathered} + 
\begin{gathered}\hspace{0.5cm}\SetScale{0.5}\text{\footnotesize\Jfggb}\end{gathered},
\end{equation}
where the dot again represents the effective Feynman 
rule for the photon propagator, defined 
in \eqn{eq:Jfgrule}. Note that we have two insertions 
of this effective Feynman rule, rendering this jet 
function to be $\mathcal{O}(\lambda^2)$. The 
insertions of subleading effective Feynman rules 
are discussed further in app.~\ref{app:seqjets}, where we 
investigate the interesting interplay between the 
$J_f$, $J_{f\gamma}$ and $J_{f\ga\ga}$ jet functions.
The full integral expression is again reported in 
app.\ \ref{app:2loopres}. The jet function is 
accompanied by a new hard function
$H_{f\gamma\gamma,\bar f}^{\mu\rho\sigma}$. 
At leading order it is defined as 
\begin{align}
H^{(0)\mu\rho\sigma}_{f\gamma\gamma,\bar f}(\hat{p}_1,\hat{p}_2,x_1,x_2) 
&= \left.ie\gamma^\mu \frac{i(-\slashed{p}_2-\slashed{\ell}_1
-\slashed{\ell}_2+m)}{(p_2+\ell_1+\ell_2)^2-m^2}
\gamma^\sigma\frac{i(-\slashed{p}_2-\slashed{\ell}_1
+m)}{(p_2+\ell_1)^2-m^2}\gamma^\rho
\right\vert_{\mathcal{O}(\lambda^0)} \nn \\
&=\frac{-ie}{(x_1+x_2)\s}\frac{1}{x_1\s}
\gamma^\mu(p_2^- \slashed{\bar{n}}
+(x_1+x_2)p_1^+\slashed{n})\gamma^\sigma
(p_2^-\slashed{\bar{n}}+x_1p_1^+\slashed{n})
\gamma^\rho.\label{eq:Hfgagafbar}
\end{align}
Note that we only had to expand up to 
$\mathcal{O}(\lambda^0)$ since the corresponding 
jet function is already at $\mathcal{O}(\lambda^2)$ 
accuracy. The diagrammatic representation is shown 
in fig.\ \ref{fig:Hfgg}. Before we can calculate 
the full $\mathcal{O}(\lambda^2)$ correction at two loops, we need the last jet function, defined in the next section.

\subsubsection{\texorpdfstring{$J_{f\!f\!\bar{f}}$}{} jet functions}
\label{sec:Jfff}
Just as for the previous jet function, the $J_{f\!f\!\bar{f}}$ 
jet enters at two loops at $\mathcal{O}(\lambda^2)$ accuracy. 
For this jet function two fermions 
and an antifermion attach to the hard function.
This can be done in two distinct ways that conserve the charge 
flow. As a result, we consider two different operator matrix 
elements contributing to this jet function. Both are needed 
to reproduce the form factors. The definitions are
\begin{align}\label{DefJfff1}
J_{f\!f\!\bar f}^{(I)}(p_1,\bar{n},\ell_1^+,\ell_2^+)
&=\int_{-\infty}^\infty \frac{d\xi_1}{2\pi}\int_{-\infty}^\infty 
\frac{d\xi_2}{2\pi}e^{-i \ell_1(\xi_1\bar{n})}
e^{-i \ell_2(\xi_2\bar{n})} \nn \\
&\hspace{-2cm}\times\bra{p_1}
\bigg[\bar\psi(0)\Phi_{\bar{n}}(0,\infty)\bigg]
\bigg[\Phi_{\bar{n}}(\infty,\xi_1 \bar{n})\psi(\xi_1\bar{n})\bigg]
\bigg[\bar\psi(\xi_2\bar{n})\Phi_{\bar{n}}(\xi_2\bar{n},\infty)\bigg]
\ket{0},
\intertext{and} \label{DefJfff2}
J_{f\!f\!\bar f}^{(I\!I)}(p_1,\bar{n},\ell_1^+,\ell_2^+)
&=\int_{-\infty}^\infty \frac{d\xi_1}{2\pi}\int_{-\infty}^\infty 
\frac{d\xi_2}{2\pi}e^{-i\ell_1(\xi_1\bar{n})}e^{-i\ell_2(\xi_2\bar{n})} \nn \\
&\hspace{-2cm}\times\bra{p_1}\bigg[\Phi_{\bar{n}}(\infty,0)\psi(0)\bigg]
\bigg[\bar\psi(\xi_1\bar{n})
\Phi_{\bar{n}}(\xi_1 \bar{n},\infty)\bigg]
\bigg[\bar\psi(\xi_2\bar{n})
\Phi_{\bar{n}}(\xi_2\bar{n},\infty)\bigg]
\ket{0},
\end{align}
which rely on the fermionic gauge-invariant 
building block $\bar\psi(x)\,\Phi_{\bar{n}}(x,\infty)$, 
already introduced in previous jet functions, 
and the corresponding complex conjugate
$\Phi_{\bar{n}}(\infty,x)\psi(x)$. Note that 
\eqns{DefJfff1}{DefJfff2} involve uncontracted 
spin indices that have to contracted in the right 
way with the hard function. We will therefore 
explicitly denote the spin indices in the 
coming equations. 

The leading order contributions for these two 
jet functions are diagrammatically given by 
\begin{align}
    J_{f\!f\!\bar f}^{(I)(2)}(p_1,\bar{n},\ell_1^+,\ell_2^+) &= \begin{gathered} \hspace{0.5cm}\SetScale{0.5}\text{\footnotesize\JfffdFurrya}\end{gathered}+\begin{gathered}\hspace{0.5cm}\SetScale{0.5}\text{\footnotesize\Jfffdf}\end{gathered}\label{eq:JfffI}\\
    J_{f\!f\!\bar f}^{(I\!I)(2)}(p_1,\bar{n},\ell_1^+,\ell_2^+) &= \begin{gathered} \hspace{0.5cm}\SetScale{0.5}\text{\footnotesize\JfffdFurryb}\end{gathered}+\begin{gathered}\hspace{0.5cm}\SetScale{0.5}\text{\footnotesize\Jfffdh}\end{gathered}\label{eq:JfffII}.
\end{align}
The full integral expressions can be found in app.\ \ref{app:2loopres}.
The first contribution to both jet functions gives rise to two diagrams with a triangle fermion loop. These diagrams cancel each other by Furry's theorem \cite{PhysRev.51.125} and can therefore be neglected. We note that the second part of the $J_{f\!f\!\bar{f}}^{(I)}$ jet function mimics diagram (f), see fig.\ \ref{fig:QED_diagrams}. According to the region result for of this diagram, the double-collinear region only contributes starting at $\mathcal{O}(\lambda^2)$. This matches well, since this jet function is also an $\mathcal{O}(\lambda^2)$ quantity. The second part of the $J_{f\!f\!\bar{f}}^{(I\!I)}$ jet function mimics the $cc'$-region of diagram (h). Compared to the $cc$-region, the $cc'$-region has a different momentum routing. The momentum routing here is fixed by the definition of the jet function. One can see this through the delta functions in \eqn{app:JfffII}. This implies that the outgoing fermions (with spinor index $b$ and $c$) carry momentum $k_1$ and $k_2$. That differs, for example, from the $J_{f\ga}$ jet, which also contributes to the double-collinear region, where the momentum $k_1$ flows through the photon by definition of the jet. As we will see later, indeed this $J_{f\!f\!\bar{f}}^{(I\!I)}$ jet contribution fully determines the $cc'$-region, while the other jet functions contribute to the $cc$-region only.

\begin{figure}[H]
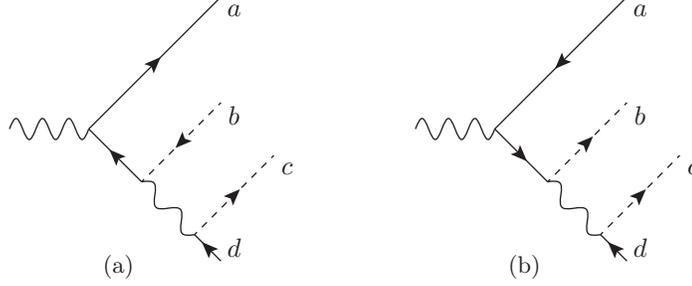

\centering
\subfloat[]{\HffffbI\label{HfffI}}
\hspace{2cm}
\subfloat[]{\HffffbII
\label{HfffII}}
\caption{Diagrammatic representation of the hard functions for the $J_{f\!f\!\bar{f}}$ jet. The dashed lines correspond to fermion lines that are connected to the hard function, through which flow collinear momenta $\ell_1$ and $\ell_2$.}
\label{fig:Hfff_diagrams}
\end{figure}

Both jet functions have their own hard functions, namely $H_{f\!f\!\bar f,\bar f}^{(I)}$ and $H_{f\!f\!\bar f,\bar f}^{(I\!I)}$, and they are depicted in fig.\ \ref{fig:Hfff_diagrams}. The expressions for the hard functions yield
\begin{align}
    H_{f\!f\!\bar{f},\bar{f}}^{(I)(0)\mu}(\hat{p}_1,\hat{p}_2,x_1,x_2) &= (-ie)^3\left(\gamma^\mu\frac{-i(\slashed{p}_2+\slashed{\ell}_1+\slashed{\ell}_2-m)}{(p_2+\ell_1+\ell_2)^2-m^2}\gamma_\nu\right)_{ab} \frac{-i}{(p_2+\ell_1)^2}(\gamma^\nu)_{cd}\nn\\
    &=\frac{-ie^3}{(x_1+x_2)\s}\left(\gamma^\mu\bigg[\hat{\slashed{p}}_2+(x_1+x_2)\hat{\slashed{p}}_1\bigg]\gamma_\nu\right)_{ab} \frac{1}{x_1\s}\left(\gamma^\nu\right)_{cd},
\intertext{and}
    H_{f\!f\!\bar{f},\bar{f}}^{(I\!I)(0)\mu}(\hat{p}_1,\hat{p}_2,x_1,x_2) &= (-ie)^3\left(\gamma_\nu\frac{i(\slashed{p}_2+\slashed{\ell}_1+\slashed{\ell}_2+m)}{(p_2+\ell_1+\ell_2)^2-m^2}\gamma^\mu\right)_{ba} \frac{-i}{(p_2+\ell_1)^2}(\gamma^\nu)_{cd}\nn\\
    &=\frac{ie^3}{(x_1+x_2)\s}\left(\gamma_\nu\bigg[\hat{\slashed{p}}_2+(x_1+x_2)\hat{\slashed{p}}_1\bigg]\gamma^\mu\right)_{ba} \frac{1}{x_1\s}\left(\gamma^\nu\right)_{cd},
\end{align}
where $x_j = \ell_j^+/p_1^+$, $j=1,2$. The spinor index $d$ is still to be contracted with the anticollinear $J_{\bar{f}}$ jet, which simply is $v(p_2)_d$ in the double-collinear limit. 
This concludes the discussion of all the jet functions with their corresponding hard functions that are needed to calculate the two-loop form factor up to $\mathcal{O}(\lambda^2)$ in the double-collinear region. Next, we will use these definitions to calculate the form factor at $\mathcal{O}(\lambda^2)$. 

\subsubsection{Form factor calculation}

We have calculated all the terms needed to check
the double-collinear region up to $\mathcal{O}(\lambda^2)$,
as given in \eqn{NLPfactorizationQQbarcc}. As a summary, 
in tab.\ \ref{tab:NLPfact_cc} we show which jet functions 
contribute to which diagrams. Again, one could check the 
factorization theorem on a diagram-by-diagram basis, with 
the understanding that the ladder and crossed-ladder 
diagram are taken together. Projecting the 
amplitude in \eqn{NLPfactorizationQQbarcc} onto
the form factors we are able to compare with the 
region calculation in \refr{terHoeve:2023ehm}. 
As discussed in sec.\ \ref{sec:f-jetfunction}, the leading power 
contribution to $F_1$ is entirely given in terms
of $V_{f,\bar f}^{\mu(2)}|_{cc}$, which in 
turn is related to the calculation of the $J_f$
jet at two loops, which we have already validated
in that section. Similarly, the $\ord(\lambda)$ 
part of the amplitude contributes to the form 
factor $F_2$; we have already validated this 
above, see discussion around \eqn{F2cc2Loops}.
The only contribution still to be validated 
is given by the $\ord(\lambda^2)$ amplitude,
which contributes at NLP to $F_1$. 
Upon projecting \eqn{NLPfactorizationQQbarcc}
onto $F_1$ we get
\begin{align}\label{F1NLPcc}
\left.F_1^{(2)}\right\vert_{\mathrm{NLP}, cc}
&=\left(\frac{\bar\mu^2}{m^2}\right)^{2\eps}\frac{m^2}{\s}
\Bigg\{-\frac{5}{2\eps^2}+\frac{1}{\eps}\left(-\frac{63}{4}+36\zeta_2\right)
-\frac{629}{8}+\frac{191\zeta_2}{2}+72\zeta_3\nn\\
&\hspace{4cm}+n_f e_f^2\left(-\frac{2}{3\eps^2}-\frac{79}{9\eps}
-\frac{2575}{54}-\frac{10\zeta_2}{3}\right) \nn\\
&\hspace{2.5cm}+\left(\frac{\mu_1^2}{\s}\right)^{\nu_1}
\left[-\frac{2}{3\eps^2}-\frac{1}{\eps}
\left(\frac{16}{\nu_1}+\frac{151}{9}\right)
-\frac{3511}{54}+42\zeta_2\right] \nn \\
&\hspace{2.5cm}+\left(\frac{\mu_2^2}{\s}\right)^{2\nu_2}
\left[\frac{2}{\eps^3}+\frac{1}{\eps^2}\left(-4+\frac{2}{\nu_2}\right)
+\frac{6}{\eps}+\frac{6+2\zeta_2}{\nu_2}
+22-20\zeta_2-\frac{10\zeta_3}{3}\right] \nn \\
&\hspace{2.5cm}+\left(\frac{\mu_3^2}{-m^2}\right)^{2\nu_2}
\bigg[-\frac{5}{\epsilon^3}-\frac{1}{\epsilon^2}
\left(\frac{2}{\nu_2}+\frac{35}{2}\right)
-\frac{1}{\epsilon}\left(\frac{279}{4}-23\zeta_2\right)
-\frac{2\zeta_2+2}{\nu_2}\nn \\ 
&\hspace{4.0cm}-\frac{2281}{8}+\frac{133\zeta_2}{2}
-144\zeta_2\log(2)+\frac{16\zeta_3}{3} \bigg] \nn \\
&\hspace{2.5cm}+\left(\frac{\mu_5}{\s}\right)^{\nu_4}
\bigg[\frac{1}{\eps^3}+\frac{1}{\eps^2}
\left(\frac{19}{2}-4\zeta_2\right)+\frac{1}{\eps}
\left(\frac{213}{4}+\frac{4}{\nu_4}+9\zeta_2-36\zeta_3\right) \nn \\
&\hspace{4.0cm}+\frac{1647}{8}+\frac{8}{\nu_4}-\frac{81\zeta_2}{2}
+\frac{298\zeta_3}{3}-\frac{356\zeta_2^2}{5}\bigg] \nn \\
&\hspace{2.5cm}+\left(\frac{\mu_4^2}{-m^2}\right)^{\nu_3}
\left(\frac{\mu_4^2}{\hat{s}}\right)^{\nu_3}
\left(\frac{\mu_5^2}{-m^2}\right)^{2\nu_4}
\bigg[\frac{20}{\epsilon^4}-\frac{1}{\epsilon^3}
\left(\frac{10}{\nu_4}+10\right) \nn \\
&\hspace{4.0cm} +\frac{1}{\epsilon^2}
\left(\frac{4}{\nu_4^2}+\frac{6}{\nu_4}-8\zeta_2-8\right)
+\frac{1}{\epsilon}\bigg(-\frac{4}{\nu_4^2} 
+\frac{2\zeta_2+6}{\nu_4} \nn \\
&\hspace{4.0cm}-\frac{4}{\nu_3}
+6\zeta_2 -\frac{112\zeta_3}{3}+16\bigg) 
+\frac{4\zeta_2-4}{\nu_4^2} +\frac{1}{\nu_4}
\bigg(\frac{80\zeta_3}{3} \nn \\
&\hspace{4.0cm}-6\zeta_2-4\bigg)-\frac{8}{\nu_3}
+46+4\zeta_2 +\frac{32\zeta_3}{3}-46\zeta_2^2\bigg]
+\mathcal{O}(\eps)\Bigg\},
\end{align}
where the last six lines correspond to the sum of diagram (d) and (h), i.e.\ the ladder and crossed-ladder diagram. The last four lines correspond precisely to the $cc'$-region of diagram (h), which is fully given by the $J_{f\!f\!\bar{f}}^{(I\!I)}$ jet. We employed the analytic regulator in the same way as in the region calculation, and our result agrees again. A similar result can be found for the combination of the $\bar{c}\bar{c}$-region and the $\bar{c}\bar{c}'$-region. 

As a final topic of this section, we address another subtlety with the analytic regulator, which appears only in diagram (e), at $\mathcal{O}(\lambda^2)$. This corresponds to the contribution in the third line of eq.~\eqref{F1NLPcc}. In the region calculation
\cite{terHoeve:2023ehm}, the analytic regulator $\nu$ is put on the propagator $\left[(k_1+k_2+p_2)^2-m^2\right]^{-1}$. This is then expanded in powers of $\lambda$, giving
\begin{equation}\label{exampleexpansion}
    \left(\frac{1}{(k_1+k_2+p_2)^2-m^2}\right)^{1+\nu} = \left(\frac{1}{2\hat{p}_2\cdot(k_1+k_2)}\right)^{1+\nu}\left[1-(1+\nu)\frac{(k_1+k_2)^2}{2\hat{p}_2\cdot(k_1+k_2)}+\mathcal{O}(\lambda^3)\right].
\end{equation}
Note the appearance of an $\mathcal{O}(\la^2\nu)$ term in the square brackets.
In the calculation of the jet functions the analytic regulator is put on the eikonal propagator. However there we do not have a term that is $\mathcal{O}(\la^2\nu)$. This mismatch results from regularizing the propagator before (in the region calculation) or after (in the factorization approach) the $\la$ expansion. The $\mathcal{O}(\la^2\nu)$ term could give a finite contribution when multiplied by $1/\nu$ arising from the first factor on the r.h.s.\ of eq.~\eqref{exampleexpansion}.\footnote{Note that $F_1$ at leading power indeed has a rapidity pole for diagram (e).} In general, one might try to avoid this regulator ambiguity by putting the regulator on a propagator that is homogeneous in $\la$, and thus needs no expansion. This proved to work for the collinear-anticollinear region where we put the regulators on the photon propagators. However, this does not work for diagram (e). In this case, the eikonal propagators need additional regularization in order to have well-defined integrals, which unavoidably leads to an $\mathcal{O}(\la^2\nu)$ term on the region side. 

A way out of this predicament is employing an analytic regulator on the integration measure \cite{Becher:2011dz}. This leads to the same expression on the factorization side, because it adds an eikonal propagator factor in the measure. In the region calculation it leads to an expression that does not lead to an $\mathcal{O}(\la^2\nu)$ term, thus avoiding the regulator ambiguity. This method works for diagram (e).

This concludes our discussion of the factorization theorem with all the relevant jet functions up to $\mathcal{O}(\la^2)$ in the power counting parameter. We checked all the contributions from these jet functions up to two loops in the coupling constant $\alpha$ with the region results and found agreement everywhere.
\begin{table}[H]
\centering
\vspace{-1cm}
\def\hMin{\hspace{-0.2cm}}
\SetScale{0.48}
\begin{tabular}{|c|c||c|c|c|c|}
\hline
\multicolumn{2}{|c||}{Diagram} & $J_f$ & $J_{f\ga}$ & $J_{f\ga\ga}$ & $J_{f\!f\!\bar{f}}$  \\
\hline
\raisebox{1.2cm}{(a)} & \hMin{\tiny\QEDda} & \hMin{\tiny\Jfda} & \hMin{\tiny\Jfgda} & & \\
\hline
\raisebox{1.2cm}{(c)} & \hMin{\tiny\QEDdc} & \hMin{\tiny\Jfdc} & \hMin{\tiny\Jfgdc} & & \\
\hline
\raisebox{1.2cm}{(e)} & \hMin{\tiny\QEDde} & \hMin{\tiny\Jfde} & \hMin{\tiny\Jfgde} & & \\
\hline
\raisebox{1.2cm}{(f)} & \hMin{\tiny\QEDdf} & & & & \hMin{\tiny\Jfffdf} \\
\hline
\raisebox{1.2cm}{(g)} & \hMin{\tiny\QEDdg} & \hMin{\tiny\Jfdg} & \hMin{\tiny\Jfgdg} & & \\
\hline
\raisebox{-0.2cm}{\hspace{-0.2cm}(d)$\!+\!$(h)\hspace{-0.1cm}} & \hMin{\tiny\QEDdd} & \hMin{\tiny\Jfdd} & \hMin{\tiny\Jfgdd} & \hMin{\tiny\Jfgga} & \\
		  & \hMin{\tiny\QEDdh} & \hMin{\tiny\Jfdh} & \hMin{\tiny\Jfgdh} & \hMin{\tiny\Jfggb} & \\
\hline
\raisebox{1.2cm}{(h)'} & \hMin{\tiny\QEDdh} & & & & \hMin{\tiny\Jfffdh}\\
\hline
\end{tabular}
\SetScale{1}
\caption{Summary of the jet functions contributing to the $cc$-region, and the $cc'$-region for diagram (h), dubbed (h)'. Note that diagram (b), see fig.~\ref{fig:QED_diagrams}, is absent, since it does not have a $cc$-region.}
\label{tab:NLPfact_cc}
\end{table}

%% file: Sections/Conclusion.tex
In this paper we investigated the factorization of the amplitude for an off-shell photon decaying to a massive fermion-antifermion pair at next-to-leading power in the fermion mass expansion. To this end we proposed new matrix-element definitions for the jet functions involved, which were anticipated in ref.~\cite{Laenen:2020nrt}. We confirmed the correctness of these definitions by a two-loop calculation of the factorization ingredients, and then comparing the results to the region calculation in ref.~\cite{terHoeve:2023ehm}.
Although we considered a specific amplitude, the jet functions are universal and can thus be part of the factorization of different amplitudes. 
The hard functions were determined by matching to the results in \refr{terHoeve:2023ehm}. 

In sec.\ \ref{sec:oneloop} we calculated the $J_f$ jet function up to NLP at one loop and proposed a definition for the $J_{f\gamma}$ and the $J_{f\partial\ga}$ jet. We verified their one-loop contributions for the collinear region. In sec.\ \ref{sec:twoloop} we performed a two-loop comparison. To this end we proposed definitions for the $J_{f\ga\ga}$ and the $J_{f\!f\!\bar{f}}$ jet functions and calculated their two-loop contributions, together with the two-loop contributions for the $J_f$, $J_{f\ga}$ and $J_{f\partial\ga}$ jet functions, all up to NLP. Indeed the proposed jet functions reproduced the corresponding region results. 

The calculations raised several subtleties. First, the employment of rapidity regulators proved to be non-trivial and it was not always possible to use the exact same regulators for the jet functions as were used for the region calculation. However, for these cases we found alternative regulator choices on both sides that led to agreement.
Second, we showed that factorization for the bare amplitude can only be achieved once self-energy corrections have been taken into account. These modify the equations of motion and, consequently, the mass-shell condition.

An interesting observation we could make in the context of the double-collinear factorization is that jet functions fix a specific momentum routing. For instance, in the region analysis a different momentum routing was needed to reveal a new contribution to the $cc$-region, dubbed $cc'$. This $cc'$-region now turns out to be fully determined by the $J_{f\!f\!\bar{f}}^{(I\!I)}$ jet function, while the other jet functions exclusively contribute to the $cc$-region. This observation suggests an alternative perspective on identifying contributing momentum regions: the factorization formula itself can serve as a tool to achieve this.
Furthermore, we found that the $J_f$, $J_{f\ga}$ and $J_{f\ga\ga}$ jet functions have an interesting interplay due to their description in terms of effective LP eikonal Feynman rules and effective photon propagator insertions at subleading powers.

The fact that factorization up to NLP in the small-mass limit holds exactly opens the door for multiple new directions. One could investigate the all-order structure of the amplitude through evolution equations. Another next step would also be to include radiation in the analysis in order to generalize our analysis to observables.

\acknowledgments

We thank Melissa van Beekveld, Max Jaarsma and Lorenzo Magnea for helpful discussions. G.W.\ and R.v.B.\ thank the University of Turin, and  L.V.\ and G.W.\ thank Nikhef for hospitality and partial support during part of the project.
E.L.\ would like to thank the Infosys Foundation for support. 
This research was also supported in part by grant NSF PHY-2309135 to the Kavli Institute for Theoretical Physics (KITP).
The research of G.W. was supported in part by ERC grant  nr.\ 101041109 (`BOSON').

%% file: Appendices/seqjets.tex
The jet functions considered in this paper are not homogeneous in the power counting and have subleading terms. 
In this appendix we show how the $J_{f}$, $J_{f\ga}$, and $J_{f\ga\ga}$ jet functions form a sequence: 
the $J_{f\ga}$ jet function cancels the $\mathcal{O}(\la)$ terms of the $J_f$ jet function in such a way that there is no overlap (or double counting) between these jet functions. In a similar fashion, we will discuss how the $J_{f\ga\ga}$ jet function cancels the $\mathcal{O}(\la^2)$ terms of the $J_f$ and $J_{f\ga}$ jet functions such that there is no overlap between these jet functions.
To show this sequence, we will start from the integral expressions of the one- and two-loop diagrams considered in this work and work our way towards the various jet functions. In particular, we discuss the structure given in tab.~\ref{tab:NLPfact_cc}. 

\subsection{Review of LP factorization}
\label{app:LPfact}
We start with a review of the LP $J_{f}$ jet function, see eq.~\eqref{eq: defJf}, following~\refr{Collins:1989bt}. 
We consider the one-loop vertex correction, given by 
\begin{align}
\bar{u}(p_1) 
(-ie\gamma^\rho)\int [dk]
\frac{i (\slashed{k}+\slashed{p}_1+m)}{k^2+2k\cdot p_1}
(-ie\gamma^\mu)
\frac{i (\slashed{k}-\slashed{p}_2+m)}{k^2-2k\cdot p_2}
(-ie\gamma^\sigma)
\frac{-i\eta_{\rho\sigma}}{k^2}
v(p_2)\,,
\label{eq:OneLoopVertexCorrection}
\end{align}
where we assume that the loop momentum $k$ is collinear to $p_1$, i.e.\ $k\sim\sqrt{s}(\la^0,\la^2,\la^1)$. 
One could now perform an expansion in $\la$ for the integrand using the method of regions, as was done in~\refr{terHoeve:2023ehm}. Doing so at leading power, it is straightforward to find the one-loop integral expression of the jet function $J_f$. 
However, when going to NLP and in particular when we consider the two-loop diagrams, working towards such integral expressions for the various jet functions turns out to be a tedious task.

We therefore take a different approach in this appendix, which was inspired by the LP derivation in \refr{Collins:1989bt}, which we review now. 
At leading power, the numerator of eq.~\eqref{eq:OneLoopVertexCorrection} can be written as 
\begin{align}
\mathcal{N}^\mu 
&= \bar{u}(p_1)\gamma^\rho(\slashed{p}_1+\slashed{k}+m)\gamma^\mu(\slashed{k}-\slashed{p}_2+m)\gamma^\sigma \eta_{\rho\sigma} v(p_2)\nn\\
&=\bar{u}(p_1)\left[-k^+\gamma^-\gamma^\rho+2(p_1^++k^+)\delta_{\rho+}\right]\gamma^\mu(k^+\gamma^--p_2^-\gamma^+)\gamma^\sigma \eta_{\rho\sigma} v(p_2)
+\cO(\la)\,,
\end{align}
where we omitted an overall factor $-e^3$.
Using that $\bar{u}(p_1)\gamma^- \sim \mathcal{O}(\lambda)$ and $\gamma^+v(p_2) \sim\mathcal{O}(\lambda)$, cf.\ eq.~\eqref{eomexpanded}, we note that the numerator is dominated by $\rho = +$, and therefore $\sigma = -$ by virtue of the metric tensor. We define
\begin{equation}
    A^\rho = \gamma^\rho(\slashed{p}_1+\slashed{k}+m),\qquad B^\rho = (-\slashed{p}_2+\slashed{k}+m)\gamma_\sigma \eta^{\sigma\rho}.
\end{equation}
We can then perform the approximations
\begin{align}
\bar{u}(p_1)A_\rho \gamma^\mu B^\rho v(p_2)
&\simeq \bar{u}(p_1)A^+\gamma^\mu B_-\, v(p_2)\nn\\
&= \bar{u}(p_1)\frac{1}{k^+}A^+\gamma^\mu k^+B_- \,v(p_2)\nn\\ 
&\simeq \bar{u}(p_1)\frac{1}{k^+}A^+ \gamma^\mu k\cdot B \,v(p_2)\nn\\
&= \bar{u}(p_1)\frac{\bar{n}\cdot  A}{\bar{n}\cdot k}\gamma^\mu k\cdot B\, v(p_2).\label{eq:GrammerYennie}
\end{align}
We can now manipulate the $k\cdot B$ term as
\begin{align}
    k\cdot B\, v(p_2) &= (-\slashed{p}_2+\slashed{k}+m)\slashed{k}\,v(p_2) =(-\slashed{p}_2+\slashed{k}+m)[(-\slashed{p}_2+\slashed{k}-m)+(\slashed{p}_2+m)]v(p_2) \nn\\
    &= \big[(p_2-k)^2-m^2\big]v(p_2)\,.\label{eq:WardIdentityTrick}
\end{align}
where we used the equations of motion. This is known as the method of Grammer and Yennie \cite{Grammer:1973db}. Effectively, we have removed one of the denominators in favour of an eikonal factor. 
The LP contribution of the vertex diagram, eq.~\eqref{eq:OneLoopVertexCorrection}, is now given by
\begin{align}
-e^3\bar{u}(p_1)\int[dk]\frac{\slashed{\bar{n}}(\slashed{p}_1+\slashed{k}+m) }{[k^2+2k\cdot p_1][k^2][\bar{n}\cdot k]}\gamma^\mu v(p_2)\,,
\end{align}
which can be expressed as the factorized expression
\begin{equation}
\frac{e^2}{16\pi^2}J_f^{(1)} H_{f,\bar{f}}^{(0)\mu} J_{\bar{f}}^{(0)}\,,
\end{equation}
with $J_{f}^{(1)}$ the one-loop expression for the $J_{f}$ jet function already given in eq.~\eqref{eq:jet_jf}, $H_{f,\bar{f}}^{(0)\mu} = -ie\gamma^\mu$ the hard function and $J_{\bar{f}}^{(0)} = v(p_2)$ the tree-level anticollinear jet function.

The above approximation generalizes to an arbitrary number of loops; for every collinear photon connecting to the hard subgraph, we can use a similar approximation as in eq.~\eqref{eq:GrammerYennie}. To be precise, we pick $A^\rho$ in eq.~\eqref{eq:GrammerYennie} to be the collinear subgraph that emits the collinear photon, whereas $B^\rho$ is the part of the hard subgraph where the photon attaches. When we sum over all ways of attaching the collinear photons to the hard part there are many cancellations. The result is that the photons decouple from the hard subgraph and end on an eikonal line instead. As a two-loop example, let us consider diagrams (d) and (h) with both photons collinear (the case that one photon is collinear and the other anti-collinear is discussed in app.~\ref{app:regtofact}). Applying the approximation eq.~\eqref{eq:GrammerYennie} to both photons yields
\begin{align}
\int[dk_1][dk_2] A_{\al\beta}(-ie\ga^\mu)
\frac{\bar{n}^\al k_1^\rho}{\bar{n}\cdot k_1} 
\frac{\bar{n}^\beta k_2^\si}{\bar{n}\cdot k_2} 
B_{\rho\si}v(p_2)\,,\label{eq:LPdiagramsdandh}
\end{align}
where we defined 
\begin{equation}
A^{\al\beta} 
= \bar{u}(p_1)
(-ie\ga^\al)
\frac{i}{\slashed{k}_1+\slashed{p}_1-m}
(-ie\ga^\beta)
\frac{i}{\slashed{k}_1+\slashed{k}_2+\slashed{p}_1-m}
\frac{-i}{k_1^2}
\frac{-i}{k_2^2}\,,
\end{equation}
and 
\begin{align}
B^{\rho\si}
=&
\frac{i}{\slashed{k}_1+\slashed{k}_2-\slashed{p_2}-m}
(-ie\ga_\rho)
\frac{i}{\slashed{k}_2-\slashed{p}_2-m}
(-ie\ga_\si)\nn\\
&+\frac{i}{\slashed{k}_1+\slashed{k}_2-\slashed{p}_2-m}
(-ie\ga_\si)
\frac{i}{\slashed{k}_1-\slashed{p}_2-m}
(-ie\ga_\rho)\,.    
\end{align}
Applying a Ward identity of the type eq.~\eqref{eq:WardIdentityTrick} for both $k_1$ and $k_2$, we can write 
\begin{equation}
\frac{\bar{n}^\al k_1^\rho}{\bar{n}\cdot k_1} 
\frac{\bar{n}^\beta k_2^\si}{\bar{n}\cdot k_2} 
B_{\rho\si}
=
\frac{-ie\bar{n}^\al}{\bar{n}\cdot k_1} 
\frac{-ie\bar{n}^\beta}{\bar{n}\cdot k_2}, 
\end{equation}
where we used the equations of motion on the spinor $v(p_2)$. Eq.~\eqref{eq:LPdiagramsdandh} therefore becomes
\begin{align}
\int[dk_1][dk_2]
&A_{\al\beta}\frac{-ie\bar{n}^\al}{\bar{n}\cdot k_1} 
\frac{-ie\bar{n}^\beta}{\bar{n}\cdot k_2} 
(-ie\ga^\mu) v(p_2)\nn\\
&= \int[dk_1][dk_2] A_{\al\beta}\bigg[\frac{-ie\bar{n}^\al}{\bar{n}\cdot (k_1+k_2)} 
\frac{-ie\bar{n}^\beta}{\bar{n}\cdot k_1} 
+  \frac{-ie\bar{n}^\al}{\bar{n}\cdot (k_1+k_2)} 
\frac{-ie\bar{n}^\beta}{\bar{n}\cdot k_2} \bigg]
(-ie\ga^\mu)v(p_2)\,,
\end{align}
where in the second line we wrote the two eikonal propagators with $k_1$ and $k_2$ attaching to a single eikonal line. This is also known as the eikonal identity. We recognize the factorized expression 
\begin{equation}\label{eq:LPfactOfDiagramsdandh2}
\left[
\begin{gathered}\SetScale{0.5}\hspace{0.5cm}\text{\footnotesize\Jfdd}\end{gathered}\hspace{0.5cm}
+
\begin{gathered}\SetScale{0.5}\hspace{0.5cm}\text{\footnotesize\Jfdh}\end{gathered}\hspace{0.5cm}\right] H_{f,\bar{f}}^{(0)\mu}\ J_{\bar{f}}^{(0)}
\end{equation}
with $H_{f,\bar{f}}^{(0)\mu} = -ie\ga^\mu$ and $J_{\bar{f}}^{(0)} = v(p_2)$, as above.

\subsection{NLP power factorization}
\label{app:NLP2L}
We just saw that the key approximation to obtain the LP contribution of the vertex diagrams was to make the following replacement for every propagator of a collinear photon\footnote{As shown in \refr{Collins:1989bt}, this replacement is not specific to Feynman gauge, but works for covariant gauges in general.}
\begin{equation}\label{eq:GrammerYennieReplacement}
\frac{-i\eta^{\rho\si}}{k^2}
\to \frac{-i}{k^2}\ \frac{\bar{n}^\rho k^\si}{\bar{n}\cdot k}\,.
\end{equation}
Let us now recall the effective Feynman rule for the photon propagator, eq.~\eqref{eq:Jfgrule}, for the jet functions $J_{f\ga}$ and $J_{f\ga\ga}$, namely
\begin{equation}\label{eq:NLPeffectivePhotonProp}
\frac{-i}{k^2}\left(\eta^{\rho\sigma}-\frac{\bar{n}^\sigma k^\rho}{\bar{n}\cdot k}\right)\delta(\bar{n}\cdot k-\ell^+)\,.
\end{equation}
Two remarks are in order. 
First, we note that in the reversed factorization approach, which was introduced in sec.~\ref{sec:JfgandJfgg}, we may omit the delta function in eq.~\eqref{eq:NLPeffectivePhotonProp}.
Second, we recognize that in eq.~\eqref{eq:NLPeffectivePhotonProp}, the normal (Feynman gauge) photon propagator is subtracted by the approximation eq.~\eqref{eq:GrammerYennieReplacement}. In other words, the LP approximation is subtracted from the full photon propagator, which of course contains the LP contribution. As a result, what is left over is an NLP term. 
Using these insights, we can easily manipulate diagrams such that we obtain the NLP factorization of the $c$-region at one loop and the $cc$-region at two loops (and therefore also $\bar{c}$ and $\bar{c}\bar{c}$-regions, the $c\bar{c}$-region is discussed separately in app.~\ref{app:regtofact}). In particular, as we will see momentarily, this explains the structure observed in tab.~\ref{tab:NLPfact_cc}.

Up to two loops, we can distinguish between one photon or two photons connecting to the hard part. 
In case only one photon connects to the hard part, as in the one-loop diagram and in diagrams (a), (c), (e) and (g), we can write for the diagrams\footnote{Here, $\cA$ should not be confused with the gauge-invariant building block of the photon field, but it should be read as the generalisation of the tensor $A$ that was used in the previous section.}
\begin{equation}\label{A16}
\cA^{(i)}_{\rho} \frac{-i \eta^{\rho\si}}{k^2} 
\left[(-ie\ga^\mu)\frac{i}{\slashed{k}-\slashed{p}_2-m}(-ie\ga_\si)\right]
v(p_2),
\end{equation}
where $\cA^{(i)}_\rho$ depends on the specific diagram $i$ under consideration. We leave out the integration over $k$ for the moment. Furthermore, we made the photon propagator together with the hard subgraph explicit, i.e.\ one recognizes that the term in square brackets is the expression for the hard function $H_{f\ga,\bar{f}}^{(0)}$, before expanding. 
Next, we can rewrite eq.~\eqref{A16} as
\begin{align}
\cA^{(i)}_{\rho} \frac{-i}{k^2}
&\left\{\frac{\bar{n}^\rho k^\si}{\bar{n}\cdot k}
+\left(\eta^{\rho\si}-\frac{\bar{n}^\rho k^\si}{\bar{n}\cdot k} \right) \right\}
\left[(-ie\ga^\mu)\frac{i}{\slashed{k}-\slashed{p}_2-m}(-ie\ga_\si)\right]
v(p_2)\label{A17}
\end{align}
where we added and subtracted a term to recognize both the photon propagators of eqs.~\eqref{eq:GrammerYennieReplacement} and~\eqref{eq:NLPeffectivePhotonProp}. 
Using the derivation of the previous subsection, we see that the first term in the curly brackets, after properly including the integration over the loop momenta, leads to the factorized expression
\begin{equation}
\left(\frac{e^2}{16\pi^2}\right)^jJ_f^{(j)(i)}\ H_{f,\bar{f}}^{(0)\mu}\ J_{\bar{f}}^{(0)},
\end{equation}
where $j$ represents the loop order. This factorized expression holds for each diagram $i$.\footnote{We already observed in sec.~\ref{sec:twoloop} that factorization, in particular in the double-collinear region, can be checked on a diagram-by-diagram basis, except for diagram (d) and (h), which will be discussed shortly.}
For the second term, we recognize the jet function $J_{f\ga}$, whose definition was given in eq.~\eqref{eq:Jfgammadef}, and this therefore leads to the factorized expression 
\begin{equation}
\left(\frac{e^2}{16\pi^2}\right)^jJ_{f\ga}^{(j)(i)}\otimes H_{f\ga,\bar{f}}^{(0)\mu} J_{\bar{f}}^{(0)}\,.
\end{equation}
This factorized expression is indeed what we observe in tab.~\ref{tab:NLPfact_cc} for diagrams (a), (c), (e) and (g).

When two photons connect to the hard part, as for diagrams (d) and (h), we can write
\begin{equation}
\cA_{\rho\al}^{(d+h)}\frac{-i \eta^{\rho\si}}{k_1^2}\frac{-i \eta^{\al\beta}}{k_2^2}
\left[(-ie\ga^\mu)\frac{i}{\slashed{k}_1+\slashed{k}_2-\slashed{p}_2-m}(-ie\ga_\si)\frac{i}{\slashed{k}_2-\slashed{p}_2-m}(-ie\ga_\beta)\right]
v(p_2)\,,
\end{equation}
where we recognize the hard function $H_{f\ga\ga,\bar{f}}^{(0)\mu}$ in the square brackets, cf.\ eq.~\eqref{eq:Hfgagafbar}. Notice that compared to eq.~\eqref{eq:LPdiagramsdandh}, we choose the momentum routing of diagrams (d) and (h) such that there is a single hard function. This will make it slightly easier to see the appearance of our factorization formula below. 
As before, we can now add and subtract a term to write the photon propagators as
\begin{equation}
\frac{-i \eta^{\rho\si}}{k_i^2} 
= \frac{-i }{k_i^2} \frac{\bar{n}^\rho k_i^{\si}}{\bar{n}\cdot k_i}
+\frac{-i}{k_i^2}\left(\eta^{\rho\si} - \frac{\bar{n}^\rho k_i^{\si}}{\bar{n}\cdot k_i} \right)\,,
\end{equation}
for $i=1,2$. 
We get four terms in total. First we have 
\begin{align}
\cA_{\rho\al}^{(d+h)}
&\frac{-i}{k_1^2}\frac{\bar{n}^\rho k_1^{\si}}{\bar{n}\cdot k_1}
\frac{-i}{k_2^2}\frac{\bar{n}^\al k_2^{\beta}}{\bar{n}\cdot k_2} 
\left[(-ie\ga^\mu)\frac{i}{\slashed{k}_1+\slashed{k}_2-\slashed{p}_2-m}(-ie\ga_\si)\frac{i}{\slashed{k}_2-\slashed{p}_2-m}(-ie\ga_\beta)\right]
v(p_2)\nn\\
&= \left[
\begin{gathered}\SetScale{0.5}\hspace{0.5cm}\text{\footnotesize\Jfdd}\end{gathered}\hspace{0.5cm}
+
\begin{gathered}\SetScale{0.5}\hspace{0.5cm}\text{\footnotesize\Jfdh}\end{gathered}\hspace{0.5cm}\right] H_{f,\bar{f}}^{(0)\mu} J_{\bar{f}}^{(0)}\,.
\end{align}
which we already derived in eq.~\eqref{eq:LPfactOfDiagramsdandh2}. Note that the loop integrations are implicit here.
Second, we have
\begin{align}
\cA_{\rho\al}^{(d+h)}
&\frac{-i}{k_1^2}\left(\eta^{\rho\si} - \frac{\bar{n}^\rho k_1^{\si}}{\bar{n}\cdot k_1} \right)
\frac{-i}{k_2^2}\left(\eta^{\al\beta} - \frac{\bar{n}^\al k_2^{\beta}}{\bar{n}\cdot k_2} \right)
\left[(-ie\ga^\mu)\frac{i}{\slashed{k}_1+\slashed{k}_2-\slashed{p}_2-m}(-ie\ga_\si)\frac{i}{\slashed{k}_2-\slashed{p}_2-m}(-ie\ga_\beta)\right]
v(p_2)\nn\\
&=\left[
\begin{gathered}\SetScale{0.5}\hspace{0.5cm}\text{\footnotesize\Jfgga}\end{gathered}\hspace{0.5cm}
+
\begin{gathered}\SetScale{0.5}\hspace{0.5cm}\text{\footnotesize\Jfggb}\end{gathered}\hspace{0.5cm}\right]\otimes H_{f\ga\ga,\bar{f}}^{(0)\mu} J_{\bar{f}}^{(0)},
\end{align}
where the term in brackets is precisely $J_{f\ga\ga}^{(2)}$, which was defined in eq.~\eqref{eq:defJfgaga}.
The last two terms are cross terms, one of which reads
\begin{align}
\cA_{\rho\al}^{(d+h)}
&\frac{-i}{k_1^2}\frac{\bar{n}^\rho k_1^{\si}}{\bar{n}\cdot k_1}
\frac{-i}{k_2^2}\left(\eta^{\al\beta} - \frac{\bar{n}^\al k_2^{\beta}}{\bar{n}\cdot k_2} \right)
\left[(-ie\ga^\mu)\frac{i}{\slashed{k}_1+\slashed{k}_2-\slashed{p}_2-m}(-ie\ga_\si)\frac{i}{\slashed{k}_2-\slashed{p}_2-m}(-ie\ga_\beta)\right]
v(p_2)\nn\\
&= \left[
\begin{gathered}\SetScale{0.5}\hspace{0.5cm}\text{\footnotesize\Jfgdd}\end{gathered}\hspace{0.5cm}
+
\begin{gathered}\SetScale{0.5}\hspace{0.5cm}\text{\footnotesize\Jfgdh}\end{gathered}\hspace{0.5cm}\right]
\otimes H_{f\ga,\bar{f}}^{(0)\mu} J_{\bar{f}}^{(0)}\nn\\
&\hspace{0.5cm}
- \cA_{\rho\al}^{(d+h)}
\frac{-i}{k_1^2}\frac{\bar{n}^\rho }{\bar{n}\cdot k_1}
\frac{-i}{k_2^2}\left(\eta^{\al\beta} - \frac{\bar{n}^\al k_2^{\beta}}{\bar{n}\cdot k_2} \right)
\left[(-ie\ga^\mu)\frac{i}{\slashed{k}_1+\slashed{k}_2-\slashed{p}_2-m}(-ie\ga_\beta)\right]
v(p_2)\,.\label{crosstermAdh}
\end{align}
The term on the last line is eventually cancelled completely by the fourth (cross) term, which is
\begin{align}
\cA_{\rho\al}^{(d+h)}
&\frac{-i}{k_1^2}\left(\eta^{\rho\si} - \frac{\bar{n}^\rho k_1^{\si}}{\bar{n}\cdot k_1} \right)
\frac{-i}{k_2^2}\frac{\bar{n}^\al k_2^{\beta}}{\bar{n}\cdot k_2}
\left[(-ie\ga^\mu)\frac{i}{\slashed{k}_1+\slashed{k}_2-\slashed{p}_2-m}(-ie\ga_\si)\frac{i}{\slashed{k}_2-\slashed{p}_2-m}(-ie\ga_\beta)\right]
v(p_2)\,,
\end{align}
leaving only the first line in eq.~\eqref{crosstermAdh}.
This explains the entries observed for diagrams (d) and (h) in tab.~\ref{tab:NLPfact_cc}.

We conclude this appendix with a few remarks. First, the general pattern for extending the jet functions to NLP is clear. Each photon that connects to an eikonal propagator in the LP jet function $J_f$ is replaced by the effective photon propagator in eq.~\eqref{eq:NLPeffectivePhotonProp}, along with a convolution with a hard function. As we have shown, this leads to the NLP jet functions $J_{f\gamma}$ and $J_{f\gamma\gamma}$ without overlap between the jet functions. This pattern can also be used to understand the factorization of the $c\bar{c}$-region, which will be discussed in the next appendix. With the results of this and the next appendix, it is moreover straightforward to verify the factorization of the $hc$-region. 
Second, at NLP, there are also the $J_{f\!f\!\bar{f}}$ jet functions, which involve multiple fermion attachments to the hard function instead of photons. The pattern described above for obtaining NLP corrections does not apply in this case and one may wonder if this might still lead to an overlap between the jet functions. However, as discussed in sec.~\ref{sec:Jfff}, the $J_{f\!f\!\bar{f}}$ jet functions correspond to momentum regions distinct from $J_f$, $J_{f\gamma}$, and $J_{f\gamma\gamma}$. Specifically, these jet functions constitute the complete $cc$-region of diagram (f) and the complete $cc'$-region of diagram (h), see tab.~\ref{tab:NLPfact_cc}.

%% file: Appendices/regtofact.tex
In this appendix we again discuss how one goes from a full diagram to its factorization in terms of jet functions by manipulation at the integrand level, but now for the $c\bar{c}$-region. We mentioned the subtleties for this region in sec.\ \ref{sec:twoloop}. In this derivation we will assume that there is no analytic regulator on one of the propagators. This is true up to $\mathcal{O}(\lambda)$ in this region, but not at $\mathcal{O}(\lambda^2)$. We comment on this issue in sec.\ \ref{sec:ccbar} and showed that the derivation performed below remains valid.

In the $c\bar c$-region, three diagrams contribute, namely (f), (g) and (h). We omit any factors of $i$ and $e$ in what follows, and work at the integrand level. The external spinors are also left implicit. The three diagrams have the following structure
\begin{align}
    \mathcal{A}_f^\mu &= \frac{S_c^\rho}{D_c}\gamma^\mu \frac{1}{\slashed{k}_1-\slashed{p}_2-m}\gamma_\sigma\frac{1}{\slashed{k}_1+\slashed{k}_2-\slashed{p}_2-m}\gamma_\rho\frac{S_{\bar c}^\sigma}{D_{\bar{c}}} \\
    \mathcal{A}_g^\mu &= \frac{S_c^\rho}{D_c}\gamma^\sigma \frac{1}{\slashed{k}_1+\slashed{k}_2+\slashed{p}_1-m}\gamma_\rho\frac{1}{\slashed{k}_2+\slashed{p}_1-m}\gamma^\mu\frac{S_{\bar c}^\sigma}{D_{\bar{c}}}\\
    \mathcal{A}_h^\mu &= \frac{S_c^\rho}{D_c}\gamma^\sigma \frac{1}{\slashed{k}_1+\slashed{k}_2+\slashed{p}_1-m}\gamma^\mu\frac{1}{\slashed{k}_1+\slashed{k}_2-\slashed{p}_2-m}\gamma_\rho\frac{S_{\bar c}^\sigma}{D_{\bar{c}}},
\end{align}
where 
\begin{align}
    S_c^\rho &= \gamma^\rho(\slashed{k}_1+\slashed{p}_1+m),\qquad D_c = k_1^2\big[(k_1+p_1)^2-m^2\big]\nn\\
     S_{\bar c}^\sigma &= (\slashed{k}_2-\slashed{p}_2+m)\gamma^\sigma,\qquad D_{\bar c} = k_2^2\big[(k_2-p_2)^2-m^2\big].
\end{align}
We are interested in the $k_1=c$ and $k_2=\bar c$ limit. Using this notation, the relevant jet functions can be written as 
\begin{align}
    J_{f} &= \frac{S_c^\rho}{D_c}\frac{2\hat{p}_{2,\rho}  }{-2\hat{p}_2\cdot k_1},\\
    J_{f\gamma}^\rho &= \frac{S_c^\rho}{D_c}-\frac{S_c^\alpha}{D_c}\frac{2\hat{p}_{2,\alpha} k_1^\rho}{-2\hat{p}_2\cdot k_1},
\end{align}
where we omitted the loop order as superscript to improve readability, and also the loop integration.
We consider diagram (g) and (h) together, resulting in
\begin{align}
    \mathcal{A}_g^\mu+\mathcal{A}_h^\mu &= \frac{S_c^\rho}{D_c}\gamma^\sigma \frac{1}{\slashed{k}_1+\slashed{k}_2+\slashed{p}_1-m}\Bigg[\gamma_\rho\frac{1}{\slashed{k}_2+\slashed{p}_1-m}\gamma^\mu+\gamma^\mu\frac{1}{\slashed{k}_1+\slashed{k}_2-\slashed{p}_2-m}\gamma_\rho\Bigg]\frac{1}{\slashed{k}_2-\slashed{p}_2-m}\gamma_\sigma\frac{1}{k_2^2}\nn\\
    &\hspace{-1cm}= J_{f\gamma}^\rho \gamma^\sigma \frac{1}{\slashed{k}_1+\slashed{k}_2+\slashed{p}_1-m}\Bigg[\gamma_\rho\frac{1}{\slashed{k}_2+\slashed{p}_1-m}\gamma^\mu+\gamma^\mu\frac{1}{\slashed{k}_1+\slashed{k}_2-\slashed{p}_2-m}\gamma_\rho\Bigg]\frac{1}{\slashed{k}_2-\slashed{p}_2-m}\gamma_\sigma\frac{1}{k_2^2}\nn\\
    &\hspace{-1cm}+\frac{S_c^\alpha}{D_c}\frac{2\hat{p}_{2,\alpha} k_1^\rho}{-2\hat{p}_2\cdot k_1}\gamma^\sigma \frac{1}{\slashed{k}_1+\slashed{k}_2+\slashed{p}_1-m}\Bigg[\gamma_\rho\frac{1}{\slashed{k}_2+\slashed{p}_1-m}\gamma^\mu+\gamma^\mu\frac{1}{\slashed{k}_1+\slashed{k}_2-\slashed{p}_2-m}\gamma_\rho\Bigg]\frac{1}{\slashed{k}_2-\slashed{p}_2-m}\gamma_\sigma\frac{1}{k_2^2}.\label{AgAh}
\end{align}
The manipulation performed here is of the same nature as we frequently did in app.~\ref{app:seqjets}, cf.\ eq.~\eqref{A17} for example.
The first line can be considered ``half factorized'', i.e.\ consisting of a jet function ($J_{f\gamma}$) times a one-loop diagram. We leave this as it is for now and focus on the second line. We contract the $\rho$ index and use Ward identities as $\slashed{k}_1 = (\slashed{k}_1+\slashed{k}_2+\slashed{p}_1-m)-(\slashed{k}_2+\slashed{p}_1-m)$ and $\slashed{k}_1 = (\slashed{k}_1+\slashed{k}_2-\slashed{p}_2-m)-(\slashed{k}_2-\slashed{p}_2-m)$, such that this second line equals
\begin{align}
    &J_{f}\gamma^\sigma \frac{1}{\slashed{k}_1+\slashed{k}_2+\slashed{p}_1-m}\Bigg[\slashed{k}_1\frac{1}{\slashed{k}_2+\slashed{p}_1-m}\gamma^\mu+\gamma^\mu\frac{1}{\slashed{k}_1+\slashed{k}_2-\slashed{p}_2-m}\slashed{k}_1\Bigg]\frac{1}{\slashed{k}_2-\slashed{p}_2-m}\gamma_\sigma\frac{1}{k_2^2} \nn\\
 &=J_{f}\gamma^\sigma\Bigg[\frac{1}{\slashed{k}_2+\slashed{p}_1-m}\gamma^\mu\frac{1}{\slashed{k}_2-\slashed{p}_2-m}-\frac{1}{\slashed{k}_1+\slashed{k}_2+\slashed{p}_1-m}\gamma^\mu\frac{1}{\slashed{k}_2-\slashed{p}_2-m}\nn\\
    &+\frac{1}{\slashed{k}_1+\slashed{k}_2+\slashed{p}_1-m}\gamma^\mu\frac{1}{\slashed{k}_2-\slashed{p}_2-m}-\frac{1}{\slashed{k}_1+\slashed{k}_2+\slashed{p}_1-m}\gamma^\mu\frac{1}{\slashed{k}_1+\slashed{k}_2-\slashed{p}_2-m}
    \Bigg] \gamma_\sigma\frac{1}{k_2^2}.
\end{align}
We note that the second and third term cancel each other. Moreover, the last term contains a scaleless integral over $k_2$ in the $k_1=c, k_2=\bar c$ limit. The only term that remains is the first term, which is simply the $J_{f}$ jet multiplied by the one-loop diagram in the anticollinear limit. Combining this result with the first line of eq.~\eqref{AgAh}, we could then write a half-factorized formula for diagram (g) and (h) in the form
\begin{equation}
    \left.\mathcal{A}_g^\mu+\mathcal{A}_h^\mu\right\vert_{c\bar c} = J_{f}\times\begin{gathered}\includegraphics[width=0.17\textwidth]{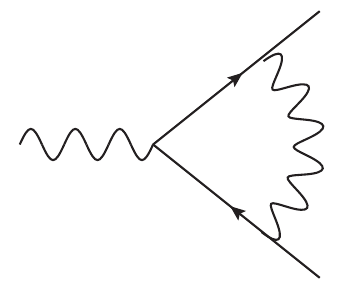}\end{gathered} +  J_{f\gamma} \otimes \Bigg[\begin{gathered}\includegraphics[width=0.17\textwidth]{Figures/NLO3_hard_function_Jfgamma-eps-converted-to.pdf}\end{gathered} + \begin{gathered}\includegraphics[width=0.17\textwidth]{Figures/NLO2_hard_function_Jfgamma-eps-converted-to.pdf}\end{gathered}\Bigg],
\end{equation}
where the $k_1$ loop has now been factorized into jet functions, while the $k_2$ loop is still unfactorized for the time being. We now turn our attention to diagram (f), which can be written as
\begin{align}
    \mathcal{A}_f^\mu &= \frac{S_c^\rho}{D_c}\gamma^\mu \frac{1}{\slashed{k}_1-\slashed{p}_2-m}\gamma_\sigma\frac{1}{\slashed{k}_1+\slashed{k}_2-\slashed{p}_2-m}\gamma_\rho\frac{S_{\bar c}^\sigma}{D_{\bar{c}}} \nn\\
    &= \bigg[J_{f\gamma}^\rho+\frac{S_c^\alpha}{D_c}\frac{2\hat{p}_{2,\alpha} k_1^\rho}{-2\hat{p}_2\cdot k_1}\bigg]\gamma^\mu \frac{1}{\slashed{k}_1-\slashed{p}_2-m}\gamma_\sigma\frac{1}{\slashed{k}_1+\slashed{k}_2-\slashed{p}_2-m}\gamma_\rho\frac{S_{\bar c}^\sigma}{D_{\bar{c}}}.
\end{align}
We can again rewrite $\slashed{k}_1$ for the second term, and obtain
\begin{align}
    &\frac{S_c^\alpha}{D_c}\frac{2\hat{p}_{2,\alpha} }{-2\hat{p}_2\cdot k_1}\gamma^\mu \frac{1}{\slashed{k}_1-\slashed{p}_2-m}\gamma_\sigma\frac{1}{\slashed{k}_1+\slashed{k}_2-\slashed{p}_2-m}\slashed{k}_1\frac{S_{\bar c}^\sigma}{D_{\bar{c}}}\nn\\
    &\hspace{1cm}= \frac{S_c^\alpha}{D_c}\frac{2\hat{p}_{2,\alpha} }{-2\hat{p}_2\cdot k_1}\gamma^\mu \frac{1}{\slashed{k}_1-\slashed{p}_2-m}\gamma_\sigma\frac{1}{\slashed{k}_2-\slashed{p}_2-m}\gamma^\sigma\frac{1}{k_2^2}\nn\\
    &\hspace{2cm}-\frac{S_c^\alpha}{D_c}\frac{2\hat{p}_{2,\alpha} }{-2\hat{p}_2\cdot k_1}\gamma^\mu \frac{1}{\slashed{k}_1-\slashed{p}_2-m}\gamma_\sigma\frac{1}{\slashed{k}_1+\slashed{k}_2-\slashed{p}_2-m}\gamma^\sigma \frac{1}{k_2^2}.
\end{align}
The second term contains a scaleless integral over $k_2$ in the $c\bar c$ limit and therefore vanishes. Hence diagram (f) has the half-factorized form
\begin{equation}\label{halffactdiagf}
    \left.\mathcal{A}_f\right\vert_{c\bar c} = J_{f\gamma}\otimes \begin{gathered}\includegraphics[width=0.17\textwidth]{Figures/NLO4_hard_function_Jfgamma-eps-converted-to.pdf}\end{gathered}  +\frac{S_c^\alpha}{D_c}\frac{2\hat{p}_{2,\alpha} }{-2\hat{p}_2\cdot k_1}\gamma^\mu \frac{1}{\slashed{k}_1-\slashed{p}_2-m}\gamma_\sigma\frac{S^\sigma_{\bar c}}{D_{\bar c}}.
\end{equation}
Note that the second term seems to be problematic in regards to the factorization approach. We set this term aside for now.

Now that we have established a half-factorized form, we continue with decomposing the anticollinear loops into the different jet functions. We start by rewriting
\begin{align}
     J_{f\gamma} \otimes \begin{gathered}\includegraphics[width=0.17\textwidth]{Figures/NLO3_hard_function_Jfgamma-eps-converted-to.pdf}\end{gathered} &= J_{f\gamma}^\rho\gamma_\sigma\frac{1}{\slashed{k}_1+\slashed{k}_2+\slashed{p}_1-m}\gamma_\rho\frac{1}{\slashed{k}_2+\slashed{p}_1-m}\gamma^\mu\left[J^\sigma_{\bar f\gamma}+\frac{2\hat{p}_{1,\beta}  k_2^\sigma}{2\hat{p}_1\cdot k_2}\frac{S^\beta_{\bar c}}{D_{\bar c}}\right].\label{eqB13}
\end{align}
The first term is exactly one of three parts of $J_{f\gamma}\otimes H_{f\gamma,\bar{f}\gamma}\otimes J_{\bar f \gamma}$, where we recall that $H_{f\ga,\bar{f}\ga}$ is the sum of three terms, see eq.~\eqref{Hfgfg}. The second term can be rewritten by applying a Ward identity, as before, and we obtain
\begin{align}
   &\left(\eta^{\alpha\rho}-\frac{2\hat{p}_2^\alpha k_1^\rho}{-2\hat{p}_2\cdot k_1}\right)\frac{1}{k_1^2}\gamma_\alpha\frac{1}{\slashed{k}_1+\slashed{p}_1-m}\slashed{k}_2\frac{1}{\slashed{k}_1+\slashed{k}_2+\slashed{p}_1-m}\gamma_\rho\frac{1}{\slashed{k}_2+\slashed{p}_1-m}\gamma^\mu\frac{2\hat{p}_{1,\beta} }{2\hat{p}_1\cdot k_2}\frac{S^\beta_{\bar c}}{D_{\bar c}}\nn\\
   &= \left(\eta^{\alpha\rho}-\frac{2\hat{p}_2^\alpha k_1^\rho}{-2\hat{p}_2\cdot k_1}\right)\frac{1}{k_1^2}\gamma_\alpha\frac{1}{\slashed{k}_1+\slashed{p}_1-m}\gamma_\rho\frac{1}{\slashed{k}_2+\slashed{p}_1-m}\gamma^\mu\frac{2\hat{p}_{1,\beta} }{2\hat{p}_1\cdot k_2}\frac{S^\beta_{\bar c}}{D_{\bar c}}\nn\\
   &-\left(\eta^{\alpha\rho}-\frac{2\hat{p}_2^\alpha k_1^\rho}{-2\hat{p}_2\cdot k_1}\right)\frac{1}{k_1^2}\gamma_\alpha\frac{1}{\slashed{k}_1+\slashed{k}_2+\slashed{p}_1-m}\gamma_\rho\frac{1}{\slashed{k}_2+\slashed{p}_1-m}\gamma^\mu\frac{2\hat{p}_{1,\beta} }{2\hat{p}_1\cdot k_2}\frac{S^\beta_{\bar c}}{D_{\bar c}}.
\end{align}
The second line contains a scaleless integral for $k_1$ and therefore vanishes. The remaining expression can be further simplified to become
\begin{align}
   &\frac{S_c^\rho}{D_c}\gamma_\rho\frac{1}{\slashed{k}_2+\slashed{p}_1-m}\gamma^\mu\frac{2\hat{p}_{1,\beta} }{2\hat{p}_1\cdot k_2}\frac{S^\beta_{\bar c}}{D_{\bar c}} - \frac{2\hat{p}_2^\alpha }{-2\hat{p}_2\cdot k_1}\frac{1}{k_1^2}\gamma_\alpha\frac{1}{\slashed{k}_1+\slashed{p}_1-m}\slashed{k}_1\frac{1}{\slashed{k}_2+\slashed{p}_1-m}\gamma^\mu\frac{2\hat{p}_{1,\beta} }{2\hat{p}_1\cdot k_2}\frac{S^\beta_{\bar c}}{D_{\bar c}}\nn\\
   &=\frac{S_c^\rho}{D_c}\gamma_\rho\frac{1}{\slashed{k}_2+\slashed{p}_1-m}\gamma^\mu\frac{2\hat{p}_{1,\beta} }{2\hat{p}_1\cdot k_2}\frac{S^\beta_{\bar c}}{D_{\bar c}} - \frac{2\hat{p}_2^\alpha }{-2\hat{p}_2\cdot k_1}\frac{1}{k_1^2}\gamma_\alpha\frac{1}{\slashed{k}_2+\slashed{p}_1-m}\gamma^\mu\frac{2\hat{p}_{1,\beta} }{2\hat{p}_1\cdot k_2}\frac{S^\beta_{\bar c}}{D_{\bar c}}\nn\\
   &+J_{f}(\slashed{p}_1-m)\frac{1}{\slashed{k}_2+\slashed{p}_1-m}\gamma^\mu\frac{2\hat{p}_{1,\beta} }{2\hat{p}_1\cdot k_2}\frac{S^\beta_{\bar c}}{D_{\bar c}}.\label{cancelstoo}
\end{align}
The first term we recognize as similar to the extra term we found in \eqn{halffactdiagf}. The second term is a scaleless $k_1$ integral and thus vanishes. The third term remains for now, but will cancel with a term arising later in the calculation.
Next, analogously to eq.~\eqref{eqB13}, we have
\begin{align}
     J_{f\gamma} \otimes \begin{gathered}\includegraphics[width=0.17\textwidth]{Figures/NLO2_hard_function_Jfgamma-eps-converted-to.pdf}\end{gathered} &= J_{f\gamma}^\rho \gamma_\sigma \frac{1}{\slashed{k}_1+\slashed{k}_2+\slashed{p}_1-m}\gamma^\mu\frac{1}{\slashed{k}_1+\slashed{k}_2-\slashed{p}_2-m}\gamma_\rho J_{\bar f \gamma}^\sigma\nn\\
&\hspace{-3cm}+\left(\eta^{\alpha\rho}-\frac{2\hat{p}_2^\alpha k_1^\rho}{-2\hat{p}_2\cdot k_1}\right)\frac{1}{k_1^2}\gamma_\alpha\frac{1}{\slashed{k}_1+\slashed{p}_1-m}\gamma_\sigma\frac{1}{\slashed{k}_1+\slashed{k}_2+\slashed{p}_1-m}\gamma^\mu\frac{1}{\slashed{k}_1+\slashed{k}_2-\slashed{p}_2-m}\gamma_\rho\frac{2\hat{p}_{1,\beta} k_2^\sigma}{2\hat{p}_1\cdot k_2}\frac{S^\beta_{\bar c}}{D_{\bar c}}.
\end{align}
The first term is again exactly one of the three parts of $J_{f\gamma}\otimes H_{f\gamma,\bar{f}\gamma}\otimes J_{\bar f \gamma}$. The second term can be rewritten as 
\begin{align}
&\left(\eta^{\alpha\rho}-\frac{2\hat{p}_2^\alpha k_1^\rho}{-2\hat{p}_2\cdot k_1}\right)\frac{1}{k_1^2}\gamma_\alpha\frac{1}{\slashed{k}_1+\slashed{p}_1-m}\slashed{k}_2\frac{1}{\slashed{k}_1+\slashed{k}_2+\slashed{p}_1-m}\gamma^\mu\frac{1}{\slashed{k}_1+\slashed{k}_2-\slashed{p}_2-m}\gamma_\rho\frac{2\hat{p}_{1,\beta} }{2\hat{p}_1\cdot k_2}\frac{S^\beta_{\bar c}}{D_{\bar c}}\nn\\
&= \left(\eta^{\alpha\rho}-\frac{2\hat{p}_2^\alpha k_1^\rho}{-2\hat{p}_2\cdot k_1}\right)\frac{1}{k_1^2}\gamma_\alpha\frac{1}{\slashed{k}_1+\slashed{p}_1-m}\gamma^\mu\frac{1}{\slashed{k}_1+\slashed{k}_2-\slashed{p}_2-m}\gamma_\rho\frac{2\hat{p}_{1,\beta} }{2\hat{p}_1\cdot k_2}\frac{S^\beta_{\bar c}}{D_{\bar c}}.\label{cancels},
\end{align}
where we omitted a scaleless integral over $k_1$. We will leave this second term for the moment and move to the third part which depends on $J_{f\gamma}$. It reads
\begin{align}
J_{f\gamma}\otimes \begin{gathered}\includegraphics[width=0.17\textwidth]{Figures/NLO4_hard_function_Jfgamma-eps-converted-to.pdf}\end{gathered} &= J^\rho_{f\gamma}\gamma^\mu \frac{1}{\slashed{k}_1-\slashed{p}_2-m}\gamma_\sigma\frac{1}{\slashed{k}_1+\slashed{k}_2-\slashed{p}_2-m}\gamma_\rho J^\sigma_{\bar f \gamma}\nn\\
&\hspace{-4cm}+\left(\eta^{\alpha\rho}-\frac{2\hat{p}_2^\alpha k_1^\rho}{-2\hat{p}_2\cdot k_1}\right)\frac{1}{k_1^2}\gamma_\alpha\frac{1}{\slashed{k}_1+\slashed{p}_1-m}\gamma^\mu \frac{1}{\slashed{k}_1-\slashed{p}_2-m}\gamma_\sigma\frac{1}{\slashed{k}_1+\slashed{k}_2-\slashed{p}_2-m}\gamma_\rho \frac{2\hat{p}_{1,\beta} k_2^\sigma}{2\hat{p}_1\cdot k_2}\frac{S^\beta_{\bar c}}{D_{\bar c}}.
\end{align}
The first term is again exactly one of the three parts of $J_{f\gamma}\otimes H_{f\gamma,\bar{f}\gamma}\otimes J_{\bar f \gamma}$. The second term can be rewritten as
\begin{align}
&\left(\eta^{\alpha\rho}-\frac{2\hat{p}_2^\alpha k_1^\rho}{-2\hat{p}_2\cdot k_1}\right)\frac{1}{k_1^2}\gamma_\alpha\frac{1}{\slashed{k}_1+\slashed{p}_1-m}\gamma^\mu \frac{1}{\slashed{k}_1-\slashed{p}_2-m}\slashed{k}_2\frac{1}{\slashed{k}_1+\slashed{k}_2-\slashed{p}_2-m}\gamma_\rho \frac{2\hat{p}_{1,\beta} }{2\hat{p}_1\cdot k_2}\frac{S^\beta_{\bar c}}{D_{\bar c}}\nn\\
&= \left(\eta^{\alpha\rho}-\frac{2\hat{p}_2^\alpha k_1^\rho}{-2\hat{p}_2\cdot k_1}\right)\frac{1}{k_1^2}\gamma_\alpha\frac{1}{\slashed{k}_1+\slashed{p}_1-m}\gamma^\mu \frac{1}{\slashed{k}_1-\slashed{p}_2-m}\gamma_\rho \frac{2\hat{p}_{1,\beta} }{2\hat{p}_1\cdot k_2}\frac{S^\beta_{\bar c}}{D_{\bar c}}\nn\\
&-\left(\eta^{\alpha\rho}-\frac{2\hat{p}_2^\alpha k_1^\rho}{-2\hat{p}_2\cdot k_1}\right)\frac{1}{k_1^2}\gamma_\alpha\frac{1}{\slashed{k}_1+\slashed{p}_1-m}\gamma^\mu\frac{1}{\slashed{k}_1+\slashed{k}_2-\slashed{p}_2-m}\gamma_\rho \frac{2\hat{p}_{1,\beta} }{2\hat{p}_1\cdot k_2}\frac{S^\beta_{\bar c}}{D_{\bar c}}.
\end{align}
Note that the second term now exactly cancels with \eqn{cancels}. Moreover, the first term is readily identified as $J_{f\gamma}\otimes H_{f\gamma,\bar{f}}J_{\bar f}$.

The final ingredient is 
\begin{align}
J_{f}\times\begin{gathered}\includegraphics[width=0.17\textwidth]{Figures/One_Loop_QED-eps-converted-to.pdf}\end{gathered} &= J_{f}\gamma_\sigma\frac{1}{\slashed{k}_2+\slashed{p}_1-m}\gamma^\mu \frac{1}{\slashed{k}_2-\slashed{p}_2-m}\gamma^\sigma \frac{1}{k_2^2} \nn\\
&= J_{f}\gamma_\sigma\frac{1}{\slashed{k}_2+\slashed{p}_1-m}\gamma^\mu J_{\bar f\gamma} + J_{f}\gamma_\sigma\frac{1}{\slashed{k}_2+\slashed{p}_1-m}\gamma^\mu\frac{2\hat{p}_{1,\beta} k_2^\sigma}{2\hat{p}_1\cdot k_2}\frac{S_{\bar c}^\beta}{D_{\bar c}}.
\end{align}
The first term is exactly equal to $J_{f} H_{f,\bar f\gamma}\otimes J_{\bar f \gamma}$. The second term can be rewritten as 
\begin{align}
J_{f}\slashed{k}_2\frac{1}{\slashed{k}_2+\slashed{p}_1-m}\gamma^\mu\frac{2\hat{p}_{1,\beta} }{2\hat{p}_1\cdot k_2}\frac{S_{\bar c}^\beta}{D_{\bar c}} = J_{f}H_{f,\bar{f}}J_{\bar f} - J_{f}(\slashed{p}_1-m)\frac{1}{\slashed{k}_2+\slashed{p}_1-m}\gamma^\mu\frac{2\hat{p}_{1,\beta} }{2\hat{p}_1\cdot k_2}\frac{S^\beta_{\bar c}}{D_{\bar c}}.
\end{align}
This last term cancels with the third term in \eqn{cancelstoo}. We therefore have all the ingredients of the jet functions, together with two diagrams that do not yet factorize. The non-factorized contribution reads
\begin{equation}\label{eq:extraterms}
    \frac{S_c^\alpha}{D_c}\frac{2\hat{p}_{2,\alpha} }{-2\hat{p}_2\cdot k_1}\gamma^\mu \frac{1}{\slashed{k}_1-\slashed{p}_2-m}\gamma_\sigma\frac{S^\sigma_{\bar c}}{D_{\bar c}} + \frac{S_c^\rho}{D_c}\gamma_\rho\frac{1}{\slashed{k}_2+\slashed{p}_1-m}\gamma^\mu\frac{2\hat{p}_{1,\beta} }{2\hat{p}_1\cdot k_2}\frac{S^\beta_{\bar c}}{D_{\bar c}}.
\end{equation}

In fact, these terms do have a place in the factorization theorem, in a way we already alluded to in sec.\ \ref{sec:twoloop}. The solution lies in properly taking into account self-energy contributions. If we consider the one-loop vertex diagram in the collinear region with a self-energy contribution on the anticollinear leg, the integrand of the amplitude is given by 
\begin{align}
    \mathcal{A}^\mu &= \frac{S_c^\rho}{D_c}\gamma^\mu\frac{1}{\slashed{k}_1-\slashed{p}_2-m}\gamma_\rho\frac{1}{k_1^2}\frac{1}{-\slashed{p}_2-m}\gamma_\sigma\frac{S^\sigma_{\bar c}}{D_{\bar c}}\nn\\
    &=\left[J_{f\gamma} + \frac{S_c^\alpha}{D_c}\frac{2\hat{p}_{2,\alpha} k_1^\rho}{-2\hat{p}_2\cdot k_1}\right]\gamma^\mu\frac{1}{\slashed{k}_1-\slashed{p}_2-m}\gamma_\rho\frac{1}{-\slashed{p}_2-m}\gamma_\sigma\frac{S^\sigma_{\bar c}}{D_{\bar c}}.
\end{align}
The first term is just the $J_{f\ga}$ jet multiplied by $H_{f\ga,\bar{f}}$, and the self energy on the anticollinear leg. The second term can be recast as
\begin{align}
    &\frac{S_c^\alpha}{D_c}\frac{2\hat{p}_{2,\alpha} }{-2\hat{p}_2\cdot k_1}\gamma^\mu\frac{1}{\slashed{k}_1-\slashed{p}_2-m}\slashed{k}_1\frac{1}{-\slashed{p}_2-m}\gamma_\sigma\frac{S^\sigma_{\bar c}}{D_{\bar c}} \nn\\
    &= J_{f}H_{f,\bar{f}}^{\mu(0)}\frac{1}{-\slashed{p}_2-m}\gamma_\sigma\frac{S^\sigma_{\bar c}}{D_{\bar c}} -\frac{S_c^\alpha}{D_c}\frac{2\hat{p}_{2,\alpha} }{-2\hat{p}_2\cdot k_1}\gamma^\mu \frac{1}{\slashed{k}_1-\slashed{p}_2-m}\gamma_\sigma\frac{S^\sigma_{\bar c}}{D_{\bar c}}.
\end{align}
Here the first term is just the $J_{f}$ jet multiplied by the appropriate hard function $H_{f,\bar{f}}$ and the self energy on the anticollinear leg. Note that the last term is identical to the first term in \eqn{eq:extraterms}, but, crucially, with an opposite sign. The second term in \eqn{eq:extraterms} similarly cancels against the one-loop vertex diagram with a self-energy contribution on the collinear leg. This shows that one can only achieve factorization once all self-energy contributions have been properly included. On the factorization side, the self-energy contribution is interpreted as being part of the $J_{f}$ jet, albeit a non-1PI contribution. This is easily checked by carrying out the Wick contractions in \eqn{eq: defJfcbar}.

Thus we have shown factorization at the integrand level. Note that in this appendix we have treated the propagator
\begin{equation}
    \frac{1}{-\slashed{p}_2-m}
\end{equation}
symbolically in order to show that factorization holds at the integrand level.
In sec.\ \ref{sec:twoloop} we included the self-energy contributions by properly performing a Dyson sum, consequently modifying the equations of motion and the mass-shell condition. These issues were already previewed in sec.\ \ref{sec:setup}. This allowed us to calculate and positively verify the factorized form of the form factors.
As a final comment, we note that the proof of factorization in the hard-collinear region can also be derived in an analysis similar to that of appendices \ref{app:seqjets} and \ref{app:regtofact}.

%% file: Appendices/2loopres.tex
We report here the integral expressions for the various jet functions at two loops. The one corresponding to the two-loop $J_{f}$ jet function, the diagrams for which are given in \eqn{eq:Jf2diag}, is

\begin{align}
    J_{f}^{(2)}(p_1,\bar{n}) &= e^4\bar{u}(p_1)\Bigg\{ \int[dk_1][dk_2]\frac{-i}{k_1^2 }\frac{-i}{k_2^2 }\gamma_\mu\frac{i(\slashed{p}_1+\slashed{k}_1+m)}{(p_1+k_1)^2-m^2 }\gamma_\nu\frac{i(\slashed{p}_1+\slashed{k}_1+\slashed{k}_2+m)}{(p_1+k_1+k_2)^2-m^2 }\nn\\
    &\hspace{-0.5cm}\quad \times \bigg[\frac{-i\bar{n}^\mu}{- \bar{n}\cdot k_1 }\frac{-i\bar{n}^\nu}{- \bar{n}\cdot(k_1+k_2) }+\frac{-i\bar{n}^\mu}{- \bar{n}\cdot k_2 }\frac{-i\bar{n}^\nu}{- \bar{n}\cdot(k_1+k_2) }\bigg]\nn\\
    &\hspace{-0.5cm}\quad+ \int[dk_1][dk_2]\frac{-i}{k_1^2 }\frac{-i}{k_2^2 }\nn\\
    &\hspace{-0.5cm}\quad\times \Bigg[\gamma_\mu \frac{i(\slashed{p}_1+\slashed{k}_1+m)}{(p_1+k_1)^2-m^2 }\gamma_\nu \frac{i(\slashed{p}_1+\slashed{k}_1+\slashed{k}_2+m)}{(p_1+k_1+k_2)^2-m^2 }\gamma^\nu \frac{i(\slashed{p}_1+\slashed{k}_1+m)}{(p_1+k_1)^2-m^2 }\frac{-i\bar{n}^\mu}{- \bar{n}\cdot k_1 }\nn\\
    &\hspace{-0.5cm}\quad +\gamma_\mu \frac{i(\slashed{p}_1+\slashed{k}_2+m)}{(p_1+k_2)^2-m^2 }\gamma_\nu \frac{i(\slashed{p}_1+\slashed{k}_1+\slashed{k}_2+m)}{(p_1+k_1+k_2)^2-m^2 }\gamma^\mu \frac{i(\slashed{p}_1+\slashed{k}_1+m)}{(p_1+k_1)^2-m^2 }\frac{-i\bar{n}^\nu}{- \bar{n}\cdot k_1 }\Bigg]\nn\\ &\quad\hspace{-.5cm} -\int[dk_1][dk_2]\left(\frac{-i}{k_2^2 }\right)^2\nn\\
    &\hspace{-0.5cm}\quad\times \Bigg[\gamma_\mu \frac{i(\slashed{p}_1+\slashed{k}_2+m)}{(p_1+k_2)^2-m^2 }\mathrm{Tr}\left[\gamma_\nu\frac{i(\slashed{k}_1+\slashed{p}_1+\slashed{k}_2+m)}{(k_1+p_1+k_2)^2-m^2 }\gamma^\mu\frac{i(\slashed{k}_1+\slashed{p}_1+m)}{(k_1+p_1)^2-m^2 }\right]\frac{-i\bar{n}^\nu}{- \bar{n}\cdot k_2 }\nn\\
    &+\gamma_\mu \frac{i(\slashed{p}_1+\slashed{k}_2+m)}{(p_1+k_2)^2-m^2 }\mathrm{Tr}\left[\gamma_\nu\frac{i(\slashed{k}_1+\slashed{p}_1+\slashed{k}_2)}{(k_1+p_1+k_2)^2 }\gamma^\mu\frac{i(\slashed{k}_1+\slashed{p}_1)}{(k_1+p_1)^2 }\right]\frac{-i\bar{n}^\nu}{- \bar{n}\cdot k_2 }\Bigg]\Bigg\}.
\end{align}
The last line constitutes the massless fermion loop and has to be multiplied by a factor of $n_f e_f^2$, for $n_f$ light fermions and $e_f$ the electric charge in terms of the positron charge. The next-to-last line represents the massive fermion loop.

For the $J_{f\ga}$ jet function, whose diagrams are given in \eqn{eq:Jfg2diag}, the integral expression yields
\begin{align}
    J_{f\gamma}^{(2)\rho}(p_1,\bar{n},\ell^+)
    &=-ie^4\bar{u}(p_1)\int [dk_1][dk_2]\delta(\bar{n}\cdot k_1-\ell^+)\times\Bigg\{ \nn\\
    &+\gamma^\rho\frac{i(\slashed{p}_1-\slashed{k}_1+m)}{(p_1-k_1)^2-m^2}\gamma_\sigma\frac{i(\slashed{p}_1-\slashed{k}_1+\slashed{k}_2+m)}{(p_1-k_1+k_2)^2-m^2}\gamma^\sigma\frac{i(\slashed{p}_1-\slashed{k}_1+m)}{(p_1-k_1)^2-m^2}\frac{-i}{k_1^2}\frac{-i}{k_2^2}\nn\\
    &\hspace{-1cm}-\gamma^\nu\frac{i(\slashed{p}_1-\slashed{k}_1+m)}{(p_1-k_1)^2-m^2}\gamma_\sigma\frac{i(\slashed{p}_1-\slashed{k}_1+\slashed{k}_2+m)}{(p_1-k_1+k_2)^2-m^2}\gamma^\sigma\frac{i(\slashed{p}_1-\slashed{k}_1+m)}{(p_1-k_1)^2-m^2}\frac{-i}{k_1^2}\frac{-i}{k_2^2}\frac{\bar{n}_\nu}{\bar{n}\cdot k_1 }k_1^\rho\nn\\ &+\gamma^\sigma\frac{i(\slashed{p}_1+\slashed{k}_2+m)}{(p_1+k_2)^2-m^2}\gamma_\rho\frac{i(\slashed{p}_1-\slashed{k}_1+\slashed{k}_2+m)}{(p_1-k_1+k_2)^2-m^2}\gamma^\sigma\frac{i(\slashed{p}_1-\slashed{k}_1+m)}{(p_1-k_1)^2-m^2}\frac{-i}{k_1^2}\frac{-i}{k_2^2}\nn\\   &\hspace{-1cm}-\gamma^\sigma\frac{i(\slashed{p}_1+\slashed{k}_2+m)}{(p_1+k_2)^2-m^2}\gamma_\nu\frac{i(\slashed{p}_1-\slashed{k}_1+\slashed{k}_2+m)}{(p_1-k_1+k_2)^2-m^2}\gamma^\sigma\frac{i(\slashed{p}_1-\slashed{k}_1+m)}{(p_1-k_1)^2-m^2}\frac{-i}{k_1^2}\frac{-i}{k_2^2}\frac{\bar{n}_\nu}{\bar{n}\cdot k_1 }k_1^\rho\nn\\
    &-\gamma^\sigma \frac{i(\slashed{p}_1-\slashed{k}_1+m)}{(p_1-k_1)^2-m^2}\mathrm{Tr}\left[\gamma^\rho\frac{i(\slashed{p}_1-\slashed{k}_1+\slashed{k}_2+m)}{(p_1-k_1+k_2)^2-m^2}\gamma_\sigma\frac{i(\slashed{p}_1+\slashed{k}_2+m)}{(p_1+k_2)^2-m^2}\right]\left(\frac{-i}{k_1^2}\right)^2\nn\\
    &\hspace{-1cm}+\gamma^\sigma \frac{i(\slashed{p}_1-\slashed{k}_1+m)}{(p_1-k_1)^2-m^2}\mathrm{Tr}\left[\gamma^\nu\frac{i(\slashed{p}_1-\slashed{k}_1+\slashed{k}_2+m)}{(p_1-k_1+k_2)^2-m^2}\gamma_\sigma\frac{i(\slashed{p}_1+\slashed{k}_2+m)}{(p_1+k_2)^2-m^2}\right]\left(\frac{-i}{k_1^2}\right)^2\frac{\bar{n}_\nu}{\bar{n}\cdot k_1 }k_1^\rho\nn\\
    &+\gamma^\rho \frac{i(\slashed{p}_1-\slashed{k}_1+m)}{(p_1-k_1)^2-m^2}\gamma_\nu \frac{i(\slashed{p}_1-\slashed{k}_1+\slashed{k}_2+m)}{(p_1-k_1+k_2)^2-m^2}\frac{-i}{k_1^2}\frac{-i}{k_2^2}\frac{-i\bar{n}^\nu}{-\bar{n}\cdot k_2 }\nn\\
    &\hspace{-1cm} -\gamma_\sigma \frac{i(\slashed{p}_1-\slashed{k}_1+m)}{(p_1-k_1)^2-m^2}\gamma_\nu \frac{i(\slashed{p}_1-\slashed{k}_1+\slashed{k}_2+m)}{(p_1-k_1+k_2)^2-m^2}\frac{-i}{k_1^2}\frac{-i}{k_2^2}\frac{-i\bar{n}^\nu}{-\bar{n}\cdot k_2 }\frac{\bar{n}^\sigma}{\bar{n}\cdot k_1 }k_1^\rho\nn\\
    &+\gamma_\nu \frac{i(\slashed{p}_1+\slashed{k}_2+m)}{(p_1+k_2)^2-m^2}\gamma^\rho \frac{i(\slashed{p}_1-\slashed{k}_1+\slashed{k}_2+m)}{(p_1-k_1+k_2)^2-m^2}\frac{-i}{k_1^2}\frac{-i}{k_2^2}\frac{-i\bar{n}^\nu}{-\bar{n}\cdot k_2 }\nn\\
    &\hspace{-1cm}-\gamma_\nu \frac{i(\slashed{p}_1+\slashed{k}_2+m)}{(p_1+k_2)^2-m^2}\gamma_\sigma \frac{i(\slashed{p}_1-\slashed{k}_1+\slashed{k}_2+m)}{(p_1-k_1+k_2)^2-m^2}\frac{-i}{k_1^2}\frac{-i}{k_2^2}\frac{-i\bar{n}^\nu}{-\bar{n}\cdot k_2 }\frac{\bar{n}^\sigma}{\bar{n}\cdot k_1 }k_1^\rho\Bigg\}.\label{eq:Jfg2int}
\end{align}
The integral expressions for the $J_{f\partial\ga}$ jet are akin to the $J_{f\ga}$ jet, in the sense that one includes an additional factor of $k_{1\perp}^\sigma$ in the two-loop integral. Hence we do not report the full expression here.

The integral expression for the $J_{f\ga\ga}$ jet contribution, as given in \eqn{eq:Jfggdiag}, reads
\begin{align}
  J_{f\gamma\gamma}^{(2)\rho\sigma}(p_1,\bar{n},
  \ell_1^+,
  \ell_2^+)&=e^4\bar{u}(p_1)\int[dk_1][dk_2]\frac{-i}{k_1^2}\frac{-i}{k_2^2}\delta(\bar{n}\cdot k_1-\ell_1^+ )\delta(\bar{n}\cdot k_2-\ell_2^+ )\,\times \Bigg\{\nn\\
    &-\bigg[\gamma^\sigma \frac{i(\slashed{p}_1-\slashed{k}_2+m)}{(p_1-k_2)^2-m^2}\gamma^\rho \frac{i(\slashed{p}_1-\slashed{k}_1-\slashed{k}_2+m)}{(p_1-k_1-k_2)^2-m^2}\nn\\
    &\hspace{.5cm}+\gamma^\rho \frac{i(\slashed{p}_1-\slashed{k}_1+m)}{(p_1-k_1)^2-m^2}\gamma^\sigma \frac{i(\slashed{p}_1-\slashed{k}_1-\slashed{k}_2+m)}{(p_1-k_1-k_2)^2-m^2}\bigg]\nn\\
    &+ \bigg[\gamma^\mu \frac{i(\slashed{p}_1-\slashed{k}_2+m)}{(p_1-k_2)^2-m^2}\gamma^\rho \frac{i(\slashed{p}_1-\slashed{k}_1-\slashed{k}_2+m)}{(p_1-k_1-k_2)^2-m^2}k_2^\sigma\nn\\
    &\hspace{.5cm}+\gamma^\rho \frac{i(\slashed{p}_1-\slashed{k}_1+m)}{(p_1-k_1)^2-m^2}\gamma^\mu \frac{i(\slashed{p}_1-\slashed{k}_1-\slashed{k}_2+m)}{(p_1-k_1-k_2)^2-m^2}k_2^\sigma\bigg]\frac{\bar{n}_{\mu}}{\bar{n}\cdot k_2}\nn\\
    &+ \bigg[\gamma^\sigma \frac{i(\slashed{p}_1-\slashed{k}_2+m)}{(p_1-k_2)^2-m^2}\gamma^\mu \frac{i(\slashed{p}_1-\slashed{k}_1-\slashed{k}_2+m)}{(p_1-k_1-k_2)^2-m^2}k_1^\rho\nn\\
    &\hspace{.5cm}+\gamma^\mu \frac{i(\slashed{p}_1-\slashed{k}_1+m)}{(p_1-k_1)^2-m^2}\gamma^\sigma \frac{i(\slashed{p}_1-\slashed{k}_1-\slashed{k}_2+m)}{(p_1-k_1-k_2)^2-m^2}k_1^\rho\bigg]\frac{\bar{n}_{\mu}}{\bar{n}\cdot k_1}\nn\\
    &-\bigg[\gamma^\mu \frac{i(\slashed{p}_1-\slashed{k}_2+m)}{(p_1-k_2)^2-m^2}\gamma^\nu \frac{i(\slashed{p}_1-\slashed{k}_1-\slashed{k}_2+m)}{(p_1-k_1-k_2)^2-m^2}k_1^\rho k_2^\sigma\nn\\
    &\hspace{.5cm}+\gamma^\mu \frac{i(\slashed{p}_1-\slashed{k}_1+m)}{(p_1-k_1)^2-m^2}\gamma^\nu \frac{i(\slashed{p}_1-\slashed{k}_1-\slashed{k}_2+m)}{(p_1-k_1-k_2)^2-m^2}k_1^\rho k_2^\sigma\bigg]\frac{\bar{n}_{\mu}}{\bar{n}\cdot k_2}\frac{\bar{n}_{\nu}}{\bar{n}\cdot k_1}\Bigg\}.
\end{align}

Finally, the integral expressions to the two $J_{f\!f\!\bar{f}}$ jet functions \eqns{eq:JfffI}{eq:JfffII} are given by 

\begin{align}
   J^{(I)(2)}_{f\!f\!\bar f}(p_1,\bar{n},\ell_1^+,\ell_2^+) &= -e^2\int[dk_1][dk_2]\,\delta(\bar{n}\cdot k_1-\ell_1^+)\delta(\bar{n}\cdot k_2-\ell_2^+)\nn\\
   &\times \left[\bar{u}(p_1)\gamma^\mu\frac{i(\slashed{p}_1-\slashed{k}_1-\slashed{k}_2+m)}{(p_1-k_1-k_2)^2-m^2}\right]_{a}\frac{-i}{(k_1+k_2)^2} \left[\frac{i(-\slashed{k}_2+m)}{k_2^2-m^2}\gamma_\mu\frac{i(\slashed{k}_1+m)}{k_1^2-m^2}\right]_{bc}\nn\\
   &+ e^2\int[dk_1][dk_2]\,\delta(\bar{n}\cdot k_1-\ell_1^+)\delta(\bar{n}\cdot k_2-\ell_2^+)\nn\\
   &\times \left[\bar{u}(p_1)\gamma^\mu \frac{i(\slashed{k}_1+m)}{k_1^2-m^2}\right]_c\frac{-i}{(p_1-k_1)^2}\left[\frac{i(-\slashed{k}_2+m)}{k_2^2-m^2}\gamma_\mu \frac{i(\slashed{p}_1-\slashed{k}_1-\slashed{k}_2+m)}{(p_1-k_1-k_2)^2-m^2}\right]_{ba},\\
    J_{f\!f\!\bar f}^{(I\!I)(2)}(p_1,\bar{n},\ell_1^+,\ell_2^+)&=-e^2\int [dk_1][dk_2]\delta(\bar{n}\cdot k_1-\ell_1^+)\delta(\bar{n}\cdot k_2-\ell_2^+)\left[\bar{u}(p_1)\gamma^\mu\frac{i(\slashed{k}_2+m)}{k_2^2-m^2}\right]_b\frac{-i}{(p_1-k_2)^2}\nn\\
    &\times \left[\frac{i(-\slashed{p}_1+\slashed{k}_1+\slashed{k}_2+m)}{(p_1-k_1-k_2)^2-m^2}\gamma_\mu \frac{i(\slashed{k}_1+m)}{k_1^2-m^2}\right]_{ac}\nn\\
    &+ e^2\int [dk_1][dk_2]\delta(\bar{n}\cdot k_1-\ell_1^+)\delta(\bar{n}\cdot k_2-\ell_2^+)\left[\bar{u}(p_1)\gamma^\mu\frac{i(\slashed{k}_1+m)}{k_1^2-m^2}\right]_c \frac{-i}{(p_1-k_1)^2}\nn\\
    &\times \left[\frac{i(-\slashed{p}_1+\slashed{k}_1+\slashed{k}_2+m)}{(p_1-k_1-k_2)^2-m^2}\gamma_\mu \frac{i(\slashed{k}_2+m)}{k_2^2-m^2}\right]_{ab},\label{app:JfffII}
\end{align}
with $a, b$ and $c$ spinor indices.

%% file: Appendices/renormalization.tex
In this appendix we discuss the factorization in the collinear-anticollinear region for the renormalized 1PI amplitude. The factorization formula of eq.~\eqref{NLPfactorization} is understood to be written in terms of the bare amplitude $\mathcal{M}_{\rm bare}$, i.e.\ defined with bare fields, coupling constant $\alpha_0$ and mass $m_0$. In QED, a bare amplitude $\mathcal{M}_{\rm bare}$ is renormalized as
\begin{align}\label{eq:renormalizationQED}
\mathcal{M}_{\rm ren}(\alpha,m) &=  \left(\Pi_i\sqrt{Z_i}\right) \mathcal{M}_{\rm bare}(\alpha_0\to Z_{\alpha}\alpha,m_0 \to Z_mm) \nonumber \\
&= \mathcal{M}_{\rm bare} + \mathcal{M}_{\rm CT,m}  + \tilde{\mathcal{M}}_{\rm CT} \,,
\end{align}
where $Z_i$ is the wave function renormalization constant of the $i$-th external leg, $Z_{\alpha}$ is the renormalization constant for the QED coupling constant and $Z_m$ is the mass renormalization constant. $\mathcal{M}_{\rm CT,m}$ receives contributions from mass renormalization and mixing terms between mass renormalization and wave function and/or coupling constant renormalization beyond NLO in $\alpha$, while $\tilde{\mathcal{M}}_{\rm CT}$ includes only the remaining wave function and coupling constant renormalization. We emphasize that eq.~\eqref{eq:renormalizationQED} is written in terms of the 1PI amplitude. Note that $Z_i$ and $Z_{\alpha}$ serve as multiplicative factors and do not change the factorization structure of the amplitude. In the following we ignore $\tilde{\mathcal{M}}_{\rm CT}$ and only consider $\mathcal{M}_{\rm CT,m}$, i.e.\ the contribution involving mass renormalization, which does not contribute at LP, but starts to contribute at $\rm {\sqrt{N}LP}$ ($\mathcal{O}(\lambda)$). Given the above consideration, the LP factorization is not affected by mass counterterms and works already at bare level. We adopt the on-shell renormalization scheme for the mass renormalization such that $Z_m$ is given by
\begin{align}
Z_m & = 1 + A\left(m^2\right) + B\left(m^2\right) + \mathcal{O}\left(\alpha^2\right) \nonumber \\
&= 1 + \frac{\alpha}{4\pi}\left(\frac{\bar{\mu}^2}{m^2}\right)^\epsilon\left[-\frac{3}{\epsilon} - 4 - \epsilon\left(8 + \frac{3}{2}\zeta_2\right) - \epsilon^2\left(16 + 2\zeta_2 -\zeta_3\right) + \mathcal{O}\left(\epsilon^3\right)\right] + \mathcal{O}\left(\alpha^2\right) \,,
\end{align}
where $A\left(m^2\right)$ and $B\left(m^2\right)$ are given in eq.~\eqref{selfEn2}. Note that giving $Z_m$ up to $\cO(\alpha)$ is sufficient in our case, because our tree-level amplitude does not depend on the fermion mass explicitly. Similarly, from the factorization point of view, the renormalized amplitude should be written as
\begin{align}\label{eq:renormalizationfac}
\mathcal{M}_{\rm ren}(\alpha,m) &=  \left(\Pi_i\sqrt{Z'_i}\right) \mathcal{M}_{\rm fac, bare}(\alpha_0\to Z_{\alpha}\alpha,m_0 \to Z_mm) \nonumber \\
&= \mathcal{M}_{\rm fac, bare} + \mathcal{M}_{\rm fac,CT,m}  + \tilde{\mathcal{M}}_{\rm fac,CT} \,,
\end{align}
where $\cM_{\rm fac,bare}$ is the bare amplitude calculated from the factorization theorem. Note that here we renormalize $\cM_{\rm fac,bare}$ as a whole and do not renormalize the hard functions and jet functions individually, such that we have $Z'_i = Z_i$. 
Again, we only consider $\mathcal{M}_{\rm fac,CT,m}$ and ignore $\tilde{\mathcal{M}}_{\rm fac,CT}$ below.

Regarding the QED process defined in eq.~\eqref{process} and according to app.~\ref{app:regtofact}, the two-loop bare 1PI amplitude $\mathcal{M}^{\mu}_{\rm bare}$ in the $c\bar{c}$-region can be reconstructed by using the factorization formula eq.~\eqref{NLPfactorization} up to two extra non-factorized contributions, which are denoted here as $\mathcal{M}^{\mu}_{\rm a, bare}$ and $\mathcal{M}^{\mu}_{\rm b, bare}$. We therefore have
\begin{align}\label{eq:amp_fac_ccbar_1}
\mathcal{M}^{\mu}_{\rm bare}&= \mathcal{M}^{\mu}_{\rm fac, bare} + \mathcal{M}^{\mu}_{\rm a, bare} + \mathcal{M}^{\mu}_{\rm b, bare} \,,
\end{align}
where 
\begin{align}
\mathcal{M}^{\mu}_{\rm a, bare} &= -ie^5\int [dk_1][dk_2]\frac{S_c^\alpha}{D_c}\frac{2\hat{p}_{2\alpha}}{-2\hat{p}_2\cdot k_1}\gamma^\mu \frac{1}{\slashed{k}_1-\slashed{p}_2-m}\gamma_\sigma\frac{S^\sigma_{\bar c}}{D_{\bar c}},
\intertext{and}
\mathcal{M}^{\mu}_{\rm b, bare} &= -ie^5 \int [dk_1][dk_2]\frac{S_c^\rho}{D_c}\gamma_\rho\frac{1}{\slashed{k}_2+\slashed{p}_1-m}\gamma^\mu\frac{2\hat{p}_{1\beta}}{2\hat{p}_1\cdot k_2}\frac{S^\beta_{\bar c}}{D_{\bar c}}
\end{align}
account for the first and second term in eq.~\eqref{eq:extraterms}, respectively. $\mathcal{M}^{\rm fac}_{\rm bare}$ is given by the factorization formula in eq.~\eqref{NLPfactorization} and shown in eq.~\eqref{eq:ccbarfact} in detail. We now insert eq.~\eqref{eq:amp_fac_ccbar_1} into eq.~\eqref{eq:renormalizationQED} and compare the result with eq.~\eqref{eq:renormalizationfac}. Provided that the factorization theorem is true at two-loop level, we have 
\begin{align}\label{eq:check_fac_ccbar_1}
\mathcal{M}^{\mu}_{\rm fac,CT,m} = \mathcal{M}^{\mu}_{\rm a, bare} + \mathcal{M}^{\mu}_{\rm b, bare} + \mathcal{M}^{\mu}_{\rm CT,m} \,,
\end{align}
It is straightforward to verify eq.~\eqref{eq:check_fac_ccbar_1} at the level of the form factor, which translates \eqn{eq:check_fac_ccbar_1} into
\begin{align}\label{eq:check_fac_ff_ccbar_1}
F^{(2),\rm fac,CT, m}_{i,c\bar{c}} = F^{(2),\rm a, bare}_{i,c\bar{c}} + F^{(2),\rm b, bare}_{i,c\bar{c}} + F^{(2),\rm CT, m}_{i,c\bar{c}} \,.
\end{align}
The contributions of $\mathcal{M}^{\mu}_{\rm a/b, bare}$ to the form factors $F_1$ and $F_2$ in the $c\bar{c}$-region are given by
\begin{align}\label{eq:ff_ab_12}
F^{(2),\rm a, bare}_{1,c\bar{c}} &= F^{(2),\rm b, bare}_{1,c\bar{c}} = \left(\frac{\bar{\mu}^2}{m^2}\right)^{2\epsilon}\left[-\frac{9}{\epsilon^3} - \frac{12}{\epsilon^2} - \frac{1}{\epsilon}\left(9\zeta_2 + 72\right) -112 - 12\zeta_2 + 6\zeta_3\right], \nonumber \\
F^{(2),\rm a, bare}_{2,c\bar{c}} &= F^{(2),\rm b, bare}_{2,c\bar{c}} = \left(\frac{\bar{\mu}^2}{m^2}\right)^{2\epsilon}\left[\frac{6}{\epsilon^3} + \frac{20}{\epsilon^2} + \frac{1}{\epsilon}\left(6\zeta_2 + 56\right) + 144 + 20\zeta_2 - 4\zeta_3\right]\,.
\end{align}

\begin{figure}[t]
\centering
\includegraphics[width=0.6\linewidth, trim = 40mm 105mm 40mm 105mm, clip]{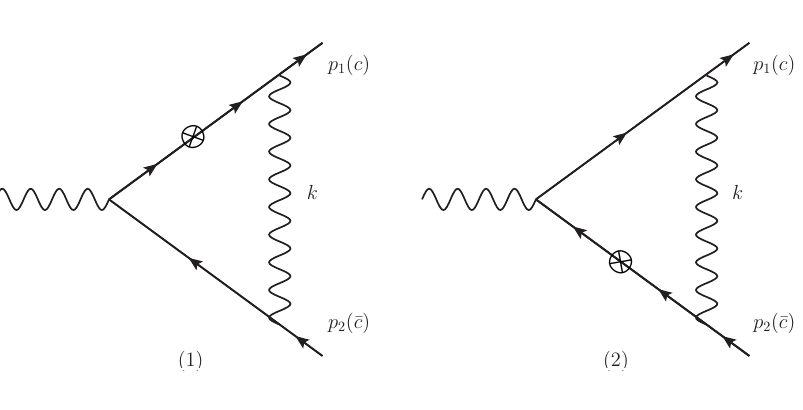}
\caption{One loop diagrams with mass counter terms insertion.}
\label{fig:oneloopCT1}
\end{figure}

The term $\mathcal{M}^{\mu}_{\rm CT,m}$ receives contributions from the two Feynman diagrams shown in fig.\ \ref{fig:oneloopCT1}, where $k$ is the loop momentum and $\otimes$ denotes the mass counterterms. If $\otimes$ is added to the collinear (anticollinear) leg as shown in diagram (1) (diagram (2)) in fig.\ \ref{fig:oneloopCT1}, it corresponds to a one-loop contribution in collinear (anticollinear) region. In addition to the mass counterterms, the two diagrams in fig.\ \ref{fig:oneloopCT1} represent the one-loop corrections induced by the loop momentum $k$. As discussed in sec.~\ref{sec:oneloop}, this loop momentum $k$ can be hard, collinear or anticollinear, which corresponds to respectively the hard, collinear and anticollinear region. Only the latter two contribute to the collinear-anticollinear region at two-loop level. Moreover, the Feynman diagram with $\otimes$ being added to the anticollinear leg and $k$ being anticollinear contributes to the double-anticollinear region, while the Feynman diagram with $\otimes$ being added to the collinear leg and $k$ being collinear contributes to double-collinear region.\footnote{In these two regions, both the l.h.s and the r.h.s of eq.~\eqref{NLPfactorization} receive the same contribution from mass renormalization, such that the UV renormalization does not change the factorization structure.} On the r.h.s.\ of eq.~\eqref{eq:check_fac_ccbar_1}, we therefore only need to consider the contribution from the anticollinear region in the Feynman diagram (1) and collinear region in the Feynman diagram (2). Their contributions to the form factor are given by
\begin{align}\label{eq:ff_ctm_12}
F^{(2),\rm CT, m}_{1,c\bar{c}} &= 2\left(\frac{\bar{\mu}^2}{m^2}\right)^{2\epsilon}\left[\frac{9}{\epsilon^3} + \frac{12}{\epsilon^2} + \frac{1}{\epsilon}\left(9\zeta_2 + 66\right) + 92 + 12\zeta_2 -6\zeta_3\right], \nonumber \\
F^{(2),\rm CT, m}_{2,c\bar{c}} &= 2\left(\frac{\bar{\mu}^2}{m^2}\right)^{2\epsilon}\left[-\frac{6}{\epsilon^3} - \frac{20}{\epsilon^2} - \frac{1}{\epsilon}\left(6\zeta_2 + 56\right) - 144 - 20\zeta_2 + 4\zeta_3\right]\,.
\end{align}
The factor of $2$ is because of symmetry between the two diagrams in fig.~\ref{fig:oneloopCT1}.
Moreover, we only need to consider the mass renormalization of hard functions on the l.h.s.\ of eq.~\eqref{eq:check_fac_ccbar_1}. Note that $H_{f,\bar{f}}^{(0)\mu}$, $H_{f\partial\gamma,\bar{f}\,\rho\sigma}^{(0)\mu}$, $H_{f,\bar{f}\partial\gamma\,\rho\sigma}^{(0)\mu}$ and $H_{f\gamma,\bar{f}\gamma\,\rho\sigma}^{(0)\mu}$ do not depend on the fermion mass $m$ explicitly. We therefore only need to consider the mass renormalization of the hard functions $H_{f\gamma,\bar{f}\,\rho}^{(1)\mu}$, defined in eq.~\eqref{eq:hardfgamma}, and $H_{f,\bar{f}\gamma\,\rho}^{(1)\mu}$. Together with the corresponding jet functions, we have 
\begin{align}
\mathcal{M}^{\mu}_{\rm fac,CT,m} &= J_{f\gamma}^{(1)\rho}\left(m_0\to m\right)H_{f\gamma,\bar{f}\,\rho}^{(1)\mu}\left(m_0\to (Z_m-1)m\right)J^{(0)}_{\bar{f}}\left(m_0\to m\right) \nonumber \\ 
&+ J^{(0)}_{f}\left(m_0\to m\right)H_{f,\bar{f}\gamma\,\rho}^{(1)\mu}\left(m_0\to (Z_m-1)m\right)J_{\bar{f}\gamma}^{(1)\rho}\left(m_0\to m\right)\,,
\end{align}
where the mass renormalization of jet functions is not involved as mentioned before. The contributions of $\mathcal{M}^{\mu}_{\rm fac,CT,m}$ to the form factors are given by
\begin{align}\label{eq:ff_ctm_fac_12}
F^{(2),\rm fac,CT, m}_{1,c\bar{c}} &= 2\left(\frac{\bar{\mu}^2}{m^2}\right)^{2\epsilon}\left( -\frac{6}{\epsilon} - 20\right) \nonumber \\
F^{(2),\rm fac,CT, m}_{2,c\bar{c}} &= 0\,.
\end{align}
Substituting eqs.~\eqref{eq:ff_ab_12},~\eqref{eq:ff_ctm_12} and~\eqref{eq:ff_ctm_fac_12} into eq.~\eqref{eq:check_fac_ff_ccbar_1}, we now conclude that our factorization formula indeed works in the $c\bar{c}$-region for the renormalized 1PI amplitude at two loops.

%% file: main.bbl
\providecommand{\href}[2]{#2}\begingroup\raggedright\begin{thebibliography}{100}

\bibitem{Low:1958sn}
F.~E. Low, {\it {Bremsstrahlung of very low-energy quanta in elementary particle collisions}},  {\em Phys. Rev.} {\bf 110} (1958) 974--977.

\bibitem{Burnett:1967km}
T.~H. Burnett and N.~M. Kroll, {\it {Extension of the low soft photon theorem}},  {\em Phys. Rev. Lett.} {\bf 20} (1968) 86.

\bibitem{DelDuca:1990gz}
V.~Del~Duca, {\it High-energy bremsstrahlung theorems for soft photons},  {\em Nucl. Phys.} {\bf B345} (1990) 369--388.

\bibitem{Laenen:2008ux}
E.~Laenen, L.~Magnea, and G.~Stavenga, {\it {On next-to-eikonal corrections to threshold resummation for the Drell-Yan and DIS cross sections}},  {\em Phys. Lett.} {\bf B669} (2008) 173--179, [\href{http://arxiv.org/abs/0807.4412}{{\tt arXiv:0807.4412}}].

\bibitem{Laenen:2010uz}
E.~Laenen, L.~Magnea, G.~Stavenga, and C.~D. White, {\it {Next-to-eikonal corrections to soft gluon radiation: a diagrammatic approach}},  {\em JHEP} {\bf 1101} (2011) 141, [\href{http://arxiv.org/abs/1010.1860}{{\tt arXiv:1010.1860}}].

\bibitem{Larkoski:2014bxa}
A.~J. Larkoski, D.~Neill, and I.~W. Stewart, {\it {Soft theorems from effective field theory}},  {\em JHEP} {\bf 1506} (2015) 077, [\href{http://arxiv.org/abs/1412.3108}{{\tt arXiv:1412.3108}}].

\bibitem{Bonocore:2014wua}
D.~Bonocore, E.~Laenen, L.~Magnea, L.~Vernazza, and C.~D. White, {\it {The method of regions and next-to-soft corrections in Drell-Yan production}},  {\em Phys. Lett.} {\bf B742} (2015) 375--382, [\href{http://arxiv.org/abs/1410.6406}{{\tt arXiv:1410.6406}}].

\bibitem{Ma:2014svb}
Y.-Q. Ma, J.-W. Qiu, G.~Sterman, and H.~Zhang, {\it {Factorized power expansion for high-$p_T$ heavy quarkonium production}},  {\em Phys. Rev. Lett.} {\bf 113} (2014), no.~14 142002, [\href{http://arxiv.org/abs/1407.0383}{{\tt arXiv:1407.0383}}].

\bibitem{Kang:2014pya}
Z.-B. Kang, Y.-Q. Ma, J.-W. Qiu, and G.~Sterman, {\it {Heavy Quarkonium Production at Collider Energies: Partonic Cross Section and Polarization}},  {\em Phys. Rev. D} {\bf 91} (2015), no.~1 014030, [\href{http://arxiv.org/abs/1411.2456}{{\tt arXiv:1411.2456}}].

\bibitem{Bonocore:2015esa}
D.~Bonocore, E.~Laenen, L.~Magnea, S.~Melville, L.~Vernazza, and C.~D. White, {\it {A factorization approach to next-to-leading-power threshold logarithms}},  {\em JHEP} {\bf 06} (2015) 008, [\href{http://arxiv.org/abs/1503.05156}{{\tt arXiv:1503.05156}}].

\bibitem{Bonocore:2016awd}
D.~Bonocore, E.~Laenen, L.~Magnea, L.~Vernazza, and C.~D. White, {\it {Non-abelian factorisation for next-to-leading-power threshold logarithms}},  {\em JHEP} {\bf 12} (2016) 121, [\href{http://arxiv.org/abs/1610.06842}{{\tt arXiv:1610.06842}}].

\bibitem{Moult:2016fqy}
I.~Moult, L.~Rothen, I.~W. Stewart, F.~J. Tackmann, and H.~X. Zhu, {\it {Subleading power corrections for N-jettiness subtractions}},  {\em Phys. Rev.} {\bf D95} (2017), no.~7 074023, [\href{http://arxiv.org/abs/1612.00450}{{\tt arXiv:1612.00450}}].

\bibitem{Feige:2017zci}
I.~Feige, D.~W. Kolodrubetz, I.~Moult, and I.~W. Stewart, {\it {A complete basis of helicity operators for subleading factorization}},  {\em JHEP} {\bf 1711} (2017) 142, [\href{http://arxiv.org/abs/1703.03411}{{\tt arXiv:1703.03411}}].

\bibitem{Moult:2017rpl}
I.~Moult, I.~W. Stewart, and G.~Vita, {\it {A subleading operator basis and matching for gg $\rightarrow$ H}},  {\em JHEP} {\bf 1707} (2017) 067, [\href{http://arxiv.org/abs/1703.03408}{{\tt arXiv:1703.03408}}].

\bibitem{Moult:2017jsg}
I.~Moult, L.~Rothen, I.~W. Stewart, F.~J. Tackmann, and H.~X. Zhu, {\it {N -jettiness subtractions for $gg\to H$ at subleading power}},  {\em Phys. Rev.} {\bf D97} (2018), no.~1 014013, [\href{http://arxiv.org/abs/1710.03227}{{\tt arXiv:1710.03227}}].

\bibitem{Gervais:2017yxv}
H.~Gervais, {\it {Soft photon theorem for high energy amplitudes in Yukawa and scalar theories}},  {\em Phys. Rev.} {\bf D95} (2017), no.~12 125009, [\href{http://arxiv.org/abs/1704.00806}{{\tt arXiv:1704.00806}}].

\bibitem{DelDuca:2017twk}
V.~Del~Duca, E.~Laenen, L.~Magnea, L.~Vernazza, and C.~D. White, {\it {Universality of next-to-leading power threshold effects for colourless final states in hadronic collisions}},  {\em JHEP} {\bf 11} (2017) 057, [\href{http://arxiv.org/abs/1706.04018}{{\tt arXiv:1706.04018}}].

\bibitem{Beneke:2017ztn}
M.~Beneke, M.~Garny, R.~Szafron, and J.~Wang, {\it {Anomalous dimension of subleading-power N-jet operators}},  {\em JHEP} {\bf 03} (2018) 001, [\href{http://arxiv.org/abs/1712.04416}{{\tt arXiv:1712.04416}}].

\bibitem{Moult:2018jjd}
I.~Moult, I.~W. Stewart, G.~Vita, and H.~X. Zhu, {\it {First Subleading Power Resummation for Event Shapes}},  {\em JHEP} {\bf 08} (2018) 013, [\href{http://arxiv.org/abs/1804.04665}{{\tt arXiv:1804.04665}}].

\bibitem{Beneke:2018rbh}
M.~Beneke, M.~Garny, R.~Szafron, and J.~Wang, {\it {Anomalous dimension of subleading-power $N$-jet operators. Part II}},  {\em JHEP} {\bf 11} (2018) 112, [\href{http://arxiv.org/abs/1808.04742}{{\tt arXiv:1808.04742}}].

\bibitem{Beneke:2018gvs}
M.~Beneke, A.~Broggio, M.~Garny, S.~Jaskiewicz, R.~Szafron, L.~Vernazza, and J.~Wang, {\it {Leading-logarithmic threshold resummation of the Drell-Yan process at next-to-leading power}},  {\em JHEP} {\bf 1903} (2019) 043, [\href{http://arxiv.org/abs/1809.10631}{{\tt arXiv:1809.10631}}].

\bibitem{Ebert:2018lzn}
M.~A. Ebert, I.~Moult, I.~W. Stewart, F.~J. Tackmann, G.~Vita, and H.~X. Zhu, {\it {Power Corrections for N-Jettiness Subtractions at ${\cal O}(\alpha_s)$}},  {\em JHEP} {\bf 12} (2018) 084, [\href{http://arxiv.org/abs/1807.10764}{{\tt arXiv:1807.10764}}].

\bibitem{Bahjat-Abbas:2019fqa}
N.~Bahjat-Abbas, D.~Bonocore, J.~Sinninghe~Damst\'e, E.~Laenen, L.~Magnea, L.~Vernazza, and C.~White, {\it {Diagrammatic resummation of leading-logarithmic threshold effects at next-to-leading power}},  {\em JHEP} {\bf 11} (2019) 002, [\href{http://arxiv.org/abs/1905.13710}{{\tt arXiv:1905.13710}}].

\bibitem{Moult:2019mog}
I.~Moult, I.~W. Stewart, and G.~Vita, {\it {Subleading Power Factorization with Radiative Functions}},  \href{http://arxiv.org/abs/1905.07411}{{\tt arXiv:1905.07411}}.

\bibitem{Beneke:2019kgv}
M.~Beneke, M.~Garny, R.~Szafron, and J.~Wang, {\it {Violation of the Kluberg-Stern-Zuber theorem in SCET}},  {\em JHEP} {\bf 09} (2019) 101, [\href{http://arxiv.org/abs/1907.05463}{{\tt arXiv:1907.05463}}].

\bibitem{Beneke:2019mua}
M.~Beneke, M.~Garny, S.~Jaskiewicz, R.~Szafron, L.~Vernazza, and J.~Wang, {\it {Leading-logarithmic threshold resummation of Higgs production in gluon fusion at next-to-leading power}},  {\em JHEP} {\bf 01} (2020) 094, [\href{http://arxiv.org/abs/1910.12685}{{\tt arXiv:1910.12685}}].

\bibitem{Liu:2019oav}
Z.~L. Liu and M.~Neubert, {\it {Factorization at subleading power and endpoint-divergent convolutions in $h\to\gamma\gamma$ decay}},  {\em JHEP} {\bf 04} (2020) 033, [\href{http://arxiv.org/abs/1912.08818}{{\tt arXiv:1912.08818}}].

\bibitem{Beneke:2019oqx}
M.~Beneke, A.~Broggio, S.~Jaskiewicz, and L.~Vernazza, {\it {Threshold factorization of the Drell-Yan process at next-to-leading power}},  {\em JHEP} {\bf 07} (2020) 078, [\href{http://arxiv.org/abs/1912.01585}{{\tt arXiv:1912.01585}}].

\bibitem{vanBeekveld:2019prq}
M.~van Beekveld, W.~Beenakker, E.~Laenen, and C.~D. White, {\it {Next-to-leading power threshold effects for inclusive and exclusive processes with final state jets}},  {\em JHEP} {\bf 03} (2020) 106, [\href{http://arxiv.org/abs/1905.08741}{{\tt arXiv:1905.08741}}].

\bibitem{vanBeekveld:2019cks}
M.~van Beekveld, W.~Beenakker, R.~Basu, E.~Laenen, A.~Misra, and P.~Motylinski, {\it {Next-to-leading power threshold effects for resummed prompt photon production}},  {\em Phys. Rev. D} {\bf 100} (2019), no.~5 056009, [\href{http://arxiv.org/abs/1905.11771}{{\tt arXiv:1905.11771}}].

\bibitem{Moult:2019uhz}
I.~Moult, I.~W. Stewart, G.~Vita, and H.~X. Zhu, {\it {The Soft Quark Sudakov}},  {\em JHEP} {\bf 05} (2020) 089, [\href{http://arxiv.org/abs/1910.14038}{{\tt arXiv:1910.14038}}].

\bibitem{Liu:2020ydl}
Z.~L. Liu and M.~Neubert, {\it {Two-Loop Radiative Jet Function for Exclusive $B$-Meson and Higgs Decays}},  {\em JHEP} {\bf 06} (2020) 060, [\href{http://arxiv.org/abs/2003.03393}{{\tt arXiv:2003.03393}}].

\bibitem{Liu:2020eqe}
Z.~L. Liu, B.~Mecaj, M.~Neubert, X.~Wang, and S.~Fleming, {\it {Renormalization and Scale Evolution of the Soft-Quark Soft Function}},  {\em JHEP} {\bf 07} (2020) 104, [\href{http://arxiv.org/abs/2005.03013}{{\tt arXiv:2005.03013}}].

\bibitem{AH:2020iki}
A.~H. Ajjath, P.~Mukherjee, and V.~Ravindran, {\it {Next to soft corrections to Drell-Yan and Higgs boson productions}},  {\em Phys. Rev. D} {\bf 105} (2022), no.~9 094035, [\href{http://arxiv.org/abs/2006.06726}{{\tt arXiv:2006.06726}}].

\bibitem{AH:2020xll}
A.~H. Ajjath, P.~Mukherjee, V.~Ravindran, A.~Sankar, and S.~Tiwari, {\it {On next to soft threshold corrections to DIS and SIA processes}},  {\em JHEP} {\bf 04} (2021) 131, [\href{http://arxiv.org/abs/2007.12214}{{\tt arXiv:2007.12214}}].

\bibitem{AH:2020qoa}
A.~H. Ajjath, P.~Mukherjee, V.~Ravindran, A.~Sankar, and S.~Tiwari, {\it {Next-to-soft corrections for Drell-Yan and Higgs boson rapidity distributions beyond N$^3$LO}},  {\em Phys. Rev. D} {\bf 103} (2021) L111502, [\href{http://arxiv.org/abs/2010.00079}{{\tt arXiv:2010.00079}}].

\bibitem{AH:2021kvg}
A.~H. Ajjath, P.~Mukherjee, V.~Ravindran, A.~Sankar, and S.~Tiwari, {\it {Next-to SV resummed Drell\textendash{}Yan cross section beyond leading-logarithm}},  {\em Eur. Phys. J. C} {\bf 82} (2022), no.~3 234, [\href{http://arxiv.org/abs/2107.09717}{{\tt arXiv:2107.09717}}].

\bibitem{Beneke:2020ibj}
M.~Beneke, M.~Garny, S.~Jaskiewicz, R.~Szafron, L.~Vernazza, and J.~Wang, {\it {Large-x resummation of off-diagonal deep-inelastic parton scattering from d-dimensional refactorization}},  {\em JHEP} {\bf 10} (2020) 196, [\href{http://arxiv.org/abs/2008.04943}{{\tt arXiv:2008.04943}}].

\bibitem{Laenen:2020nrt}
E.~Laenen, J.~Sinninghe~Damst\'e, L.~Vernazza, W.~Waalewijn, and L.~Zoppi, {\it {Towards all-order factorization of QED amplitudes at next-to-leading power}},  {\em Phys. Rev. D} {\bf 103} (2021), no.~3 034022, [\href{http://arxiv.org/abs/2008.01736}{{\tt arXiv:2008.01736}}].

\bibitem{Liu:2020tzd}
Z.~L. Liu, B.~Mecaj, M.~Neubert, and X.~Wang, {\it {Factorization at subleading power, Sudakov resummation, and endpoint divergences in soft-collinear effective theory}},  {\em Phys. Rev. D} {\bf 104} (2021), no.~1 014004, [\href{http://arxiv.org/abs/2009.04456}{{\tt arXiv:2009.04456}}].

\bibitem{vanBeekveld:2021hhv}
M.~van Beekveld, E.~Laenen, J.~Sinninghe~Damst\'e, and L.~Vernazza, {\it {Next-to-leading power threshold corrections for finite order and resummed colour-singlet cross sections}},  {\em JHEP} {\bf 05} (2021) 114, [\href{http://arxiv.org/abs/2101.07270}{{\tt arXiv:2101.07270}}].

\bibitem{Broggio:2021fnr}
A.~Broggio, S.~Jaskiewicz, and L.~Vernazza, {\it {Next-to-leading power two-loop soft functions for the Drell-Yan process at threshold}},  {\em JHEP} {\bf 10} (2021) 061, [\href{http://arxiv.org/abs/2107.07353}{{\tt arXiv:2107.07353}}].

\bibitem{vanBeekveld:2021mxn}
M.~van Beekveld, L.~Vernazza, and C.~D. White, {\it {Threshold resummation of new partonic channels at next-to-leading power}},  {\em JHEP} {\bf 12} (2021) 087, [\href{http://arxiv.org/abs/2109.09752}{{\tt arXiv:2109.09752}}].

\bibitem{Bonocore:2021qxh}
D.~Bonocore, A.~Kulesza, and J.~Pirsch, {\it {Classical and quantum gravitational scattering with Generalized Wilson Lines}},  {\em JHEP} {\bf 03} (2022) 147, [\href{http://arxiv.org/abs/2112.02009}{{\tt arXiv:2112.02009}}].

\bibitem{Engel:2021ccn}
T.~Engel, A.~Signer, and Y.~Ulrich, {\it {Universal structure of radiative QED amplitudes at one loop}},  {\em JHEP} {\bf 04} (2022) 097, [\href{http://arxiv.org/abs/2112.07570}{{\tt arXiv:2112.07570}}].

\bibitem{Beneke:2022obx}
M.~Beneke, M.~Garny, S.~Jaskiewicz, J.~Strohm, R.~Szafron, L.~Vernazza, and J.~Wang, {\it {Next-to-leading power endpoint factorization and resummation for off-diagonal \textquotedblleft{}gluon\textquotedblright{} thrust}},  {\em JHEP} {\bf 07} (2022) 144, [\href{http://arxiv.org/abs/2205.04479}{{\tt arXiv:2205.04479}}].

\bibitem{Bell:2022ott}
G.~Bell, P.~B\"oer, and T.~Feldmann, {\it {Muon-electron backward scattering: a prime example for endpoint singularities in SCET}},  {\em JHEP} {\bf 09} (2022) 183, [\href{http://arxiv.org/abs/2205.06021}{{\tt arXiv:2205.06021}}].

\bibitem{Liu:2022ajh}
Z.~L. Liu, M.~Neubert, M.~Schnubel, and X.~Wang, {\it {Factorization at next-to-leading power and endpoint divergences in gg \textrightarrow{} h production}},  {\em JHEP} {\bf 06} (2023) 183, [\href{http://arxiv.org/abs/2212.10447}{{\tt arXiv:2212.10447}}].

\bibitem{Engel:2023ifn}
T.~Engel, {\it {The LBK theorem to all orders}},  {\em JHEP} {\bf 07} (2023) 177, [\href{http://arxiv.org/abs/2304.11689}{{\tt arXiv:2304.11689}}].

\bibitem{Engel:2023rxp}
T.~Engel, {\it {Multiple soft-photon emission at next-to-leading power to all orders}},  {\em JHEP} {\bf 03} (2024) 004, [\href{http://arxiv.org/abs/2311.17612}{{\tt arXiv:2311.17612}}].

\bibitem{Broggio:2023pbu}
A.~Broggio, S.~Jaskiewicz, and L.~Vernazza, {\it {Threshold factorization of the Drell-Yan quark-gluon channel and two-loop soft function at next-to-leading power}},  \href{http://arxiv.org/abs/2306.06037}{{\tt arXiv:2306.06037}}.

\bibitem{vanBeekveld:2023gio}
M.~van Beekveld, A.~Danish, E.~Laenen, S.~Pal, A.~Tripathi, and C.~D. White, {\it {Next-to-soft radiation from a different angle}},  {\em Phys. Rev. D} {\bf 109} (2024), no.~7 074005, [\href{http://arxiv.org/abs/2308.12850}{{\tt arXiv:2308.12850}}].

\bibitem{vanBeekveld:2023liw}
M.~van Beekveld, L.~Vernazza, and C.~D. White, {\it {Exponentiation of soft quark effects from the replica trick}},  {\em JHEP} {\bf 07} (2024) 109, [\href{http://arxiv.org/abs/2312.11606}{{\tt arXiv:2312.11606}}].

\bibitem{Balsach:2023ema}
R.~Balsach, D.~Bonocore, and A.~Kulesza, {\it {Soft-photon spectra and the Low-Burnett-Kroll theorem}},  {\em Phys. Rev. D} {\bf 110} (2024), no.~1 016029, [\href{http://arxiv.org/abs/2312.11386}{{\tt arXiv:2312.11386}}].

\bibitem{Beneke:2024cpq}
M.~Beneke, Y.~Ji, and X.~Wang, {\it {Renormalization of the next-to-leading-power \ensuremath{\gamma}\ensuremath{\gamma} \textrightarrow{} h and gg \textrightarrow{} h soft quark functions}},  {\em JHEP} {\bf 05} (2024) 246, [\href{http://arxiv.org/abs/2403.17738}{{\tt arXiv:2403.17738}}].

\bibitem{Beneke:2025ufd}
M.~Beneke, Y.~Ji, E.~S\"underhauf, and X.~Wang, {\it {Renormalization of the Next-to-Leading-Power Soft Function for the Drell-Yan Off-diagonal Channel}},  \href{http://arxiv.org/abs/2502.01973}{{\tt arXiv:2502.01973}}.

\bibitem{Mueller:1979ih}
A.~H. Mueller, {\it On the asymptotic behavior of the sudakov form-factor},  {\em Phys. Rev.} {\bf D20} (1979) 2037.

\bibitem{Collins:1980ih}
J.~C. Collins, {\it Algorithm to compute corrections to the sudakov form- factor},  {\em Phys. Rev.} {\bf D22} (1980) 1478.

\bibitem{Collins:1981uk}
J.~C. Collins and D.~E. Soper, {\it Back to back jets in qcd},  {\em Nucl. Phys.} {\bf B193} (1981) 381.

\bibitem{Sen:1981sd}
A.~Sen, {\it Asymptotic behavior of the sudakov form-factor in qcd},  {\em Phys. Rev.} {\bf D24} (1981) 3281.

\bibitem{Collins:1989bt}
J.~C. Collins, {\it {Sudakov form-factors}},  {\em Adv.Ser.Direct.High Energy Phys.} {\bf 5} (1989) 573--614, [\href{http://arxiv.org/abs/hep-ph/0312336}{{\tt hep-ph/0312336}}].

\bibitem{Dixon:2008gr}
L.~J. Dixon, L.~Magnea, and G.~F. Sterman, {\it {Universal structure of subleading infrared poles in gauge theory amplitudes}},  {\em JHEP} {\bf 0808} (2008) 022, [\href{http://arxiv.org/abs/0805.3515}{{\tt arXiv:0805.3515}}].

\bibitem{Aybat:2006mz}
S.~M. Aybat, L.~J. Dixon, and G.~F. Sterman, {\it {The Two-loop soft anomalous dimension matrix and resummation at next-to-next-to leading pole}},  {\em Phys. Rev.} {\bf D74} (2006) 074004, [\href{http://arxiv.org/abs/hep-ph/0607309}{{\tt hep-ph/0607309}}].

\bibitem{Gardi:2009qi}
E.~Gardi and L.~Magnea, {\it {Factorization constraints for soft anomalous dimensions in QCD scattering amplitudes}},  {\em JHEP} {\bf 0903} (2009) 079, [\href{http://arxiv.org/abs/0901.1091}{{\tt arXiv:0901.1091}}].

\bibitem{Becher:2009qa}
T.~Becher and M.~Neubert, {\it {On the Structure of Infrared Singularities of Gauge-Theory Amplitudes}},  {\em JHEP} {\bf 0906} (2009) 081, [\href{http://arxiv.org/abs/0903.1126}{{\tt arXiv:0903.1126}}].

\bibitem{Magnea:1990zb}
L.~Magnea and G.~F. Sterman, {\it {Analytic continuation of the Sudakov form-factor in QCD}},  {\em Phys. Rev.} {\bf D42} (1990) 4222--4227.

\bibitem{Contopanagos:1996nh}
H.~Contopanagos, E.~Laenen, and G.~F. Sterman, {\it {Sudakov factorization and resummation}},  {\em Nucl. Phys. B} {\bf 484} (1997) 303--330, [\href{http://arxiv.org/abs/hep-ph/9604313}{{\tt hep-ph/9604313}}].

\bibitem{Kidonakis:1997gm}
N.~Kidonakis and G.~Sterman, {\it Resummation for qcd hard scattering},  {\em Nucl. Phys.} {\bf B505} (1997) 321--348, [\href{http://arxiv.org/abs/hep-ph/9705234}{{\tt hep-ph/9705234}}].

\bibitem{Kidonakis:1998bk}
N.~Kidonakis, G.~Oderda, and G.~Sterman, {\it Threshold resummation for dijet cross-sections},  {\em Nucl. Phys.} {\bf B525} (1998) 299, [\href{http://arxiv.org/abs/hep-ph/9801268}{{\tt hep-ph/9801268}}].

\bibitem{Magnea:2000ss}
L.~Magnea, {\it {Analytic resummation for the quark form-factor in QCD}},  {\em Nucl. Phys. B} {\bf 593} (2001) 269--288, [\href{http://arxiv.org/abs/hep-ph/0006255}{{\tt hep-ph/0006255}}].

\bibitem{Catani:2000ef}
S.~Catani, S.~Dittmaier, and Z.~Trocsanyi, {\it {One loop singular behavior of QCD and SUSY QCD amplitudes with massive partons}},  {\em Phys. Lett. B} {\bf 500} (2001) 149--160, [\href{http://arxiv.org/abs/hep-ph/0011222}{{\tt hep-ph/0011222}}].

\bibitem{Penin:2005eh}
A.~A. Penin, {\it {Two-loop photonic corrections to massive Bhabha scattering}},  {\em Nucl. Phys. B} {\bf 734} (2006) 185--202, [\href{http://arxiv.org/abs/hep-ph/0508127}{{\tt hep-ph/0508127}}].

\bibitem{Penin:2005kf}
A.~A. Penin, {\it {Two-loop corrections to Bhabha scattering}},  {\em Phys. Rev. Lett.} {\bf 95} (2005) 010408, [\href{http://arxiv.org/abs/hep-ph/0501120}{{\tt hep-ph/0501120}}].

\bibitem{Mitov:2006xs}
A.~Mitov and S.~Moch, {\it {The Singular behavior of massive QCD amplitudes}},  {\em JHEP} {\bf 05} (2007) 001, [\href{http://arxiv.org/abs/hep-ph/0612149}{{\tt hep-ph/0612149}}].

\bibitem{Becher:2007cu}
T.~Becher and K.~Melnikov, {\it {Two-loop QED corrections to Bhabha scattering}},  {\em JHEP} {\bf 06} (2007) 084, [\href{http://arxiv.org/abs/0704.3582}{{\tt arXiv:0704.3582}}].

\bibitem{Czakon:2007ej}
M.~Czakon, A.~Mitov, and S.~Moch, {\it {Heavy-quark production in massless quark scattering at two loops in QCD}},  {\em Phys. Lett. B} {\bf 651} (2007) 147--159, [\href{http://arxiv.org/abs/0705.1975}{{\tt arXiv:0705.1975}}].

\bibitem{Czakon:2007wk}
M.~Czakon, A.~Mitov, and S.~Moch, {\it {Heavy-quark production in gluon fusion at two loops in QCD}},  {\em Nucl. Phys. B} {\bf 798} (2008) 210--250, [\href{http://arxiv.org/abs/0707.4139}{{\tt arXiv:0707.4139}}].

\bibitem{Mitov:2009sv}
A.~Mitov, G.~F. Sterman, and I.~Sung, {\it {The Massive Soft Anomalous Dimension Matrix at Two Loops}},  {\em Phys. Rev.} {\bf D79} (2009) 094015, [\href{http://arxiv.org/abs/0903.3241}{{\tt arXiv:0903.3241}}].

\bibitem{Becher:2009kw}
T.~Becher and M.~Neubert, {\it {Infrared singularities of QCD amplitudes with massive partons}},  {\em Phys. Rev.} {\bf D79} (2009) 125004, [\href{http://arxiv.org/abs/0904.1021}{{\tt arXiv:0904.1021}}].

\bibitem{Banerjee:2020rww}
P.~Banerjee, T.~Engel, A.~Signer, and Y.~Ulrich, {\it {QED at NNLO with McMule}},  {\em SciPost Phys.} {\bf 9} (2020) 027, [\href{http://arxiv.org/abs/2007.01654}{{\tt arXiv:2007.01654}}].

\bibitem{Wang:2023qbf}
G.~Wang, T.~Xia, L.~L. Yang, and X.~Ye, {\it {On the high-energy behavior of massive QCD amplitudes}},  {\em JHEP} {\bf 05} (2024) 082, [\href{http://arxiv.org/abs/2312.12242}{{\tt arXiv:2312.12242}}].

\bibitem{Wang:2024pmv}
G.~Wang, T.~Xia, L.~L. Yang, and X.~Ye, {\it {Two-loop QCD amplitudes for $ t\overline{t}H $ production from boosted limit}},  {\em JHEP} {\bf 07} (2024) 121, [\href{http://arxiv.org/abs/2402.00431}{{\tt arXiv:2402.00431}}].

\bibitem{Sterman:1987aj}
G.~Sterman, {\it Summation of large corrections to short distance hadronic cross-sections},  {\em Nucl. Phys.} {\bf B281} (1987) 310.

\bibitem{Catani:1989ne}
S.~Catani and L.~Trentadue, {\it {Resummation of the QCD perturbative series for hard processes}},  {\em Nucl. Phys.} {\bf B327} (1989) 323.

\bibitem{Catani:1990rp}
S.~Catani and L.~Trentadue, {\it Comment on qcd exponentiation at large x},  {\em Nucl. Phys.} {\bf B353} (1991) 183--186.

\bibitem{Korchemsky:1993xv}
G.~P. Korchemsky and G.~Marchesini, {\it Structure function for large x and renormalization of wilson loop},  {\em Nucl. Phys.} {\bf B406} (1993) 225--258, [\href{http://arxiv.org/abs/hep-ph/9210281}{{\tt hep-ph/9210281}}].

\bibitem{Korchemsky:1993uz}
G.~P. Korchemsky and G.~Marchesini, {\it {Resummation of large infrared corrections using Wilson loops}},  {\em Phys. Lett.} {\bf B313} (1993) 433--440.

\bibitem{Bauer:2000yr}
C.~W. Bauer, S.~Fleming, D.~Pirjol, and I.~W. Stewart, {\it An effective field theory for collinear and soft gluons: Heavy to light decays},  {\em Phys. Rev.} {\bf D63} (2001) 114020, [\href{http://arxiv.org/abs/hep-ph/0011336}{{\tt hep-ph/0011336}}].

\bibitem{Bauer:2001yt}
C.~W. Bauer, D.~Pirjol, and I.~W. Stewart, {\it {Soft collinear factorization in effective field theory}},  {\em Phys. Rev. D} {\bf 65} (2002) 054022, [\href{http://arxiv.org/abs/hep-ph/0109045}{{\tt hep-ph/0109045}}].

\bibitem{Beneke:2002ph}
M.~Beneke, A.~P. Chapovsky, M.~Diehl, and T.~Feldmann, {\it {Soft collinear effective theory and heavy to light currents beyond leading power}},  {\em Nucl. Phys. B} {\bf 643} (2002) 431--476, [\href{http://arxiv.org/abs/hep-ph/0206152}{{\tt hep-ph/0206152}}].

\bibitem{Moult:2015aoa}
I.~Moult, I.~W. Stewart, F.~J. Tackmann, and W.~J. Waalewijn, {\it {Employing Helicity Amplitudes for Resummation}},  {\em Phys. Rev. D} {\bf 93} (2016), no.~9 094003, [\href{http://arxiv.org/abs/1508.02397}{{\tt arXiv:1508.02397}}].

\bibitem{Kolodrubetz:2016uim}
D.~W. Kolodrubetz, I.~Moult, and I.~W. Stewart, {\it {Building blocks for subleading helicity operators}},  {\em JHEP} {\bf 1605} (2016) 139, [\href{http://arxiv.org/abs/1601.02607}{{\tt arXiv:1601.02607}}].

\bibitem{Qiu:1990xxa}
J.-w. Qiu and G.~F. Sterman, {\it {Power corrections in hadronic scattering. 1. Leading 1/Q**2 corrections to the Drell-Yan cross-section}},  {\em Nucl. Phys. B} {\bf 353} (1991) 105--136.

\bibitem{Qiu:1990xy}
J.-w. Qiu and G.~F. Sterman, {\it {Power corrections to hadronic scattering. 2. Factorization}},  {\em Nucl. Phys. B} {\bf 353} (1991) 137--164.

\bibitem{Bernreuther:2004ih}
W.~Bernreuther, R.~Bonciani, T.~Gehrmann, R.~Heinesch, T.~Leineweber, P.~Mastrolia, and E.~Remiddi, {\it {Two-loop QCD corrections to the heavy quark form-factors: The Vector contributions}},  {\em Nucl. Phys. B} {\bf 706} (2005) 245--324, [\href{http://arxiv.org/abs/hep-ph/0406046}{{\tt hep-ph/0406046}}].

\bibitem{Gluza:2009yy}
J.~Gluza, A.~Mitov, S.~Moch, and T.~Riemann, {\it {The QCD form factor of heavy quarks at NNLO}},  {\em JHEP} {\bf 07} (2009) 001, [\href{http://arxiv.org/abs/0905.1137}{{\tt arXiv:0905.1137}}].

\bibitem{Blumlein:2020jrf}
J.~Bl\"umlein, A.~De~Freitas, C.~Raab, and K.~Sch\"onwald, {\it {The $O(\alpha^2)$ initial state QED corrections to $e^+e^- \rightarrow \gamma^*/Z_0^*$}},  {\em Nucl. Phys. B} {\bf 956} (2020) 115055, [\href{http://arxiv.org/abs/2003.14289}{{\tt arXiv:2003.14289}}].

\bibitem{Henn:2016kjz}
J.~M. Henn, A.~V. Smirnov, and V.~A. Smirnov, {\it {Analytic results for planar three-loop integrals for massive form factors}},  {\em JHEP} {\bf 12} (2016) 144, [\href{http://arxiv.org/abs/1611.06523}{{\tt arXiv:1611.06523}}].

\bibitem{Henn:2016tyf}
J.~Henn, A.~V. Smirnov, V.~A. Smirnov, and M.~Steinhauser, {\it {Massive three-loop form factor in the planar limit}},  {\em JHEP} {\bf 01} (2017) 074, [\href{http://arxiv.org/abs/1611.07535}{{\tt arXiv:1611.07535}}].

\bibitem{Ablinger:2017hst}
J.~Ablinger, A.~Behring, J.~Bl\"umlein, G.~Falcioni, A.~De~Freitas, P.~Marquard, N.~Rana, and C.~Schneider, {\it {Heavy quark form factors at two loops}},  {\em Phys. Rev. D} {\bf 97} (2018), no.~9 094022, [\href{http://arxiv.org/abs/1712.09889}{{\tt arXiv:1712.09889}}].

\bibitem{Lee:2018nxa}
R.~N. Lee, A.~V. Smirnov, V.~A. Smirnov, and M.~Steinhauser, {\it {Three-loop massive form factors: complete light-fermion corrections for the vector current}},  {\em JHEP} {\bf 03} (2018) 136, [\href{http://arxiv.org/abs/1801.08151}{{\tt arXiv:1801.08151}}].

\bibitem{Lee:2018rgs}
R.~N. Lee, A.~V. Smirnov, V.~A. Smirnov, and M.~Steinhauser, {\it {Three-loop massive form factors: complete light-fermion and large-N$_{c}$ corrections for vector, axial-vector, scalar and pseudo-scalar currents}},  {\em JHEP} {\bf 05} (2018) 187, [\href{http://arxiv.org/abs/1804.07310}{{\tt arXiv:1804.07310}}].

\bibitem{Ablinger:2018yae}
J.~Ablinger, J.~Bl\"umlein, P.~Marquard, N.~Rana, and C.~Schneider, {\it {Heavy quark form factors at three loops in the planar limit}},  {\em Phys. Lett. B} {\bf 782} (2018) 528--532, [\href{http://arxiv.org/abs/1804.07313}{{\tt arXiv:1804.07313}}].

\bibitem{Blumlein:2018tmz}
J.~Bl\"umlein, P.~Marquard, and N.~Rana, {\it {Asymptotic behavior of the heavy quark form factors at higher order}},  {\em Phys. Rev. D} {\bf 99} (2019), no.~1 016013, [\href{http://arxiv.org/abs/1810.08943}{{\tt arXiv:1810.08943}}].

\bibitem{Blumlein:2019oas}
J.~Bl\"umlein, P.~Marquard, N.~Rana, and C.~Schneider, {\it {The Heavy Fermion Contributions to the Massive Three Loop Form Factors}},  {\em Nucl. Phys. B} {\bf 949} (2019) 114751, [\href{http://arxiv.org/abs/1908.00357}{{\tt arXiv:1908.00357}}].

\bibitem{Fael:2022rgm}
M.~Fael, F.~Lange, K.~Sch\"onwald, and M.~Steinhauser, {\it {Massive Vector Form Factors to Three Loops}},  {\em Phys. Rev. Lett.} {\bf 128} (2022), no.~17 172003, [\href{http://arxiv.org/abs/2202.05276}{{\tt arXiv:2202.05276}}].

\bibitem{Fael:2022miw}
M.~Fael, F.~Lange, K.~Sch\"onwald, and M.~Steinhauser, {\it {Singlet and nonsinglet three-loop massive form factors}},  {\em Phys. Rev. D} {\bf 106} (2022), no.~3 034029, [\href{http://arxiv.org/abs/2207.00027}{{\tt arXiv:2207.00027}}].

\bibitem{Fael:2023zqr}
M.~Fael, F.~Lange, K.~Sch\"onwald, and M.~Steinhauser, {\it {Massive three-loop form factors: Anomaly contribution}},  {\em Phys. Rev. D} {\bf 107} (2023), no.~9 094017, [\href{http://arxiv.org/abs/2302.00693}{{\tt arXiv:2302.00693}}].

\bibitem{Blumlein:2023uuq}
J.~Bl\"umlein, A.~De~Freitas, P.~Marquard, N.~Rana, and C.~Schneider, {\it {Analytic results on the massive three-loop form factors: Quarkonic contributions}},  {\em Phys. Rev. D} {\bf 108} (2023), no.~9 094003, [\href{http://arxiv.org/abs/2307.02983}{{\tt arXiv:2307.02983}}].

\bibitem{terHoeve:2023ehm}
J.~ter Hoeve, E.~Laenen, C.~Marinissen, L.~Vernazza, and G.~Wang, {\it {Region analysis of QED massive fermion form factor}},  {\em JHEP} {\bf 02} (2024) 024, [\href{http://arxiv.org/abs/2311.16215}{{\tt arXiv:2311.16215}}].

\bibitem{Honemann:2018mrb}
I.~H\"onemann, K.~Tempest, and S.~Weinzierl, {\it {Electron self-energy in QED at two loops revisited}},  {\em Phys. Rev. D} {\bf 98} (2018), no.~11 113008, [\href{http://arxiv.org/abs/1811.09308}{{\tt arXiv:1811.09308}}]. [Erratum: Phys.Rev.D 110, 059901 (2024)].

\bibitem{Hill:2002vw}
R.~J. Hill and M.~Neubert, {\it {Spectator interactions in soft collinear effective theory}},  {\em Nucl. Phys. B} {\bf 657} (2003) 229--256, [\href{http://arxiv.org/abs/hep-ph/0211018}{{\tt hep-ph/0211018}}].

\bibitem{Magnea:2018ebr}
L.~Magnea, E.~Maina, G.~Pelliccioli, C.~Signorile-Signorile, P.~Torrielli, and S.~Uccirati, {\it {Factorisation and Subtraction beyond NLO}},  {\em JHEP} {\bf 12} (2018) 062, [\href{http://arxiv.org/abs/1809.05444}{{\tt arXiv:1809.05444}}].

\bibitem{Becher:2009th}
T.~Becher and M.~D. Schwartz, {\it {Direct photon production with effective field theory}},  {\em JHEP} {\bf 02} (2010) 040, [\href{http://arxiv.org/abs/0911.0681}{{\tt arXiv:0911.0681}}].

\bibitem{Becher:2010pd}
T.~Becher and G.~Bell, {\it {The gluon jet function at two-loop order}},  {\em Phys. Lett. B} {\bf 695} (2011) 252--258, [\href{http://arxiv.org/abs/1008.1936}{{\tt arXiv:1008.1936}}].

\bibitem{PhysRev.51.125}
W.~H. Furry, {\it A symmetry theorem in the positron theory},  {\em Phys. Rev.} {\bf 51} (Jan, 1937) 125--129.

\bibitem{Klappert:2020nbg}
J.~Klappert, F.~Lange, P.~Maierh\"ofer, and J.~Usovitsch, {\it {Integral reduction with Kira 2.0 and finite field methods}},  {\em Comput. Phys. Commun.} {\bf 266} (2021) 108024, [\href{http://arxiv.org/abs/2008.06494}{{\tt arXiv:2008.06494}}].

\bibitem{Maierhofer:2017gsa}
P.~Maierh\"ofer, J.~Usovitsch, and P.~Uwer, {\it {Kira\textemdash{}A Feynman integral reduction program}},  {\em Comput. Phys. Commun.} {\bf 230} (2018) 99--112, [\href{http://arxiv.org/abs/1705.05610}{{\tt arXiv:1705.05610}}].

\bibitem{Vermaseren:2000nd}
J.~A.~M. Vermaseren, {\it New features of form},  \href{http://arxiv.org/abs/math-ph/0010025}{{\tt math-ph/0010025}}.

\bibitem{Ruijl:2017dtg}
B.~Ruijl, T.~Ueda, and J.~Vermaseren, {\it {FORM version 4.2}},  \href{http://arxiv.org/abs/1707.06453}{{\tt arXiv:1707.06453}}.

\bibitem{Vermaseren:1998uu}
J.~A.~M. Vermaseren, {\it {Harmonic sums, Mellin transforms and integrals}},  {\em Int. J. Mod. Phys. A} {\bf 14} (1999) 2037--2076, [\href{http://arxiv.org/abs/hep-ph/9806280}{{\tt hep-ph/9806280}}].

\bibitem{Czakon:2005rk}
M.~Czakon, {\it {Automatized analytic continuation of Mellin-Barnes integrals}},  {\em Comput. Phys. Commun.} {\bf 175} (2006) 559--571, [\href{http://arxiv.org/abs/hep-ph/0511200}{{\tt hep-ph/0511200}}].

\bibitem{Ochman:2015fho}
M.~Ochman and T.~Riemann, {\it {MBsums - a Mathematica package for the representation of Mellin-Barnes integrals by multiple sums}},  {\em Acta Phys. Polon.} {\bf B46} (2015), no.~11 2117, [\href{http://arxiv.org/abs/1511.01323}{{\tt arXiv:1511.01323}}].

\bibitem{Huber:2005yg}
T.~Huber and D.~Maitre, {\it {HypExp: A Mathematica package for expanding hypergeometric functions around integer-valued parameters}},  {\em Comput. Phys. Commun.} {\bf 175} (2006) 122--144, [\href{http://arxiv.org/abs/hep-ph/0507094}{{\tt hep-ph/0507094}}].

\bibitem{Becher:2011dz}
T.~Becher and G.~Bell, {\it {Analytic Regularization in Soft-Collinear Effective Theory}},  {\em Phys. Lett. B} {\bf 713} (2012) 41--46, [\href{http://arxiv.org/abs/1112.3907}{{\tt arXiv:1112.3907}}].

\bibitem{Gritschacher:2013tza}
S.~Gritschacher, A.~Hoang, I.~Jemos, and P.~Pietrulewicz, {\it {Two loop soft function for secondary massive quarks}},  {\em Phys. Rev. D} {\bf 89} (2014), no.~1 014035, [\href{http://arxiv.org/abs/1309.6251}{{\tt arXiv:1309.6251}}].

\bibitem{Grammer:1973db}
J.~Grammer, G. and D.~Yennie, {\it {Improved treatment for the infrared divergence problem in quantum electrodynamics}},  {\em Phys. Rev.} {\bf D8} (1973) 4332--4344.

\end{thebibliography}\endgroup
